\RequirePackage{pdf14}

\documentclass[a4paper,11pt]{article}
\pdfoutput=1

\usepackage{array,setspace,mathrsfs,amsfonts,amsmath,yfonts,dsfont,bbm,colonequals,amscd}
\usepackage{relsize,suffix,mathtools,cancel,bbm,tikz-cd}
\usepackage{subcaption}

\usepackage{jheppub} 

\usepackage[T1]{fontenc} 
\usepackage{subcaption}
\graphicspath{ {graphs/} }

\usepackage{xcolor}

\preprint{PUPT-2588,\\
	\phantom{~} \hfill  TCDMATH 19-XX}

\abstract{We develop new techniques to compute five-point correlation functions from IIB supergravity on $AdS_5\times S^5$. Our methods rely entirely on symmetry and general consistency conditions, and eschew detailed knowledge of the supergravity effective action. We demonstrate our methods by computing the five-point function of the $\mathbf{20'}$ operator, which is the superconformal primary of the stress tensor multiplet. We also develop systematic methods to compute the five-point conformal blocks in series expansions. Using the explicit expressions of the conformal blocks, we perform an Euclidean OPE analysis of the $\mathbf{20'}$ five-point function. We find expected agreement with non-renormalized quantities  and also extract new CFT data at strong coupling.}

\title{\boldmath \Large
$20'$ Five-Point Function from $AdS_5\times S^5$ Supergravity 
 }

\author[a]{Vasco Gon\c{c}alves,}
\author[b]{Raul Pereira,}
\author[c]{Xinan Zhou}

\affiliation[a]{ICTP South American Institute for Fundamental Research, IFT-UNESP, \\S\~ao Paulo, SP Brazil 01440-070}
\affiliation[b]{School of Mathematics and Hamilton Mathematics Institute, Trinity College Dublin,\\
Dublin 2, Ireland}
\affiliation[c]{Princeton Center for Theoretical Science, Princeton University, \\Princeton, NJ 08544, U.S.A.}

\emailAdd{vasco.dfg@gmail.com}
\emailAdd{raul@maths.tcd.ie}
\emailAdd{xinanz@princeton.edu}

\begin{document}
\maketitle
\flushbottom

\section{Introduction}\label{introduction}
Recent years have witnessed a resurgence of activity in studying holographic correlation functions using the AdS/CFT correspondence.  An abundance of interesting new results has been obtained by leveraging modern techniques, thanks to an inflow of ideas and technologies from the conformal bootstrap and the scattering amplitude program. The progress is especially evident in the paradigmatic example of 4d $\mathcal{N}=4$ Super Yang-Mills theory, which is dual to IIB string theory on $AdS_5\times S^5$. At the level of two-derivative supergravity, {\it all} the four-point functions of one-half BPS  operators have been obtained at subleading order in $1/N$ by solving an algebraic bootstrap problem in Mellin space \cite{Rastelli:2016nze,Rastelli:2017udc}.\footnote{See \cite{Arutyunov:2017dti,Arutyunov:2018neq,Arutyunov:2018tvn} for several highly nontrivial checks of this result by explicit supergravity calculations.} The  complete set of tree-level four-point correlators contains a wealth of physical information. However, extracting the data is still highly nontrivial since the double-trace operators in the operator product expansion have degenerate contributions. One therefore needs to solve the associated mixing problem by exploiting the knowledge of all four-point functions. In \cite{Alday:2017xua,Aprile:2017bgs,Aprile:2017xsp} machinery for performing a systematic analysis was developed, and the complete anomalous dimension spectrum of double-trace operators has been obtained \cite{Aprile:2018efk}. The tree-level data in turn allows one to further obtain one-loop results, by using ``AdS unitarity methods''  \cite{Aharony:2016dwx,Alday:2017xua,Aprile:2017bgs,Aprile:2017xsp,Alday:2017vkk,Aprile:2017qoy}. More precisely, this is achieved by feeding the anomalous dimensions and OPE coefficients into the crossing equation, and focusing on the double-discontinuity \cite{Caron-Huot:2017vep}. The one-loop four-point functions are rather cumbersome in position space. However, the corresponding Mellin amplitudes look remarkably simple \cite{Alday:2018kkw}, suggesting that the Mellin representation remains a natural language beyond tree level. Furthermore, four-point functions receive higher-derivative corrections from AdS string theory, which are suppressed by inverse powers of the 't Hooft coupling. These stringy effects have recently been studied in \cite{Alday:2018pdi,Binder:2019jwn} at both tree level and at one loop\footnote{See also \cite{Goncalves:2014ffa} for earlier discussion at the tree level.}, showing an interesting interplay between Mellin amplitudes and flat space scattering amplitudes. These results  shed new light on quantum gravity from the CFT perspective, and constitute new precision tests of the AdS/CFT correspondence. While the $AdS_5\times S^5$ background has attracted most attention,  many interesting results have been obtained for other string theory/M-theory backgrounds as well. See \cite{Zhou:2017zaw,Rastelli:2017ymc,Heslop:2017sco,Chester:2018lbz,Zhou:2018ofp,Chester:2018aca,Chester:2018dga,Binder:2018yvd,Giusto:2018ovt,Abl:2019jhh,Rastelli:2019gtj,Giusto:2019pxc} for some recent developments.

In this paper we will initiate a systematic study of five-point functions from tree-level IIB supergravity on $AdS_5\times S^5$, as a first step towards extending the above program to arbitrary $n$-point functions. There are several motivations for considering higher-point correlators. First of all, a very practical reason to study holographic correlators is to extract CFT data at strong coupling. Considering higher-point correlation functions of one-half BPS operators allows us to access new unprotected data not contained in their four-point functions. This becomes especially clear when we look at OPE limits. For example, by taking the OPE limit for one pair of operators in the five-point function, we can obtain four-point functions with one unprotected double-trace operator. These four-point functions encode infinitely many new unprotected three-point functions, which can be extracted after taking another OPE limit. Secondly, previous studies of four-point functions suggested an intricate relation between holographic correlators and scattering amplitudes in  flat space. Many aspects of holographic correlators appear to be analogous to the ones in flat space. We would like to further explore these connections and sharpen the analogies, by studying five-point functions. In particular, we will demonstrate how factorization, an important tool for flat space amplitudes, can be used to understand the structure of correlation functions from AdS supergravity. Finally, the study of correlators at strong coupling is motivated by the possibility of discovering unexpected structures. Recently, it was observed that tree-level one-half BPS four-point functions from $AdS_5\times S^5$ exhibit a hidden ten dimensional conformal symmetry \cite{Caron-Huot:2018kta}.\footnote{See also \cite{Rastelli:2019gtj} for an analogous story in $AdS_3\times S^3$ where a hidden six dimensional conformal symmetry emerges in the tree-level supergravity four-point correlators.} In terms of this symmetry, four-point functions of different conformal dimensions can all be related to each other. It is interesting to see if such a symmetry also exists in higher-point correlation functions. This may shed some light on its mysterious origin. 

Results of five-point functions at strong coupling are scarce. To classify them, it is useful to grade the correlators by ascending extremality $E$, which is defined by $2E=\sum_{i=1}^{4}k_i-k_5$. Here $k_i$ are the scaling dimensions of the operators and we have assumed $k_5$ is the largest. R-symmetry selection rules require $E$ to take integer values. For $E=0$ and $E=1$, the five-point functions are called {\it extremal} and {\it next-to-extremal}. In these cases, it is known that the five-point functions are protected by non-renormalization theorems \cite{DHoker:1999jke,Bianchi:1999ie,Eden:1999kw,Erdmenger:1999pz,Eden:2000gg} and therefore can be obtained from the free theory. If we further increase $E$ by one, the five-point functions are no longer protected and start to become nontrivial. It was argued in \cite{DHoker:2000xhf} that such five-point functions should have factorized structures and can be expressed in terms of lower-point correlators. These near-extremal correlators ($E=0,1,2$) however are very special, and the derivation of these results (from the bulk side) rely heavily on the fact that extremal couplings vanish.\footnote{The vanishing of extremal couplings is a self-consistency condition. This is because extremal contact Witten diagrams are divergent but the effective action should be finite.} When the extremality is further increased, {\it i.e.}, $E\geq 3$, one encounters the generic case and no such simplification exists. One would imagine that examples of generic five-point functions may have been computed using the traditional algorithm of Witten diagram expansion. However the traditional algorithm is too complicated to be a practical recipe. Implementing this method requires inputting all the precise vertices, which can be in principle obtained from expanding the effective supergravity Lagrangian to the quintic order. Such an expansion is devilishly complicated and has never been attempted in the literature.

In this paper, we will develop new techniques for computing five-point correlators with arbitrary extremality. Since a brute force approach is not viable, our strategy is to avoid the details of the effective Lagrangian as much as possible. We accomplish this by using  superconformal symmetry and self-consistency conditions, in the same spirit of \cite{Rastelli:2016nze,Rastelli:2017udc}. Let us sketch the methods and state our main results.  For simplicity and concreteness, we will focus on the five-point function of the $\mathbf{20'}$ operator, which is the bottom component of the stress tensor multiplet.\footnote{The five-point function $\langle\mathcal{O}_{\mathbf{20'}}\mathcal{O}_{\mathbf{20'}}\mathcal{O}_{\mathbf{20'}}\mathcal{O}_{\mathbf{20'}}\mathcal{O}_{\mathbf{20'}}\rangle$ has extremality $E=3$, and therefore belongs to the generic case.
} Although the methods will be phrased in this particular context, it will be clear that they can be applied to general five-point correlators after some obvious modifications. The starting point of our method is an ansatz which splits into a singular part and a regular part. The singular part includes all possible exchange Witten diagrams, and the regular part contains all possible contact Witten diagrams. The coefficient of each diagram could be computed if the vertices were known, but we will leave them as undetermined coefficients. To solve this ansatz, we use superconformal symmetry and self-consistency conditions. The singular part can be uniquely fixed by using factorization in AdS space. Roughly stated, the factorization condition means that the ``residue'' of the five-point function at an internal bulk-to-bulk propagator is a ``product'' of three-point functions and four-point functions. This can be stated more precisely in Mellin space \cite{Goncalves:2014rfa}. To fix the regular part, we use the chiral algebra twist \cite{Beem:2013sza}, which predicts that the twisted five-point function is the same as in the free theory. This fixes all but one coefficient in the ansatz, which multiplies a structure insensitive to the chiral algebra constraint. The last coefficient can be determined by further using an independent topological twist \cite{Drukker:2009sf} which involves the entire $SO(6)$ R-symmetry group.
This gives the complete answer to the $\mathbf{20'}$ five-point function from AdS supergravity, and is one of the main results of this paper. We will also discuss a variation of this method which starts with an ansatz in Mellin space. The alternative method avoids certain position space calculations and is more suitable for generalizing to higher-weight five-point functions. The final result is expressed as a Mellin amplitude in \eqref{MellinResult}, and takes a very compact form. As a technical development, we have also set up systematic methods to compute five-point conformal blocks in series expansions. This allows us to perform a conformal block decomposition for the $\mathbf{20'}$ five-point function and  extract new data. For simplicity, we looked at the Euclidean OPE and restricted our attention to the singular and leading regular terms. By taking a single OPE limit, we obtain a new four-point function with three $\mathbf{20'}$ one-half BPS operators and one unprotected double-trace operator.
The result can be compactly written as a combination of $D$-functions, which is presented in \eqref{new4pt}. By taking a double OPE limit, we extract various three-point functions. The protected three-point functions we found are in perfect agreement with their free theory values, which constitute nontrivial consistency checks of our result. We also extract a new unprotected three-point function  \eqref{OCC} involving one $\mathbf{20'}$ one-half BPS operator and two operators from semi-short multiplets. 
The unprotected three- and four-point functions give new predictions of $\mathcal{N}=4$ SYM at strong coupling. We hope these results can one day be compared with the integrability program.

The rest of the paper is organized as follows. In Section \ref{kinematics} we discuss the superconformal kinematics of the five-point function. In Section \ref{Mellinfactorization} we review the Mellin representation and the factorization of Mellin amplitudes. After these preparations, we introduce our position space method in Section \ref{5ptcorrelator} and compute the five-point function of the $\mathbf{20'}$ operator. In Section \ref{Mellinansatz} we point out an alternative approach using Mellin space, which simplifies some calculations in position space. The result for the five-point function is analyzed in Section \ref{CFTdata}, where we perform consistency checks and extract new CFT data. Various technical details are relegated to the appendices.

\section{Superconformal kinematics}\label{kinematics}
The $\mathbf{20'}$ operator $\mathcal{O}_\mathbf{20'}^{IJ}=tr (\Phi^{\{I}\Phi^{J\}})$  has protected conformal dimension $\Delta=2$ and transforms in the rank-2 symmetric traceless representation of $SO(6)_R$. It is the superconformal primary of the 1/2-BPS multiplet which also contains the R-symmetry current $\mathcal{J}_\mu^{[IJ]}$  and the stress tensor $\mathcal{T}_{\mu\nu}$. Our primary object of study is the five-point correlation function of $\mathbf{20'}$ operators
\begin{equation}
\langle \mathcal{O}_\mathbf{20'}^{I_1J_1}(x_1) \mathcal{O}_\mathbf{20'}^{I_2J_2}(x_2) \mathcal{O}_\mathbf{20'}^{I_3J_3}(x_3) \mathcal{O}_\mathbf{20'}^{I_4J_4}(x_4) \mathcal{O}_\mathbf{20'}^{I_5J_5}(x_5)\rangle\;.
\end{equation}
It is convenient to absorb the R-symmetry indices by contracting with null R-symmetry vectors $t^I$
\begin{equation}
\mathcal{O}_\mathbf{20'}(x,t)\equiv \mathcal{O}_\mathbf{20'}^{IJ}t^It^J\;,\quad t^It^I=0\;.
\end{equation}
The contraction automatically projects the operator into the symmetric traceless representation and turns the five-point correlator into a scalar function which depends not only on the spacetime coordinates but also the R-symmetry coordinates
\begin{equation}
G_5(x_i,t_i)=\langle\mathcal{O}_\mathbf{20'}(x_1,t_1)\mathcal{O}_\mathbf{20'}(x_2,t_2)\mathcal{O}_\mathbf{20'}(x_3,t_3)\mathcal{O}_\mathbf{20'}(x_4,t_4)\mathcal{O}_\mathbf{20'}(x_5,t_5) \rangle\;.
\end{equation}
It is easy to see that the null vectors can only appear in $G_5(x_i,t_i)$
as polynomials of $t_{ij}\equiv t_i\cdot t_j$. Moreover, $G_5(x_i,t_i)$ is subject to the homogeneity condition that under $t_i\to \lambda_i t_i$
\begin{equation}
G_5(x_i,t_i)\to \lambda_1^2\lambda_2^2\lambda_3^2\lambda_4^2\lambda_5^2 G_5(x_i,t_i)
\end{equation}
where the $\lambda_i$ are independent. A basis of R-symmetry structures  for $G_5$ is then given by all the monomials of $t_{ij}$ satisfying the homogeneity condition. 
\begin{figure}
	\centering
	\begin{tikzpicture}
	\draw (180:1)  node[label=left : $i$] {} --  (108:1)  node[label=above : $j$] {} -- (36:1)  node[label=right : $k$] {} -- (-36:1)  node[label=right : $l$] {} -- (-108:1)  node[label=below : $m$] {} -- cycle ;
	\node at (0.2,-2) {$A_{(ijklm)}$};
	\draw[xshift=5cm] (-180:1)   node[label=left : $i$] {} -- (60:1)   node[label=above : $j$] {} -- (-60:1)   node[label=below : $k$] {} -- cycle ;
	\node at (5.6,-2) {$A_{(ijk)(lm)}$};
	\draw[xshift=6.2 cm] (0, .90)   node[label=above : $l$] {}  .. controls (-.2,0.5) and (-.2,-0.5) .. (0,-.95)  node[label=below : $m$] {} ;
	\draw[xshift=6.2 cm] (0,.90) .. controls (.2,.5) and (.2,-.5) .. (0,-.95);
	\end{tikzpicture}
	\caption{Each line between points $i$ and $j$ corresponds to a factor of $t_{ij}$. We can then see that there are two types of structures, $A_{(ijklm)}$ and $A_{(ijk)(lm)}$, with one and two closed cycles respectively. Note that each cycle is invariant under cyclic permutations and reflection, so the number of independent structures is given by the number of ways to distribute the points into the cycles, modulo those symmetries.}\label{Rbasis}
\end{figure}
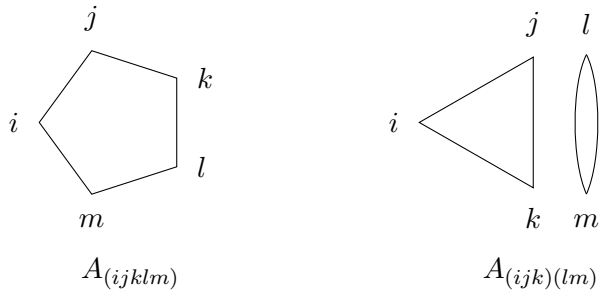
There are 22 such terms, which can be dividided according to the lengths of their cycles, as explained in Figure \ref{Rbasis}
\begin{align}\label{RstructuresA}
	A_{(ijklm)}&= t_{ij} t_{jk} t_{kl} t_{lm} t_{mi} \,, \nonumber\\
	A_{(ijk)(lm)}&= t_{ij} t_{jk} t_{ki} t_{lm}^2\,.
\end{align}
The basis vectors are in one-to-one correspondence with the inequivalent Wick contractions of the $\Phi^I$ fields in the free theory limit. 
We can parameterize the five-point function such that each R-symmetry structure is multiplied with a function of the spacetime coordinates. However, the full correlator is invariant under permutations of the five external operators. As we can easily check, crossing symmetry permutes separately the R-symmetry structures $\{A_{(ijklm)}\}$ and  $\{A_{(ijk)(lm)}\}$. Therefore the various functions multiplying the different R-symmetry monomials in the same group are interrelated under crossing, and in the end there are only two independent functions of spacetime coordinates in the five-point function. 
We can further exploit the conformal  
covariance to extract a kinematic factor from the correlator
\begin{equation}\label{G5cr}
G_5 = \frac{x_{13}^2}{x_{12}^4 x_{35}^4 x_{14}^2 x_{34}^2} \mathcal G_5(V_i ; t_i) \;.
\end{equation}
The kinematic factor takes care of the covariance under conformal 
transformations, and the five-point correlator becomes a function of the five conformal cross ratios 
\footnote{Similarly, the correlator depends also on five R-symmetry cross ratios, which can be chosen as
	\begin{equation}
	\sigma_1=\frac{t_{25} t_{34}}{t_{24} t_{35}}\;,\quad \sigma_2=\frac{t_{31} t_{45}}{t_{35} t_{14}}\;,\quad \sigma_3=\frac{t_{24} t_{15}}{t_{14} t_{25}}\;,\quad \sigma_4=\frac{t_{12} t_{35}}{t_{25} t_{13}}\;,\quad \sigma_5=\frac{t_{14} t_{23}}{t_{13} t_{24}}\;.
	\end{equation}
	}
\begin{equation}\label{CrossRatios}
V_1= \frac{x_{12}^2 x_{34}^2}{x_{13}^2 x_{24}^2}\,,\quad V_2= \frac{x_{14}^2 x_{23}^2}{x_{13}^2 x_{24}^2}\,,\quad
V_3= \frac{x_{14}^2 x_{35}^2}{x_{13}^2 x_{45}^2}\,,\quad V_4= \frac{x_{15}^2 x_{34}^2}{x_{13}^2 x_{45}^2}\,,\quad
V_5= \frac{x_{12}^2 x_{35}^2}{x_{13}^2 x_{25}^2}\,.
\end{equation}

So far we have exploited only the bosonic part of the superconformal group $PSU(2,2|4)$. The fermionic charges impose further constraints on the correlator. One such constraint comes from the special properties of 1/2-BPS correlation functions under twisting a subalgebra $\mathfrak{su}(1,1|2)\subset \mathfrak{psu}(2,2|4)$, known as the chiral algebra twist \cite{Beem:2013sza}. In order to perform the twist, we restrict all five operators to a two-dimensional plane inside $\mathbb{R}^4$. This allows us to parameterize the positions of these operators in terms of the holomorphic and anti-holomorphic coordinates $z_i$, $\bar{z}_i$. The $t^5$ and $t^6$ components of the six-dimensional null vector $t^I$ are set to zero, reducing the null vector into a four-dimensional one denoted as $t^\mu$. The vector $t^\mu$ can be further written as the product of a pair of spinors
\begin{equation}
t^\mu=\sigma^\mu_{\alpha\dot{\alpha}}\,v^\alpha \bar{v}^{\dot{\alpha}}\;.
\end{equation}
Rescaling $v$ and $\bar{v}$ amounts to multiplying $t$ with a number, which does not change the null vector since it is defined modulo rescaling.  Both $v^\alpha$ and $\bar{v}^{\dot{\alpha}}$ therefore have only one degree of freedom, and can be written as 
\begin{equation}
v_i=\left(\begin{array}{c}1 \\ y_i \end{array}\right)\;,\quad \bar{v}_i=\left(\begin{array}{c}1 \\ \bar{y}_i \end{array}\right)\;.
\end{equation}
When the R-symmetry orientations of the operators are correlated with the positions on the plane
\begin{equation}
\bar{y}_i=\bar{z}_i\;,
\end{equation}
the construction of \cite{Beem:2013sza} dictates that the twisted five-point function becomes a holomorphic function of the $z_i$ only 
\begin{equation}
G_5(z_i,\bar{z}_i;y_i,\bar{y}_i=\bar{z}_i)=g(z_i;y_i)\;.
\end{equation}
Moreover, the twisted correlator is independent of the marginal coupling, and therefore equal to its free field value
\begin{equation}
g(z_i;y_i)=g_{free}(z_i;y_i)\;.
\end{equation}
The holomorphic requirement of the twisted correlator imposes nontrivial constraints on the structure of the correlation function $G_5$. A similar twist applies to the holomorphic variables. 

Another important constraint comes from performing an independent topological twist which probes the full $SO(6)$ $R$-symmetry group \cite{Drukker:2009sf}. Unlike in the chiral algebra twist, operators are inserted at generic points $x_i\in \mathbb{R}^4$, with position-dependent polarizations 
\begin{equation}\label{SO6twist}
t=(2ix^1,2ix^2,2ix^3,2ix^4,i(1-(x^\mu)^2),1+(x^\mu)^2)\;.
\end{equation} 
Such twisted $n$-point correlation functions preserve two common supercharges\footnote{When there are $n<5$ points there is more supersymmetry preserved.}. Moreover, the twisted translations and the exactly marginal deformation are exact with respect to the preserved the supercharges. The $SO(6)$ twisted correlators are therefore topological and protected. All in all, the topological twist imposes the constraint that
\begin{equation}\label{5ptSO6twisted}
G_5 (x_i, t_{ij}= x_{ij}^2) = \frac{20 \sqrt 2}{N} + \frac{48 \sqrt 2}{N^3}
\end{equation}
where the two-point function of $\mathcal{O}_{\bf 20'}$ is unit normalized, and the twisted five-point function is computed in the free theory.

A small comment is in order. For correlation functions of two, three and four 1/2-BPS operators, it is possible to show that the constraints derived from the chiral algebra twist have exhausted the full constraining power of superconformal symmetry. In particular, the requirement of the twisted four-point function being a holomorphic function is equivalent to the superconformal Ward identity \cite{Eden:2000bk,Nirschl:2004pa}. On the other hand, the chiral algebra twist leads only to a subset of the full superconformal constraints for correlation functions with five points or more.  The $SO(6)$ twist of \cite{Drukker:2009sf} imposes extra constraints which are not captured by the chiral algebra twist. It is an interesting question for the future to explore the full consequence of superconformal symmetry on five-point and higher-point correlation functions.

\section{Mellin representation and factorization}\label{Mellinfactorization}
The goal of this section is to give a brief review on the Mellin representation formalism \cite{Mack:2009mi,Mack:2009gy,Penedones:2010ue} and the factorization properties of Mellin amplitudes \cite{Fitzpatrick:2011ia,Goncalves:2014rfa}. Mellin amplitudes for scalar operators  are defined as an integral transform of the correlation function
\begin{align}
\langle \mathcal{O}_1\dots \mathcal{O}_n \rangle = \int [d\gamma]M(\gamma_{ij})\prod_{1\leq i <j \leq n}\Gamma(\gamma_{ij})(x_{ij}^2)^{-\gamma_{ij}} \;,
\end{align} 
where the integration variables satisfy the constraint $\sum_{i}\gamma_{ij}=0$, with $\gamma_{ii}=-\Delta_i$, ensuring the correct scaling of the external operators. 

Correlation functions for operators with spin are more easily expressed using the embedding space formalism for CFTs, see, {\it e.g.}, \cite{Costa:2011mg}  for a detailed account.  In this formalism, each point in $\mathbb{R}^d$ is mapped  to a null ray through the origin in $\mathbb{R}^{d+1,1}$, and the action of the conformal group is linearized as the Lorentz rotations in the embedding space $\mathbb{R}^{d+1,1}$. A primary operator with dimension $\Delta$ and spin $J$ in $\mathbb{R}^d$ is mapped to a field in $\mathbb{R}^{d+1,1}$ depending on both null rays $P$ and $Z$ and satisfying 
\begin{align}
\mathcal{O}(\lambda P,\alpha Z) = \lambda^{-\Delta}\alpha^J\mathcal{O}(P,Z), \ \ \ \  Z\cdot P = 0\;, 
\end{align}
and 
\begin{align}
\mathcal{O}(P,Z+\beta P)=\mathcal{O}(P,Z)\label{eq:transverse}.
\end{align}
These properties guarantee that the operator is symmetric, traceless and transverse. Now we can define the Mellin amplitude for one operator with spin $J$ and $n$ scalar operators as
\begin{align}
&\langle \mathcal{O}(P,Z) \mathcal{O}_1(P_1)\dots \mathcal{O}_n(P_n) \rangle\nonumber\\
& = \sum_{a_1,\dots,a_J=1}^n \left(\prod_{k=1}^J Z\cdot P_{a_k}\right)\int [d\gamma] M^{\{a\}}\prod_{\substack{i,j=1\\i<j}}^n\frac{\Gamma(\gamma_{ij})}{(-2P_i\cdot P_j)^{\gamma_{ij}}}\prod_{i=1}^n\frac{\Gamma(\gamma_i+\{a\}_i)}{(-2P_i\cdot P)^{\gamma_i+\{a\}_i}}\label{eq:MellindefinitionSpin}
\end{align}
where $\{a\}$ stands for the set of indices $a_1\dots a_J$,  and $\{a\}_i$ counts the number of times that the index $i$ appears in the set $a_1,\dots a_J$
\begin{align}
\{a\}_i\equiv{ \delta}_{a_1}^i+\dots+\delta_{a_J}^i\;.
\end{align}
Moreover, we impose  
\begin{align}\label{gammas}
\gamma_i=-\sum_{j=1}^n\gamma_{ij}\,, \qquad \gamma_{ii}=-\Delta_i \,,\qquad \sum_{i,j=1}^n\gamma_{ij}=J-\Delta \,,
\end{align}
such that the correlator has the correct scalings.

The transverse property (\ref{eq:transverse}) is not automatically satisfied by the Mellin amplitude $M^{\{a\}}$, instead it implies that
\begin{align}
\sum_{a_1=1}^n(\gamma_{a_1}+\delta_{a_1}^{a_2}+\delta_{a_1}^{a_3}+\dots+\delta_{a_1}^{a_J})M^{a_1a_2\dots a_J}=0 \;.\label{eq:FactorizationTransversality}
\end{align}
In the case where the spinning operator is conserved, {\it i.e.}, $\Delta=d-2+J$,  the Mellin amplitude has to satisfy one further constraint 
\begin{align}
2J\sum_{a,b=1}^{n}\gamma_{ab}[M^{ac_2\dots c_J}]^{ab}=(J-1)\sum_{a,b=1}^n\gamma_{ab}[M^{abc_3\dots c_J}]^{ab} \;,
\end{align}
with 
\begin{align}
[M(\gamma_{ij})]^{ab}\equiv M(\gamma_{ij}+\delta_{i}^{a}\delta_{j}^{b}+\delta_{j}^a\delta_i^{b}). 
\end{align}

The operator product expansion is one of the most important properties of a conformal field theory as it allows to write a product of $k$ local operators at different positions in terms of an infinite sum of local operators
\begin{align}
\mathcal{O}_1(x_1)\dots \mathcal{O}_k(x_k)=\sum_p C_{\mu_1\dots \mu_J}^{(1\dots k,p)}(x_1,\dots,x_k,y,\partial_{y})\mathcal{O}_p^{\mu_1\dots \mu_J}(y)
\end{align}
where the position $y$ is arbitrary as long as it stays within a sphere that encircles all  $k$ local operators. This expansion can be used inside a correlation function,  effectively rewriting an $n$-point correlation function as a sum of products of $(k+1)$- and $(n-k+1)$-point functions.

This property implies that the Mellin amplitude is an analytic function of the Mellin variables $\gamma_{ij}$ with at most simple poles 
\begin{align}
M\approx \frac{\mathcal{Q}^J_{m}}{\gamma_{LR}-(\Delta-J+2m)},\, \ \ \ m=0,1,2,\dots, \,\,\, \gamma_{LR}=\sum_{a=1}^k\sum_{i=k+1}^n \gamma_{ai}\;,
\end{align}
where the residue $\mathcal{Q}^J_m$ depends on the product of lower-point Mellin amplitudes. Each pole is associated with the contribution of an exchanged primary operator ($m=0$), or a descendant ($m>0$) with twist $\tau=(\Delta-J)+m$. For instance, for an exchanged scalar operator it is given by
\begin{align}
&\mathcal{Q}^0_0=-2\Gamma(\Delta)M_L(\gamma_{ab})M_R(\gamma_{ij})
\end{align}
where we only spelled out the $m=0$ since it will be enough for this work. The Mellin amplitudes $M_L$ and $M_R$ are defined as
\begin{align}
\langle \mathcal{O}_1(P_1)\dots \mathcal{O}_k(P_k)\mathcal{O}(P_0)\rangle &=\int [d\lambda] M_L(\lambda_{ab})\prod_{1\leq a <b \leq k}\frac{\Gamma(\lambda_{ab})}{P_{ab}^{\lambda_{ab}}}\prod_{1\leq a \leq k}\frac{\Gamma(\lambda_a)}{P_{a0}^{\lambda_a}} 
\\
\langle \mathcal{O}_1(P_{k+1})\dots \mathcal{O}_k(P_n)\mathcal{O}(P_0)\rangle &=\int [d\rho] M_L(\rho_{ij})\prod_{k+1\leq i<j \leq n}\frac{\Gamma(\rho_{ij})}{P_{ij}^{\rho_{ij}}}\prod_{k+1\leq i \leq n}\frac{\Gamma(\rho_i)}{P_{i0}^{\rho_i}}
\end{align}
where 
\begin{align}
\lambda_a=-\sum_{b=1}^{k}\lambda_{ab}, \,\,\lambda_{aa}=-\Delta_a\,\,\,\, \sum_{a,b=1}^k\lambda_{ab}=-\Delta
\end{align}
and analogously for $\rho$. We also use the notation where $a,b$ label the first $k$ operators while $i,j$ label the remaining $n-k$ operators.

The residue $\mathcal{Q}^J_m$ associated with the exchange of an operator with spin depends on mixed Mellin variables where both types of indices appear. For the exchange of a vector operator the residue is given by
\begin{align}
\mathcal{Q}^1_0=\sum_{a=1}^k\sum_{i=k+1}^n\gamma_{ai}\,M^{a}_L \,M_R^{i} \,,
\end{align}
while for the exchange of a spin $2$ operator it is given by
\begin{align}
\mathcal{Q}^2_0=-\frac{(\Delta+1)\Gamma(\Delta-1)}{2}\sum_{a,b=1}^k\sum_{i,j=k+1}^n\gamma_{ai}(\gamma_{bj}+\delta_{b}^a\delta_{j}^i)\,M_L^{ab}\,M_R^{ij}.
\end{align}
The residues for any $m$ and up to spin $2$ as well as any spin and $m=0$ have been obtained in \cite{Goncalves:2014rfa} but they are not needed for this work.

\section{$\mathbf{20'}$ five-point function from supergravity}\label{5ptcorrelator}
\subsection{Outline of strategy}
\begin{figure}[h]
 
\begin{subfigure}{0.32\textwidth}
\includegraphics[width=0.8\textwidth]{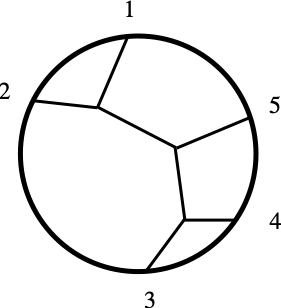} 
\caption{}
\label{fig:ssDE}
\end{subfigure}
\begin{subfigure}{0.32\textwidth}
\includegraphics[width=0.8\textwidth]{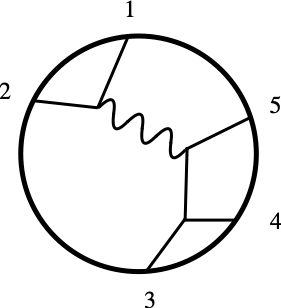}
\caption{}
\label{fig:svDE}
\end{subfigure}
\begin{subfigure}{0.32\textwidth}
\includegraphics[width=0.8\textwidth]{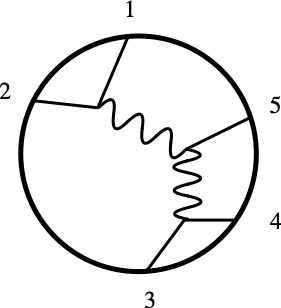}
\caption{}
\label{fig:vvDE}
\end{subfigure}
\begin{subfigure}{0.32\textwidth}
\includegraphics[width=0.8\textwidth]{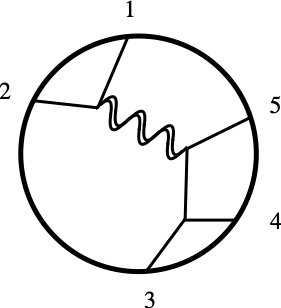}
\caption{}
\label{fig:gsDE}
\end{subfigure}

\caption{The four types of double-exchange Witten diagrams allowed by R-symmetry selection rules. The straight, curly and double curly lines correspondingly represent the scalar, graviphoton and graviton field.}
\label{fig:DE}
\end{figure}

\begin{figure}[h]
 
\begin{subfigure}{0.32\textwidth}
\includegraphics[width=0.8\textwidth]{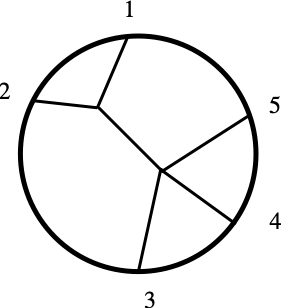} 
\caption{}
\label{fig:sSE}
\end{subfigure}
\begin{subfigure}{0.32\textwidth}
\includegraphics[width=0.8\textwidth]{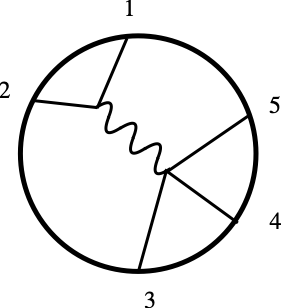}
\caption{}
\label{fig:vSE}
\end{subfigure}
\begin{subfigure}{0.32\textwidth}
\includegraphics[width=0.8\textwidth]{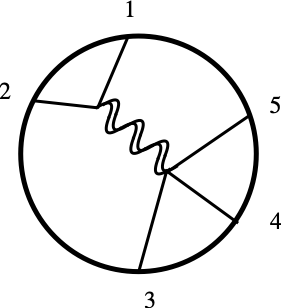}
\caption{}
\label{fig:gSE}
\end{subfigure}
 
\caption{The three types of single-exchange Witten diagrams  allowed by R-symmetry selection rules. Here we have suppressed the derivative information in the quartic vertices.}
\label{fig:SE}
\end{figure}

\begin{figure}[hbp]
\begin{center}
\includegraphics[width=0.256\textwidth]{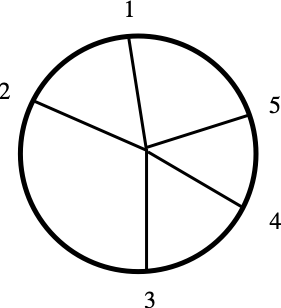}
\caption{A contact Witten diagram. The information of derivatives in the quintic vertex is also suppressed in the diagram.}
\label{fig:CON}
\end{center}
\end{figure}

Using the holographic dictionary, correlators of the boundary theory can be computed from IIB supergravity on $AdS_5\times S^5$ by performing a sum over all the possible Witten diagrams. This is the traditional algorithm of computing holographic correlators. The connected component of the five-point correlator receives leading contribution from the tree-level Witten diagrams at the order $\mathcal{O}(1/N^3)$ (Figure \ref{fig:DE}, \ref{fig:SE}, \ref{fig:CON}).\footnote{There is also a disconnected part of order $\mathcal{O}(1/N)$, which consists of products of two-point functions with three-point functions. The disconnected component is trivial to compute, since it coincides with the free field value thanks to the non-renormalization theorems of 1/2-BPS two and three-point functions.} When the external operators are the $\mathbf{20'}$ operators, the only relevant bulk fields are a scalar field $s^I$, the graviphoton $V^a_\mu$ and the graviton $\varphi_{\mu\nu}$, thanks to the AdS selection rules (see Section 2 of \cite{Rastelli:2017udc} for a detailed account), while all massive KK modes decouple.  Equivalently, the tree-level correlator of $\mathbf{20'}$  operators can be  computed from the 5d $\mathcal{N}=8$ gauged supergravity, which is a consistent truncation of the KK-reduced IIB supergravity theory. These fields have the quantum numbers displayed in Table \ref{sugrafields}, and are respectively dual to the $\mathbf{20'}$ scalar $\mathcal{O}_\mathbf{20'}$, the R-symmetry current $\mathcal{J}_\mu$ and the stress tensor $\mathcal{T}_{\mu\nu}$ of the boundary theory. The tree-level Witten diagrams are classified according to the number of internal lines and consist of double-exchange diagrams (Figure \ref{fig:DE}), single-exchange diagrams (Figure \ref{fig:SE}) and contact diagrams (Figure \ref{fig:CON}). 

\begin{table}[htp]
\begin{center}
\begin{tabular}{|c|c|c|c|}\hline
supergravity fields & dimension $\Delta$ & spin $\ell$& $SU(4)_R$ representation\\
\hline
scalar: $s^I$& 2&0&$[0,2,0]$\\
\hline
graviphoton: $V^a_\mu$& 3&1&$[1,0,1]$\\
\hline
graviton: $\varphi_{\mu\nu}$ & 4& 2& $[0,0,0]$\\
\hline
\end{tabular}
\caption{The relevant supergravity fields and their quantum numbers.}
\label{sugrafields}
\end{center}
\end{table}

The major difficulty of following this recipe is in obtaining the precise interaction vertices. To compute the five-point functions one needs to expand the supergravity effective action to the quintic order. This is extremely tedious and nonetheless unnecessary as we will see. Instead our plan is to use an ``on-shell'' approach which works directly with the five-point correlator. By working with the correlator, we can shortcut through the intermediate complexities that one encounters starting from the off-shell effective Lagrangian. Moreover, correlators are constrained by superconformal symmetry, and satisfy nontrivial self-consistency conditions. Among them is factorization, which relates higher-point correlation functions to the lower-point ones.  By exploiting symmetries and self-consistency conditions we bootstrap the supergravity correlator and eschew the details of the effective Lagrangian altogether.

Our concrete line of attack comes in three steps. We outline the procedure below.

\subsubsection*{Step 1: computing the singular part of the correlator using factorization.}
We divide the five-point function into two parts according to their behaviors in the OPE limits
\begin{equation}
G_5=G_5^{sing}+G_5^{reg}\;.
\end{equation}
The singular part $G_5^{sing}$ consists of all the double and single-exchange diagrams, and the regular part $G_5^{reg}$ contains only the contact Witten diagrams.  

The various contributing Witten diagrams can be evaluated by generalizing the method of \cite{DHoker:1999mqo}, however the coefficient of each diagram is not fixed. To fix these coefficients, the key ingredient of our method is the factorization of the supergravity five-point correlator. For example, we can collect all the exchange Witten diagrams with a scalar exchange in the 12 channel. The Mellin amplitude of this collection of diagrams has a simple pole at $\gamma_{12}=1$. Factorization then dictates that the residue of the Mellin amplitude at $\gamma_{12}=1$ equals to the product of the Mellin amplitudes of the three-point function $\langle\mathcal{O}_\mathbf{20'}(x_1)\mathcal{O}_\mathbf{20'}(x_2)\mathcal{O}_\mathbf{20'}(x_6) \rangle$ and the four-point function $\langle\mathcal{O}_\mathbf{20'}(x_6)\mathcal{O}_\mathbf{20'}(x_3)\mathcal{O}_\mathbf{20'}(x_4)\mathcal{O}_\mathbf{20'}(x_5) \rangle$. Similarly, the factorization of all the  graviphoton exchange diagrams in the 12 channel relates the Mellin amplitude residue to the three- and four-point Mellin amplitudes of $\langle\mathcal{O}_\mathbf{20'}(x_1)\mathcal{O}_\mathbf{20'}(x_2)\mathcal{J}_\mu(x_6) \rangle$ and $\langle\mathcal{J}_\mu(x_6)\mathcal{O}_\mathbf{20'}(x_3)\mathcal{O}_\mathbf{20'}(x_4)\mathcal{O}_\mathbf{20'}(x_5) \rangle$; the factorization of all the graviton exchange diagrams in the 12 channel expresses the Mellin amplitude residue in terms of the Mellin amplitudes of $\langle\mathcal{O}_\mathbf{20'}(x_1)\mathcal{O}_\mathbf{20'}(x_2)\mathcal{T}_{\mu\nu}(x_6) \rangle$ and $\langle\mathcal{T}_{\mu\nu}(x_6)\mathcal{O}_\mathbf{20'}(x_3)\mathcal{O}_\mathbf{20'}(x_4)\mathcal{O}_\mathbf{20'}(x_5) \rangle$. The spinning three-point functions are non-renormalized and take the free theory values. Their Mellin amplitudes therefore can be easily obtained. On the other hand, the spinning four-point correlators are coupling-dependent but are related to the scalar four-point function via superconformal Ward identities \cite{Belitsky:2014zha}. It requires some work to extract their Mellin amplitudes and we will discuss its details in Appendix \ref{spinningcorrelator}. It turns out that factorization uniquely fixes the singular part of the correlator $G_5^{sing}$ which contains all the double-exchange and single-exchange diagrams.

\subsubsection*{Step 2: computing the regular part of the correlator by taking the chiral algebra twist.}
Factorization is agnostic about the regular part of the correlator $G_5^{reg}$ since the regular part does not contribute to  factorization.  To fix it, we first write down the most general ansatz for $G_5^{reg}$ which contains contact Witten diagrams with all R-symmetry structures and up to two derivatives. The upper bound on the number of derivatives comes from the fact that 5d $\mathcal{N}=8$ gauged supergravity is a two-derivative theory. We then take the chiral algebra twist of the total correlator $G_5^{sing}+G_5^{reg}$. The requirement that the ansatz should reduce to the same holomorphic function as obtained from the free theory imposes nontrivial constraints on the unknown coefficients in $G_5^{reg}$. After the dust settles, we 
find that $G_5^{reg}$ is fixed up to a single undetermined coefficient, which multiplies the following zero-derivative contact term
\begin{equation}\label{vanishunderchiral}
\frac{\lambda_c}{\pi^2 N^3} \left(\sum A_{(ijklm)}-\sum A_{(ijk)(lm)}\right)D_{22222}\;.
\end{equation}
The R-symmetry factor vanishes identically under the chiral algebra twist, and therefore the coefficient $\lambda_c$ remains unfixed at this stage. 

\subsubsection*{Step 3: fixing the remaining coefficient by taking the $SO(6)$ twist.}
To fix the remaining coefficient, we exploit the $SO(6)$ twist which sets
\begin{equation}
t_{ij}=x_{ij}^2\;,
\end{equation} 
with  generic insertion points $x_i\in\mathbb{R}^4$. The analysis of \cite{Drukker:2009sf} dictates that the twisted five-point function is topological and protected. Note that the combination (\ref{vanishunderchiral}), which vanishes under the chiral algebra twist, does not vanish under the $SO(6)$ twist. This implies $\lambda_c$ can be fixed by comparing with the free theory.

Our final results for $G^{sing}_5$ and $G^{reg}_5$ are given respectively by (\ref{G5singfixed}) and (\ref{G5regonly0der}).\footnote{A Mathematica notebook with the full position space five-point function is also included in the online version of the paper.} In the following subsections we spell out the details of the above procedure.
 
\subsection{Singular part of the correlator}
It is not difficult to see that the diagrams in Figures \ref{fig:DE} and \ref{fig:SE}, under permutations of the external labels, exhaust all the possibilities of exchange diagrams allowed by R-symmetry selection rules. Double-exchange diagrams involving one graviton and one graviphoton or two gravitons, for example, are not allowed\footnote{$[0,2,0]\otimes[0,0,0]=[0,2,0]$.}. The allowed exchange diagrams constitute the singular part of the correlator. In this subsection, we fix the coefficients of these exchange diagrams by using the factorization properties of the five-point function. 

\subsubsection{Factorization on an internal graviton line}
\begin{figure}[htbp]
\begin{center}
\includegraphics[width=0.85\textwidth]{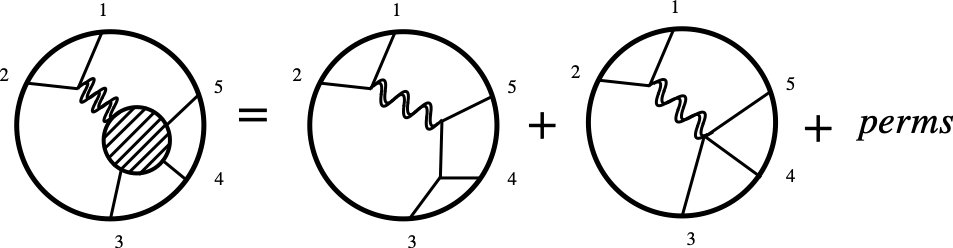}
\caption{Factorization on an internal graviton line. Here ``{\it perms}'' denotes the other inequivalent diagrams obtained by permuting the external legs 3, 4 and 5. Upon factorizing the five-point function on the internal graviton line, we obtain a three-point function $\langle \mathcal{O}_\mathbf{20'}\mathcal{O}_\mathbf{20'}\mathcal{T}_{\mu\nu}\rangle$ and a four-point function $\langle\mathcal{T}_{\mu\nu}\mathcal{O}_\mathbf{20'}\mathcal{O}_\mathbf{20'}\mathcal{O}_\mathbf{20'}\rangle$.}
\label{facintg}
\end{center}
\end{figure}

We start from the factorization of the correlator on an internal graviton line. Without loss of generality, we choose the exchanged graviton field to be in the 12 channel. This isolates the diagrams of type \ref{fig:gsDE} and \ref{fig:gSE} (see Figure \ref{facintg}). Because there is a unique solution to the R-symmetry Casimir equation for exchanging the singlet representation in the 12 channel, all the exchange Witten diagrams have the same R-symmetry polynomial.
We can therefore forget about the R-symmetry polynomial in intermediate steps and only multiply it back in the end.

Let us denote the double-exchange diagram \ref{fig:gsDE} as $W^{\varphi_{[12]},s_{[34]}}$. The other two double-exchange diagrams can be obtained by permuting the external labels 3, 4, 5, and are denoted as $W^{\varphi_{[12]},s_{[35]}}$ and $W^{\varphi_{[12]},s_{[45]}}$. In $W^{\varphi_{[12]},s_{[34]}}$, the graviton field is minimally coupled to the scalar field, {\it i.e.}, the cubic vertex has the form 
\begin{equation}
\int_{AdS_5} \varphi_{\mu\nu} T^{\mu\nu}
\end{equation}
where $T_{\mu\nu}$ is the energy-stress tensor
\begin{equation}
T_{\mu\nu}=\triangledown^\mu s^I \triangledown^\nu s^I-\frac{1}{2}g^{\mu\nu}(\triangledown^\rho s^I \triangledown_\rho s^I+m_s^2 s^I s^I)\;,
\end{equation}
with $m_s^2=\Delta_s(\Delta_s-4)=-4$. By using the AdS Feynman rules, the diagram $W^{\varphi_{[12]},s_{[34]}}$ takes the following form 
\begin{equation}
\begin{split}\label{Wgs}
W^{\varphi_{[12]},s_{[34]}}=\int \frac{dz^5}{z_{0}^5} \frac{dy^5}{y_{0}^5} \frac{dw^5}{w_{0}^5}{}&T^{(12)}_{\mu\nu}(x_1,x_2;z)G_{graviton}^{\mu\nu;\rho\sigma}(z,y)T^{(5)}_{\rho\sigma}(x_5;y,w)\\ {}&
\times G_{B\partial}^{\Delta=2}(w;x_3)G_{B\partial}^{\Delta=2}(w;x_4)
\end{split}
\end{equation}
where $T^{(12)}_{\mu\nu}(x_1,x_2;z)$, $T^{(5)}_{\rho\sigma}(x_5;y,w)$ are obtained from $T_{\mu\nu}$ by replacing the scalar field with the scalar bulk-to-bulk and bulk-to-boundary propagators as prescribed by the diagram
\begin{eqnarray}
\nonumber T^{(12)}_{\mu\nu}(x_1,x_2;z)&=&\triangledown^{(\mu} G^{\Delta=2}_{B\partial}(z;x_1) \triangledown^{\nu)} G^{\Delta=2}_{B\partial}(z;x_2)-\frac{1}{2}g^{\mu\nu}\triangledown^\rho G^{\Delta=2}_{B\partial}(z;x_1) \triangledown_\rho G^{\Delta=2}_{B\partial}(z;x_2)\\
&&-\frac{1}{2}g^{\mu\nu}m^2 G^{\Delta=2}_{B\partial}(z;x_1)G^{\Delta=2}_{B\partial}(z;x_2)\;,\\
\nonumber T^{(5)}_{\mu\nu}(x_5;y,w)&=&\triangledown^{(\mu} G^{\Delta=2}_{B\partial}(y;x_5) \triangledown^{\nu)} G^{\Delta=2}_{BB}(y,w)-\frac{1}{2}g^{\mu\nu}\triangledown^\rho G^{\Delta=2}_{B\partial}(y;x_5) \triangledown_\rho G^{\Delta=2}_{BB}(y;w)\\
&&-\frac{1}{2}g^{\mu\nu}m^2 G^{\Delta=2}_{B\partial}(y;x_5)G^{\Delta=2}_{BB}(y;w)\;.
\end{eqnarray}
The evaluation of this diagram has an important subtlety: the source 
\begin{equation}\label{IT5}
\mathcal{I}^{(5)}_{\rho\sigma}(y;x_3,x_4,x_5)=\int\frac{dw^5}{w_0^5} T^{(5)}_{\rho\sigma}(x_5;y,w)G_{B\partial}^{\Delta=2}(w;x_3)G_{B\partial}^{\Delta=2}(w;x_4)
\end{equation}
 coupled to one end of the graviton bulk-to-bulk propagator is {\it not} gauge invariant. In fact, by using the equation of motion identity of the bulk-to-bulk propagator
 \begin{equation}
 (-\square+m^2_s)G^{\Delta=2}_{BB}(y,w)=\delta^{(5)}(y,w)\;,
 \end{equation} 
we find that the source $\mathcal{I}_{\rho\sigma}^{(5)}$ has a nonzero divergence
\begin{equation}
\triangledown^\rho_y\, \mathcal{I}_{\rho\sigma}^{(5)}(y;x_3,x_4,x_5)=-\frac{1}{2}\triangledown_{y,\sigma} G^{\Delta=2}_{B\partial}(y;x_5)\, G^{\Delta=2}_{B\partial}(y;x_3) G^{\Delta=2}_{B\partial}(y;x_4)\;.
\end{equation}
This seems to create problems because gauge fields can only couple to conserved sources, and also renders the method of \cite{DHoker:1999mqo} inapplicable. However we should notice that gauge invariance is not necessarily achieved by an individual diagram, but only the sum of diagrams.\footnote{More precisely, these are the diagrams with a graviton exchange in the 12 channel. We can require gauge invariance of this collection of diagrams because the factorization on the internal graviton line gives a physical three-point function and a physical four-point function, which are gauge invariant.}

To fix this problem, we must also include the single-exchange diagrams \ref{fig:gSE}. The sum of all double-exchange diagrams introduces a source with divergence 
\begin{equation}
\triangledown^\rho_y\, \mathcal{I}_{\rho\sigma}^{(3)}+\triangledown^\rho_y\, \mathcal{I}_{\rho\sigma}^{(4)}+\triangledown^\rho_y\, \mathcal{I}_{\rho\sigma}^{(5)}=-\frac{1}{2} \triangledown_{y,\sigma} \left(G^{\Delta=2}_{B\partial}(y;x_5)\, G^{\Delta=2}_{B\partial}(y;x_3) G^{\Delta=2}_{B\partial}(y;x_4)\right)\;.
\end{equation}
The minimal choice to cancel this divergence is to introduce a single-exchange diagram which is derived from a quartic coupling of the form 
\begin{equation}\label{vertexsss}
\int_{AdS_5}\varphi_{\mu\nu}g^{\mu\nu} s^I s^I s^K c_{IJK}\;.
\end{equation}
Here $c_{IJK}$ is an R-symmetry invariant tensor that makes the vertex a singlet.
Denoting the single-exchange diagram as $W^{\varphi_{[12]}}$, we have
\begin{equation}\small\label{WgSE}
W^{\varphi_{[12]}}=\int \frac{dz^5}{z_{0}^5} \frac{dy^5}{y_{0}^5}T^{(12)}_{\mu\nu}(x_1,x_2;z)G_{graviton}^{\mu\nu;\rho\sigma}(z,y)g_{\rho\sigma}(y)G^{\Delta=2}_{B\partial}(y;x_3)G^{\Delta=2}_{B\partial}(y;x_4)G^{\Delta=2}_{B\partial}(y;x_5)\;.
\end{equation}
It is easy to verify that the sum of diagrams
\begin{equation}\label{graviton12sum}
W^{graviton_{12}}_{\bf tot}=\lambda_\varphi\,  R^{(0,0),(1,1)}_{12|34} \left(W^{\varphi_{[12]},s_{[34]}}+W^{\varphi_{[12]},s_{[35]}}+W^{\varphi_{[12]},s_{[45]}}+\frac{1}{2}W^{\varphi_{[12]}}\right)\;,
\end{equation}
is gauge invariant. The  $ R^{(0,0),(1,1)}_{12|34}$ is the overall R-symmetry factor defined in Appendix \ref{Rsymmpoly} and $\lambda_\varphi$ is an overall coefficient.

The double-exchange diagrams and the single-exchange diagram can be evaluated using the method of \cite{DHoker:1999mqo},  pretending the coupling to the graviton is conserved in each diagram. This prescription can be justified since the total coupling in (\ref{graviton12sum}) is conserved and the extra contributions from each non-vanishing divergence cancel in the end. Details of the evaluation are discussed in Appendix \ref{integratingout} and the corresponding results of the exchange diagrams are given by (\ref{Wgsvalue}) and (\ref{WgSEvalue}).

Having obtained the gauge invariant combination (\ref{graviton12sum}), it is straightforward to go to Mellin space and check that Mellin factorization on the internal graviton line is satisfied. We find the residue of the Mellin amplitude at $\gamma_{12}=1$ is correctly related to the Mellin amplitudes of $\langle\mathcal{O}_\mathbf{20'}(x_1)\mathcal{O}(x_2)_\mathbf{20'}\mathcal{T}_{\mu\nu}(x_6) \rangle$ and $\langle\mathcal{T}_{\mu\nu}(x_6)\mathcal{O}_\mathbf{20'}(x_3)\mathcal{O}_\mathbf{20'}(x_4)\mathcal{O}_\mathbf{20'}(x_5) \rangle$. A more careful analysis of the normalizations could also fix $\lambda_\varphi$, but we will leave it undetermined for the moment and fix it when we consider the factorization on an internal scalar line. We therefore have fixed all the exchange diagrams involving a graviton internal line up to an overall normalization.

\subsubsection{Factorization on an internal graviphoton line}
Let us now proceed to the factorization on an internal graviphoton line. We first focus on the double-exchange diagrams, which turn out to consist of only two types. 

One type of double-exchange diagrams is \ref{fig:vvDE} which involves two internal graviphotons. The graviphoton couples to the scalars via the minimal coupling
\begin{equation}
\int_{AdS_5} V^{a,\mu} J_{a,\mu}
\end{equation}
where $a=[I,J]$ and
\begin{equation}
J_{a,\mu}=-s_I{\overset{\leftrightarrow}{\triangledown}}_\mu s_J\;.
\end{equation}
The graviphoton-graviphoton-scalar coupling is given by 
\begin{equation}
\int_{AdS_5} s^I F^a_{\mu\nu}F^{b,\mu\nu}d_{I,ab}
\end{equation}
where $F^a_{\mu\nu}$ is the field strength of the graviphoton field and $d_{I,ab}$ is a tensor that makes the vertex a singlet. Denoting \ref{fig:vvDE} as $W^{V_{[12]},V_{[34]}}$, we have
\begin{equation}\label{Wvv}
\begin{split}
W^{V_{[12]},V_{[34]}}=\int \frac{dz^5}{z_{0}^5} \frac{dy^5}{y_{0}^5}\frac{dw^5}{w_{0}^5}{}&J^{(12)}_{\mu}(x_1,x_2;z)\triangledown_y^\sigma G_{vector}^{\mu;\rho}(z,y) g_{\sigma\lambda}(y)g_{\rho\kappa}(y)
\\{}&\times \triangledown_y^{[\lambda} G_{vector}^{\kappa];\nu}(y,w)J^{(34)}_{\nu}(x_3,x_4;w) G^{\Delta=2}_{B\partial}(y;x_5)
\end{split}
\end{equation}
where
\begin{equation}
\begin{split}
J^{(12)}_{\mu}(x_1,x_2;z)=&\triangledown_{z,\mu} G^{\Delta=2}_{B\partial}(z;x_1)\, G^{\Delta=2}_{B\partial}(z;x_2)-G^{\Delta=2}_{B\partial}(z;x_1)\triangledown_{z,\mu}G^{\Delta=2}_{B\partial}(z;x_2)\;,\\
J^{(34)}_{\mu}(x_3,x_4;w)=&\triangledown_{w,\mu} G^{\Delta=2}_{B\partial}(w;x_3)\, G^{\Delta=2}_{B\partial}(w;x_4)-G^{\Delta=2}_{B\partial}(w;x_3)\triangledown_{w,\mu}G^{\Delta=2}_{B\partial}(w;x_4)\;.
\end{split}
\end{equation}
It is not difficult to check that this diagram is already gauge invariant by itself. The diagram is evaluated in Appendix \ref{integratingout}, and the explicit expression is given by (\ref{Wvvvalue}). Moreover, this diagram comes with an R-symmetry factor $R^{(1,0),(1,0)}_{12|34}$, defined in Appendix \ref{Rsymmpoly}. The symmetric combination 
\begin{equation}\label{WVtot1}
W^{graviphoton_{12}}_{{\bf tot},1}=\lambda_{V,1} \left(R^{(1,0),(1,0)}_{12|34}W^{V_{[12]},V_{[34]}}+R^{(1,0),(1,0)}_{12|35} W^{V_{[12]},V_{[35]}}+R^{(1,0),(1,0)}_{12|45} W^{V_{[12]},V_{[45]}} \right)
\end{equation}
can be obtained from the first term via permuting the external labels.

The other type of double-exchange Witten diagram is \ref{fig:svDE}. We denote \ref{fig:svDE} as $W^{V_{[12]},s_{[34]}}$. The diagram reads
\begin{equation}\label{Wvs}
\begin{split}
W^{V_{[12]},s_{[34]}}={}&\int \frac{dz^5}{z_{0}^5} \frac{dy^5}{y_{0}^5} J^{(12)}_{\mu}(x_1,x_2;z) G_{vector}^{\mu;\nu}(z,y)\mathcal{I}^{(34;5)}_\mu(y;x_3,x_4,x_5)\\
\end{split}
\end{equation}
where the source is
\begin{equation}
\begin{split}
\mathcal{I}^{(34;5)}_\mu(y;x_3,x_4,x_5)=\int \frac{dw^5}{w_{0}^5} {}&G^{\Delta=2}_{B\partial}(y;x_5)\triangledown_{y,\mu}\left(G^{\Delta=2}_{BB}(y;w)\right) G^{\Delta=2}_{B\partial}(w;x_3) G^{\Delta=2}_{B\partial}(w;x_4)\\
{}&-\triangledown_{y,\mu} \left(G^{\Delta=2}_{B\partial}(y;x_5)\right)G^{\Delta=2}_{BB}(y;w) G^{\Delta=2}_{B\partial}(w;x_3) G^{\Delta=2}_{B\partial}(w;x_4)\;.
\end{split}
\end{equation}
It is easy to check that the source is not conserved 
\begin{equation}
\triangledown_y^\mu \mathcal{I}^{(34;5)}_\mu(y;x_3,x_4,x_5)=-G^{\Delta=2}_{B\partial}(y;x_3) G^{\Delta=2}_{B\partial}(y;x_4)G^{\Delta=2}_{B\partial}(y;x_5)\;,
\end{equation}
therefore the diagram is not gauge invariant. However when we multiply the diagram with the R-symmetry polynomial $R^{(1,0),(1,1)}_{12|34}$ and sum over all the permutations of 3, 4, 5, the combination 
\begin{equation}\label{WVtot2}
W^{graviphoton_{12}}_{{\bf tot},2}=\lambda_{V,2} \left(R^{(1,0),(1,1)}_{12|34} W^{V_{[12]},s_{[34]}}+R^{(1,0),(1,1)}_{12|35} W^{V_{[12]},s_{[35]}}+R^{(1,0),(1,1)}_{12|45} W^{V_{[12]},s_{[45]}} \right)
\end{equation}
is gauge invariant, since
\begin{equation}
\triangledown_y^\mu \left(R^{(1,0),(1,1)}_{12|34}\mathcal{I}^{(34;5)}_\mu+R^{(1,0),(1,1)}_{12|35}\mathcal{I}^{(35;4)}_\mu+R^{(1,0),(1,1)}_{12|45}\mathcal{I}^{(45;3)}_\mu\right)=0\;.
\end{equation}
The diagram is easy to evaluate and the result is given by (\ref{Wvsvalue}).

\begin{figure}[htbp]
\begin{center}
\includegraphics[width=0.85\textwidth]{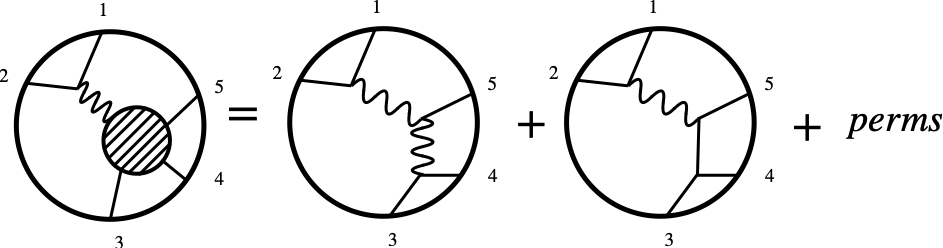}
\caption{Factorization on an internal graviphoton line. Here ``{\it perms}'' denotes the other inequivalent diagrams obtained by permuting the external legs 3, 4 and 5. Upon factorizing the five-point function on the internal graviphotonline, we obtain a three-point function $\langle \mathcal{O}_\mathbf{20'}\mathcal{O}_\mathbf{20'}\mathcal{J}_{\mu}\rangle$ and a four-point function $\langle\mathcal{J}_{\mu}\mathcal{O}_\mathbf{20'}\mathcal{O}_\mathbf{20'}\mathcal{O}_\mathbf{20'}\rangle$.}
\label{facintv}
\end{center}
\end{figure}

We can check if these two gauge invariant combinations of diagrams can already reproduce the factorization. It turns out that
\begin{equation}\label{Wgraviphotontot}
W^{graviphoton_{12}}_{{\bf tot}}=W^{graviphoton_{12}}_{{\bf tot},1}+W^{graviphoton_{12}}_{{\bf tot},2}\;,
\end{equation}
with 
\begin{equation}
\frac{\lambda_{V,1}}{\lambda_{V,2}}=\frac{1}{2}\;,
\end{equation}
gives the correct answer. The overall normalization could also be determined from factorization but we will defer it until later. From the factorization analysis, we can conclude that no graviphoton single-exchange diagrams of \ref{fig:vSE} appear. The factorization of the five-point function on an internal graviphoton line is illustrated in Figure \ref{facintv}.

\subsubsection{Factorization on an internal scalar line}
Finally let us look at the factorization on an internal scalar line. The relevant double-exchange diagrams are $W^{s_{[12]},\varphi_{[34]}}$, $W^{s_{[12]},V_{[34]}}$, $W^{s_{[12]},s_{[34]}}$ and their permutations of 3, 4, 5. The diagrams $W^{s_{[12]},\varphi_{[34]}}$ and $W^{s_{[12]},V_{[34]}}$ have already been discussed in the previous subsections, and are simply related to  $W^{\varphi_{[12]},s_{[34]}}$ and $W^{V_{[12]},s_{[34]}}$ by exchanging 12 with 34.
The double-exchange diagram $W^{s_{[12]},s_{[34]}}$ is constructed from the cubic vertex $s^I s^J s^K c_{IJK}$, and is given by the integral  
\begin{equation}\label{Wss}
\begin{split}
W^{s_{[12]},s_{[34]}}=\int \frac{dz^5}{z_{0}^5} \frac{dy^5}{y_{0}^5}\frac{dw^5}{w_{0}^5}{}& G^{\Delta=2}_{B\partial}(z;x_1) G^{\Delta=2}_{B\partial}(z;x_2)G^{\Delta=2}_{BB}(z;y)\\
{}&\times 
G^{\Delta=2}_{B\partial}(y;x_5)G^{\Delta=2}_{BB}(y;w)G^{\Delta=2}_{B\partial}(w;x_3) G^{\Delta=2}_{B\partial}(w;x_4)\;.
\end{split}
\end{equation}
This diagram can be easily evaluated using the method of Appendix \ref{integratingout} and the result reads
\begin{equation}
W^{s_{[12]},s_{[34]}}=\frac{D_{11112}}{16x_{12}^2x_{34}^2}\;.
\end{equation}
The diagram $W^{s_{[12]},s_{[34]}}$ is associated with an R-symmetry factor $R^{(1,1),(1,1)}_{12|34}$, which can be found in Appendix \ref{Rsymmpoly}.

There are also scalar single-exchange diagrams \ref{fig:sSE}, which can have zero or two derivatives in the quartic coupling. There cannot be more than two derivatives because the 5d $\mathcal{N}=8$ supergravity contains only two derivatives. The zero-derivative single-exchange diagram is denoted by $W^{s_{[12]}}_{\rm 0-der}$ and evaluates to 
\begin{equation}\label{Ws0der}
W^{s_{[12]}}_{\rm 0-der}=\frac{D_{11222}}{4x_{12}^2}\;.
\end{equation}
For the two-derivative type, we have a basis of diagrams where the pair of derivatives are on $\{3,4\}$, $\{3,5\}$ and $\{4,5\}$. These diagrams are denoted respectively by $W^{s_{[12]},(5)}_{\rm 2-der}$, $W^{s_{[12]},(4)}_{\rm 2-der}$, $W^{s_{[12]},(3)}_{\rm 2-der}$, and are related to each other by permuting the external labels 3, 4, 5. The diagram $W^{s_{[12]},(5)}_{\rm 2-der}$ reads
\begin{equation}
W^{s_{[12]},(5)}_{\rm 2-der}=\frac{1}{x_{12}^2}\left(D_{11222}-2x_{34}^2D_{11332}\right)\;.
\end{equation}
The scalar single-exchange diagrams can have 6 independent R-symmetry structures (which can be seen by solving the 12 channel R-symmetry Casimir equation alone). We can pick a  basis of solutions as (see Appendix \ref{Rsymmpoly} for the definition of $\mathcal{A}_i$, $\mathcal{D}_i$, $\mathcal{E}_i$, $\mathcal H$ and $\mathcal I$)
\begin{align}
r_1&=\mathcal{E}_1\;,&  r_2&=\mathcal{E}_2\;, &r_3&=\mathcal{I}\;,\nonumber\\
r_4&=\frac{\mathcal A_1 + \mathcal A_2}{2}-\frac{\mathcal H}{6}\;,& r_5&=\frac{\mathcal D_1 + \mathcal D_2}{2}-\frac{\mathcal H}{6} \;, &r_6&=\frac{\mathcal A_3 + \mathcal A_4}{2}-\frac{\mathcal H}{6}\;.
\end{align}

\begin{figure}[htbp]
\begin{center}
\includegraphics[width=0.85\textwidth]{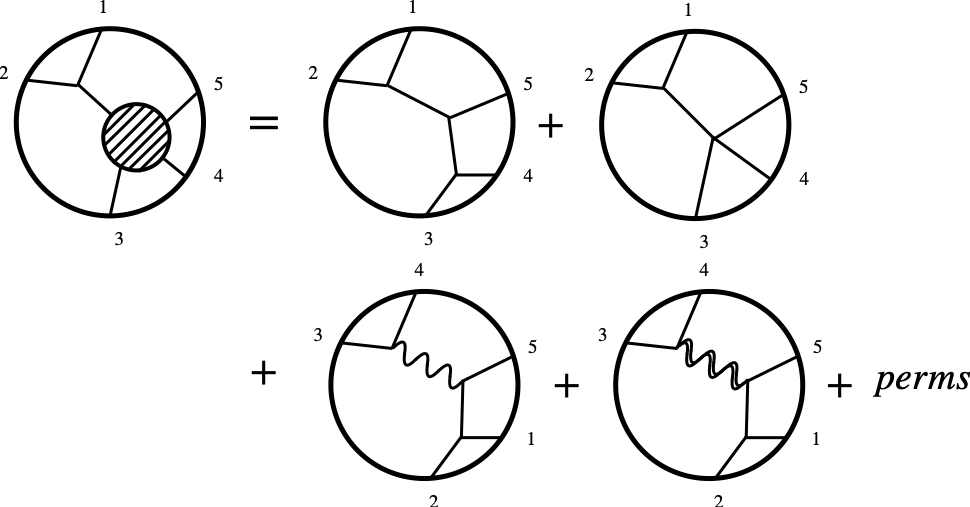}
\caption{Factorization on an internal scalar line. Here ``{\it perms}'' denotes the other inequivalent diagrams obtained by permuting the external legs 3, 4 and 5. The scalar single-exchange diagram represents both the zero-derivative diagram and the two-derivative diagram. Upon factorizing the five-point function on the internal scalar, we obtain a three-point function $\langle \mathcal{O}_\mathbf{20'}\mathcal{O}_\mathbf{20'}\mathcal{O}_\mathbf{20'}\rangle$ and a four-point function $\langle\mathcal{O}_\mathbf{20'}\mathcal{O}_\mathbf{20'}\mathcal{O}_\mathbf{20'}\mathcal{O}_\mathbf{20'}\rangle$.}
\label{facints}
\end{center}
\end{figure}

Let us now collect all the exchange diagrams containing a scalar internal line in the 12 channel (Figure \ref{facints}). We have the following ansatz
\begin{eqnarray}
\nonumber W^{scalar_{12}}_{{\bf tot}}=&&\lambda_s \left(R^{(1,1),(1,1)}_{12|34}  W^{s_{[12]},s_{[34]}}+R^{(1,1),(1,1)}_{12|35} W^{s_{[12]},s_{[35]}}+R^{(1,1),(1,1)}_{12|45} W^{s_{[12]},s_{[45]}}\right)\\
\nonumber&&+\lambda_{V,2}\left(R^{(1,1),(1,0)}_{12|34} W^{s_{[12]},V_{[34]}}+R^{(1,1),(1,0)}_{12|35} W^{s_{[12]},V_{[35]}}+R^{(1,1),(1,0)}_{12|45} W^{s_{[12]},V_{[45]}}\right)\\
\nonumber&&+\lambda_\varphi \left(R^{(1,1),(0,0)}_{12|34}W^{s_{[12]},\varphi_{[34]}}+R^{(1,1),(0,0)}_{12|35}W^{s_{[12]},\varphi_{[35]}}+R^{(1,1),(0,0)}_{12|45}W^{s_{[12]},\varphi_{[45]}}\right)\\
\nonumber&&+\sum_{i=1}^6 \lambda_i^{{\rm 2-der},(5)} r_i W^{s_{[12]},(5)}_{\rm 2-der}+\sum_{i=1}^6 \lambda_i^{{\rm 2-der},(4)} r_i W^{s_{[12]},(4)}_{\rm 2-der}+\sum_{i=1}^6 \lambda_i^{{\rm 2-der},(3)} r_i W^{s_{[12]},(3)}_{\rm 2-der}\\
&&+\sum_{i=1}^6 \lambda_i^{\rm 0-der} r_i W^{s_{[12]}}_{\rm 0-der}
\end{eqnarray}
where $\lambda_\varphi$ and $\lambda_{V,2}$ showed up previously in (\ref{graviton12sum}) and (\ref{Wgraviphotontot}). We also require permutation symmetry among the external legs 3, 4 and 5. 
The Mellin amplitude of $W^{scalar_{12}}_{{\bf tot}}$ contains a simple pole at $\gamma_{12}=1$. Factorization of the five-point correlator requires that the residue at the simple pole should give the product of the Mellin amplitudes of the three-point function $\langle\mathcal{O}_\mathbf{20'}\mathcal{O}_\mathbf{20'}\mathcal{O}_\mathbf{20'}\rangle$ and the four-point function $\langle\mathcal{O}_\mathbf{20'}\mathcal{O}_\mathbf{20'}\mathcal{O}_\mathbf{20'}\mathcal{O}_\mathbf{20'}\rangle$. Together with permutation symmetry, this gives rise to a set of linear equations for the unknown coefficients. Solving these constraints, we have\footnote{\label{footnotehomosolfac} The linear equations do not fix all coefficients, meaning that there exists homogenous solutions to factorization. These homogenous solutions always appear with a multiplicative factor which can be written as the five-point zero-derivative contact diagram $D_{22222}$. Their existence just reflects the ambiguity in separating $G_5$ into $G^{sing}_5$ and $G^{reg}_5$, and their contribution can be combined into the ansatz for the latter. We have set these free parameters to zero without loss of generality.}
\begin{equation}\label{Wscalarcoeffs}
\begin{split}
{}&\lambda_s=\frac{64 \sqrt 2}{\pi^2 N^3}\;,\quad\quad \lambda_{V,2}=-\frac{16 \sqrt 2}{\pi^2 N^3}\;,\quad\quad \lambda_\varphi=-\frac{2\sqrt 2}{\pi^2 N^3}\;,\\
{}&\lambda_1^{\rm 0-der}=\lambda_2^{\rm 0-der}=\lambda_3^{\rm 0-der}=\frac{8 \sqrt 2}{\pi^2N^3}\;,\quad \quad \lambda_4^{\rm 0-der}=\lambda_5^{\rm 0-der}=\lambda_6^{\rm 0-der}=-\frac{32 \sqrt 2}{\pi^2N^3}\;,\\
{}& \lambda_6^{{\rm 2-der},(3)}= \lambda_4^{{\rm 2-der},(4)}= \lambda_5^{{\rm 2-der},(5)}=\frac{4 \sqrt 2}{\pi^2N^3}\;,
\end{split}
\end{equation}
and all the other coefficients are zero.

We have now computed the singular part of the five-point correlation function $G^{sing}_5$. The result is the following
\begin{equation}\label{G5singfixed}
\begin{split}
G^{sing}_5={}&sym\left[W^{graviton_{12}}_{\bf tot}\right]+\frac{1}{2}sym\left[W^{graviphoton_{12}}_{{\bf tot},1}\right]+sym\left[W^{graviphoton_{12}}_{{\bf tot},2}\right]\\
{}&+\frac{1}{2}sym\left[W^{scalar_{12}}_{{\bf tot},1}\right]+sym\left[W^{scalar_{12}}_{{\bf tot},2}\right]
\end{split}
\end{equation}
where 
\begin{equation}
W^{scalar_{12}}_{{\bf tot},1}=\lambda_s \left(R^{(1,1),(1,1)}_{12|34}  W^{s_{[12]},s_{[34]}}+R^{(1,1),(1,1)}_{12|35} W^{s_{[12]},s_{[35]}}+R^{(1,1),(1,1)}_{12|45} W^{s_{[12]},s_{[45]}}\right)\;,
\end{equation}
\begin{equation}
\begin{split}
W^{scalar_{12}}_{{\bf tot},2}={}& \lambda_5^{{\rm 2-der},(5)} r_5 W^{s_{[12]},(5)}_{\rm 2-der}+\lambda_4^{{\rm 2-der},(4)} r_4 W^{s_{[12]},(4)}_{\rm 2-der}+ \lambda_6^{{\rm 2-der},(3)} r_6 W^{s_{[12]},(3)}_{\rm 2-der}\\
{}&+\sum_{i=1}^6 \lambda_i^{\rm 0-der} r_i W^{s_{[12]}}_{\rm 0-der}\;.
\end{split}
\end{equation}
The expressions for $W^{graviton_{12}}_{\bf tot}$, $W^{graviphoton_{12}}_{{\bf tot},1}$, $W^{graviphoton_{12}}_{{\bf tot},2}$ were respectively given in (\ref{graviton12sum}), (\ref{WVtot1}) and (\ref{WVtot2}). We can evaluate them in terms of $D$-functions, and explicit expressions can be found in Appendix \ref{integratingout}. The various coefficients are given by (\ref{Wscalarcoeffs}). The operation $sym$ means to symmetrize with respect to the external labels, {\it i.e.},   
\begin{equation}
\begin{split}
sym[A]={}&A+A|_{13245}+A|_{14325}+A|_{15342}+A|_{23145}\\
{}&+A|_{24315}+A|_{25341}+A|_{34125}+A|_{35142}+A|_{45312}
\end{split}
\end{equation}
where $A|_{a_1a_2a_3a_4a_5}$ means to map the labels 1, 2, 3, 4, 5 to $a_1$, $a_2$, $a_3$, $a_4$, $a_5$. The factors $\frac{1}{2}$ appear because the double exchange diagrams $W^{V_{[12]},V_{[34]}}$, $W^{s_{[12]},s_{[34]}}$ have an extra $\mathbb{Z}_2$ symmetry under exchanging 12 with 34. The symmetrization is such that all the diagrams have strength 1.

\subsection{Regular part of the correlator}\label{regular}
We now solve the regular part of the ansatz. The regular part $G^{reg}_5$ consists only of contact Witten diagrams with zero and two derivatives
\begin{equation}
G^{reg}_5=\left(\sum_{I=1}^{22} \lambda_I^{\{1,2\},(2)} A_I \, x_{12}^2 D_{33222}+perms\right)+\sum_{I=1}^{22} \lambda_I^{(0)} A_I D_{22222}\;.
\end{equation}
Here $A_I$ with $I=1,\ldots,22$ are the 22 R-symmetry structures defined in (\ref{RstructuresA}), and $\lambda_I^{\{i,j\},(2)}$, $\lambda_I^{(0)}$ are undetermined coefficients. We require the ansatz $G^{reg}_5$ to be invariant under crossing. 

To fix the coefficients, we first use the chiral algebra twist as was reviewed in Section \ref{kinematics}. The five operators are now restricted on a plane,  parameterized by the 2d coordinates $z_i$, $\bar{z}_i$. The R-symmetry polarizations are restricted to rotate under only an $SO(4)$ subgroup of $SO(6)_R$, and the null vectors are parameterized as $t_i^\mu=\sigma^\mu_{\alpha\dot{\alpha}}v_i^\alpha\bar{v}_i^{\dot{\alpha}}$ with $v_i=(1,y_i)$, $\bar{v}_i=(1,\bar{y}_i)$. The chiral algebra twist amounts to setting $\bar{y}_i=\bar{z}_i$, and the non-renormalization of chiral algebra requires that
\begin{equation}\label{5ptchiralalgebratwist}
\left(G^{sing}_5+G^{reg}_5\right)\big|_{\bar{y}_i=\bar{z}_i}=G^{free}_5\big|_{\bar{y}_i=\bar{z}_i}
\end{equation}
where the $G^{free}_5$ is the correlator computed in the free theory and is given by 
\begin{equation}
G^{free}_5=\frac{2\sqrt{2}}{N}\sum \frac{A_{(ijklm)}}{x_{ij}^2x_{jk}^2x_{kl}^2x_{lm}^2x_{mi}^2}+\frac{4\sqrt{2}}{N^3}\sum \frac{A_{(ijk)(lm)}}{x_{ij}^2x_{jk}^2x_{ki}^2x_{lm}^4}
\end{equation}
Note that the r.h.s. is a simple rational function of the holomorphic coordinates. On the other hand, the l.h.s. comes from a complicated sum of $D$-functions where each $D$-function has transcendental degree 2 and is far from being a rational function. This means that the unknown coefficients in the l.h.s. must be fine-tuned to reproduce a rational function,  therefore imposing strong constraints on the unknown coefficients.  

The condition (\ref{5ptchiralalgebratwist}) is not yet in a form that is ready for use. Extracting the constraints on the coefficients from (\ref{5ptchiralalgebratwist}) still requires some nontrivial work. Our strategy is to find a basis to decompose the l.h.s.. 
Using the differential recursion relations in Appendix \ref{propertyDfunction}, all these $D$-functions can be related to the basic $D$-function $D_{11112}$ (and its permutations) by taking derivatives. The function $D_{11112}$ can be evaluated in closed form in terms of one-loop scalar box diagrams \cite{Bern:1992em,Bern:1993kr}
\begin{equation}
D_{11112}=\frac{4\pi^2}{x_{14}^2x_{35}^2x_{25}^2}\sum_{i=1}^5\frac{\eta_{i5}\hat{I}^{(i)}_4}{N_5}\;.
\end{equation}
Here $\eta_{i5}$, $N_5$ are rational functions of the conformal cross ratios, and $\hat{I}^{(i)}_4$ are one-loop box diagrams (also denoted as $\Phi$ in Appendix \ref{propertyDfunction}) where the $i^{th}$ point is omitted. When the insertion points $x_i$ are generic, {\it i.e.}, not lying on a two-dimensional plane, the five box diagrams $\hat{I}^{(i)}_4$, $i=1,\ldots,5$ are independent. Taking derivatives with respect to $x_{ij}^2$, one can obtain $D$-functions of higher weights. Since the box diagrams obey differential recursion relations (\ref{Phidiffrecur}), one finds that all the $D$-functions can be uniquely decomposed into a basis spanned by $\hat{I}^{(i)}_4$, logarithms and 1, with rational coefficient functions. Apparently, the ansatz $G^{sing}_5+G^{reg}_5$ also admits such a unique decomposition under this basis with rational coefficient functions. However, to use the chiral algebra twist condition (\ref{5ptchiralalgebratwist}), we need to further restrict the five insertions on a plane. This gives rise to subtleties which require some extra care. The problem is that some elements of the basis develop relations. For example, the five one-loop box diagrams are now linearly dependent\footnote{However the denominator $N_5$ also becomes zero at the same rate so $D_{11112}$ remains  finite (and nonzero) when all the points are put on plane.}
\begin{equation}
\sum_{i=1}^5\eta_{i5}\hat{I}^{(i)}_4\big|_{x_i\in \mathbb{R}^2}=0\;,
\end{equation}
which follows from the identity \cite{Abel}
\begin{equation}\small
Li_2\left(\frac{zw}{(1-z)(1-w)}\right)=Li_2\left(\frac{z}{1-w}\right)+Li_2\left(\frac{w}{1-z}\right)-Li_2(z)-Li_2(w)-\log(1-z)\log(1-w) \label{eq:Abel}.
\end{equation}
After properly taking care of the relations among the basis vectors, we find the following basis of independent functions
\begin{align}
&\Phi\left(z,\bar{z}\right),\, \ \ \ \ \Phi\left(w,\bar{w}\right),\, \ \ \ \ \Phi\left(\frac{z}{w},\frac{\bar{z}}{\bar{w}}\right),\, \ \ \ \ \Phi\left(\frac{1-z}{1-w},\frac{1-\bar{z}}{1-\bar{w}}\right),\, \ \  \ln z\bar{z},\, \ \  \ln w\bar{w},\, \ \\\
&  \ln (1-z)(1-\bar{z}),\, \ \ \ln (1-w)(1-\bar{w}),\, \ \  \ln (w-z)(\bar{w}-\bar{z}),\, \ \ 1,\nonumber
\end{align}
where $z$ and $w$ are the complex coordinates of the two insertion points not fixed by conformal symmetry, with $\bar{z}$, $\bar{w}$ being their complex conjugates. They are related to the cross ratios $V_i$ defined in (\ref{CrossRatios}) via 
\begin{equation}
\begin{split}
{}&V_1=z\bar{z}\;,\quad V_2=(1-z)(1-\bar z)\;,\\
{}&V_3=(1-w)(1-\bar{w})\;,\quad V_4=w\bar{w}\;,\quad V_5=\frac{z\bar{z}(1-w)(1-\bar{w})}{(w-z)(\bar{w}-\bar{z})}\;.
\end{split}
\end{equation}
Decomposing the supergravity ansatz into this basis gives coefficient functions which are rational in the cross ratios. Equating the coefficients in  (\ref{5ptchiralalgebratwist}) gives a set of linear equations for the unknown coefficients.

The constraints turn out to be remarkably constraining. We find that all the two-derivative vertices vanish
\begin{equation}
\lambda_I^{\{i,j\},(2)}=0\;.
\end{equation}
Moreover, all but one of the zero-derivative coefficients is fixed, yielding
\begin{equation}\label{G5regonly0der}
G^{reg}_5= \frac{1}{\pi^2 N^3}\left(\frac{11 \sqrt 2}{3 } \sum A_{(ijk)(lm)}+\lambda_c \left(\sum A_{(ijklm)}-\sum A_{(ijk)(lm)}\right) \right)D_{22222}\;.
\end{equation}
Chiral algebra is incapable of fixing $\lambda_c$ because the multiplied R-symmetry polynomial vanishes automatically under twisting. 

To determine the remaining coefficient, we use the $SO(6)$ twist as we reviewed in Section \ref{kinematics}. This uniquely fixes the coefficient to be
\begin{equation}
\lambda_c=6 \sqrt 2\;.
\end{equation}

Before we end this section, let us make a comment about the contact Witten diagrams which contribute to $G^{reg}_5$. These five-point contact interactions in fact are {\it not intrinsic} in the sense that they can be absorbed into $G^{sing}_5$ by redefining certain vertices of the exchange Witten diagrams. We have already noticed such an ambiguity in footnote \ref{footnotehomosolfac}. More precisely, we can rewrite $G^{reg}_5$ in such a way that it can be absorbed in the scalar single-exchange Witten diagrams $W^{scalar_{12}}_{{\bf tot},2}$ (and all other diagrams by permutations) while keeping the quartic vertices in \ref{fig:sSE} symmetric and with no more than two derivatives. To see this, let us define the scalar single-exchange Witten diagrams $\widetilde{W}^{s_{[12]},(i)}_{2-{\rm der}}$ for which the two derivatives act on the external leg $i=3,4,5$, and the internal leg $I$. By using the equation of motion identities of the propagators and integration by parts, one can show
\begin{equation}
\widetilde{W}^{s_{[12]},(i)}_{2-{\rm der}}=W^{s_{[12]},(i)}_{2-{\rm der}}+\frac{1}{2}D_{22222}\;
\end{equation}
These identities can be used to make the two ways of distributing the derivatives, {\it i.e.}, $(3,4)$, $(4,5)$, $(3,5)$ and $(3,I)$, $(4,I)$, $(5,I)$, appear symmetrically in the solution at the cost of generating some new $D_{22222}$. The total collection of $D_{22222}$ with different R-symmetry structures can then be reinterpreted as scalar single-exchange Witten diagrams with derivatives on the same leg. Let us note that when the derivatives are on the same external leg, the diagram is simply 
\begin{equation}
-4W^{s_{[12]}}_{0-{\rm der}}\;,
\end{equation}
by the equation of motion. When both derivatives act on the internal leg, there is an extra delta function in the equation of motion and therefore equals to
\begin{equation}
-4W^{s_{[12]}}_{0-{\rm der}}-D_{22222}\;.
\end{equation}
The latter term allows us to absorb all the $D_{22222}$ into $G^{sing}_5$.

\section{An alternative approach using Mellin space}\label{Mellinansatz}
An alternative approach to the previous section is to start from an ansatz in Mellin space and then solve it by imposing constraints. The construction of the ansatz is facilitated in Mellin space, thanks to the simple analytic structure of Mellin amplitudes. As reviewed in section \ref{Mellinfactorization}, the poles of the Mellin amplitude are determined by the twists of the exchanged operators. In the tree level supergravity limit, only single-trace one-half BPS operators and multi-trace operators constructed from them are present. The polar information of the latter is already captured by the Gamma function factors, and the former is manifested as the simple poles in the Mellin amplitude. In the case of $\mathbf{20'}$ five-point functions, the exchanged single-trace operators are the $\mathbf{20'}$ operator, the $R$-symmetry current and the stress tensor. They give rise to leading simple poles at $\gamma_{ij}=1$. On the other hand, by using a similar $1/N$ argument as in Section 3.2 of \cite{Rastelli:2017udc}, we can conclude that there are no satellite poles associated with the exchange of the single-trace operators. It is  instructive to look at the factorization of the Mellin amplitude. For example, in the $12$ channel, the Mellin amplitude is expected to have the following structure
\begin{align}
\mathcal{M} &= \sum_{a,b=1}^{2}\sum_{i,j=3}^{5}\frac{\gamma_{ai}(\gamma_{bj}+\delta_{b}^a\delta_{j}^i)\mathcal{M}_{3,\mathcal{T}}^{ab}\mathcal{M}_{4,\mathcal{T}}^{ij}}{\gamma_{12}-1}+\sum_{a}^{2}\sum_{i}^{5}\frac{\gamma_{ai}\mathcal{M}_{3,\mathcal{J}}^{a}\mathcal{M}_{4,\mathcal{J}}^{i}}{\gamma_{12}-1}\nonumber\\
&+\frac{\mathcal{M}_{3,\mathcal{O}_{\mathbf{20'}}}\mathcal{M}_{4,\mathcal{O}_{\mathbf{20'}}}}{\gamma_{12}-1}+\mathcal{M}_{reg,12}
\label{eq:factorizationformulaMellinansatz}
\end{align}
where $\mathcal{M}_{3,\mathcal{T}}^{ab}$, $\mathcal{M}_{4,\mathcal{T}}^{ij}$ are respectively the three and four-point Mellin amplitudes of $\mathcal{O}_{\mathbf{20'}}$ with one stress tensor, and $\mathcal{M}_{3,\mathcal{J}}^{a}$, $\mathcal{M}_{4,\mathcal{J}}^{a}$ are the Mellin amplitudes with one R-symmetry current.  The term $\mathcal{M}_{reg,12}$ is regular with respect to $\gamma_{12}$. However it must contain singularities in other independent $\gamma_{ij}$ such that the five-point Mellin amplitude $\mathcal{M}$ is crossing symmetric. Note that there can be at most two simultaneous poles in the Mellin amplitude, which correspond to the double-exchange Witten diagrams. The simultaneous poles involving $\gamma_{12}$ can be explicitly seen from the above formula where the other pole is supplied by the four-point Mellin amplitudes.

This motivates us to write down the following ansatz for the five-point Mellin amplitude 
\begin{align}
\mathcal{M}_{ansatz}(\gamma_{ij})=\sum_{(ij)\neq(i'j')}\frac{\mathcal{P}_2^{ij,i'j'}(\gamma_{ml})}{(\gamma_{ij}-1)(\gamma_{i'j'}-1)}+\sum_{(ij)}\frac{\mathcal{P}_1^{ij}(\gamma_{ml})}{\gamma_{ij}-1}+\mathcal{P}_{0}(\gamma_{ml})\;,
\end{align}
which has the structure of a sum of simultaneous poles, single poles and a regular piece. The residues $\mathcal{P}_2^{ij,i'j'}$, $\mathcal{P}_1^{ij}$ and $\mathcal{P}_{0}$ are polynomials in the Mandelstam variables $\gamma_{ml}$. They are also polynomials in $t_{ij}$, but we will suppress the R-symmetry dependence for the moment and focus on the $\gamma_{ml}$ dependence. We will assume that $\mathcal{P}_2^{ij,i'j'}$ are degree 2 polynomials of $\gamma_{ml}$, while both $\mathcal{P}_1^{ij}$ and $\mathcal{P}_{0}$ are of degree 1. The degrees of these polynomials can be justified as follows. For the residues $\mathcal{P}_2^{ij,i'j'}$, the degree simply follows from the counting of the total number of derivatives in the cubic vertices. Just as in flat space, the residue has degree $L$ if the vertices contain in total $2L$ derivatives. For the double-exchange diagrams involving one stress tensor, it follows from R-symmetry selection rule that the other exchange field can only be the scalar field. Such diagrams have only four derivatives in all vertices. In the vector-vector double-exchange diagrams, the total number of derivatives is also four. For the vector-scalar and scalar-scalar double-exchange diagrams, the total numbers of derivatives are two and zero.\footnote{Note the derivatives in the scalar cubic coupling have been removed by nonlinear redefinition of the scalar fields \cite{Lee:1998bxa}.} This leads us to conclude that $\mathcal{P}_2^{ij,i'j'}$ are degree 2 polynomials of $\gamma_{ml}$. It is tempting to apply the same argument on the single pole residues $\mathcal{P}_1^{ij}$. However the counting holds only for the vector and scalar single-exchange diagrams where at most two derivatives are present. For the graviton single-exchange diagrams, the residue appears to have degree 2 since there could be in total four derivatives. The leading degree 2 terms would correspond to a constant piece in the term $\mathcal{M}_{4,\mathcal{T}}^{ij}$ of the factorization formula (\ref{eq:factorizationformulaMellinansatz}).\footnote{The Mellin amplitude $\mathcal{M}_{4,\mathcal{T}}^{ij}$ contains simple poles with constant residues which are due to the exchange of scalar fields. There is also an additional constant piece which is due to the quartic interactions. As is shown in Appendix \ref{spinningcorrelator},  this constant term is completely determined by the singular terms via transversality.} However, a closer look at $\mathcal{M}_{4,\mathcal{T}}^{ij}$ reveals that the contribution of the constant term to  (\ref{eq:factorizationformulaMellinansatz}) vanishes after the summation. Therefore, the single pole residues $\mathcal{P}_1^{ij}$ are degree 1 polynomials. Finally, $\mathcal{P}_{0}$ receives contribution from the five-point contact vertices. Since the gauged supergravity contains at most two derivatives, it follows that the degree of $\mathcal{P}_{0}$ is at most 1. 

Let us now be more explicit about the R-symmetry dependence. We write the residues as 
\begin{align}
\mathcal{P}_{2}^{ij,i'j'}(\gamma)&= \sum_{(ml),(m'l')\neq (ij),(i'j')}\sum_{I_2,I_2'=\mathbf{20'},\mathbf{15},\mathbf{1}} c_{ij,i'j'| ml,m'l'}^{I_2,I_2'}R^{I_2,I_2'}_{ij|i'j'}\gamma_{ml}\gamma_{m'l'}\nonumber\\
&+\sum_{(ml)\neq (ij),(i'j')}\sum_{I_2,I_2'=\mathbf{20'},\mathbf{15},\mathbf{1}}c_{ij,i'j'| ml}^{I_2,I_2'} R^{I_2,I_2'}_{ij|i'j'} \gamma_{ml}+\sum_{I_2,I_2'=\mathbf{20'},\mathbf{15},\mathbf{1}}c_{ij,i'j'}^{I_2,I_2'}R^{I_2,I_2'}_{ij|i'j'}\;,\\
\mathcal{P}_{1}^{ij}(\gamma)&=\sum_{(ml)\neq (ij)}\sum_{I_1=\mathbf{20'},\mathbf{15},\mathbf{1}}\sum_a d_{ij|ml,a}^{I_1}R^{I_1}_{ij,a}\gamma_{ml}+\sum_{I_1=\mathbf{20'},\mathbf{15},\mathbf{1}}\sum_a d_{ij,a}^{I_1}R^{I_1}_{ij,a}\;,\\
\mathcal{P}_{0}(\gamma)&=\sum_{(ml)}\sum_{I_0}e_{ml}^{I_0}A^{I_0}\gamma_{ml}+\sum_{I_0}e^{I_0}A^{I_0}\;
\end{align} 
where various coefficients $c$, $d$, $e$ parameterize the degrees of freedom in the ansatz. The R-symmetry polynomials $R^{I_2,I_2'}_{ij|i'j'}$ are the solutions to the double R-symmetry Casimir equation where the representation $I_2$, $I_2'$ are exchanged in the channels $(i,j)$ and $(i',j')$ respectively. The polynomials $R^{I_1}_{ij,a}$ are solutions to the single R-symmetry Casimir equation in the $(i,j)$ channel where the exchanged representation is $I_1$. The  index $a$ labels the different solutions to the single Casimir equation, of which a basis can be obtained from $R^{I_1,I_1'}_{ij|i'j'}$ where $(i',j')$ is any other compatible channel and $I_1'$ is over all possible R-symmetry representations.
 Note that the sums over the representations $I_2$, $I_2'$ in the simultaneous pole residues are restricted to $\mathbf{20'}$, $\mathbf{15}$ and $\mathbf{1}$, in correspondence to the R-symmetry representations of the exchanged single-trace fields. Similarly, the sums over $I_1$ in the single pole residues are also restricted to $\mathbf{20'}$, $\mathbf{15}$ and $\mathbf{1}$, which correspond to the single-exchange Witten diagrams. On the other hand, in the regular part we sum over all 22 R-symmetry structures $A^{I_0}$ defined in \eqref{RstructuresA}, since all of them can appear.

The Mellin amplitude ansatz $\mathcal{M}_{ansatz}$ is further constrained by three other consistency conditions. First, the ansatz $\mathcal{M}_{ansatz}$ should be crossing symmetric. In implementing this constraint, it is important to take into account the linear constraints satisfied by $\gamma_{ij}$ which leaves only five independent variables. Second, the correlator needs to satisfy the chiral algebra condition (\ref{5ptchiralalgebratwist}). Unfortunately this condition is not straightforward to implement in Mellin space. This is essentially because the independent Mandelstam variables $\gamma_{ij}$ are dual to the independent conformal cross ratios for generic configurations. To perform the chiral algebra twist, one needs to restrict the five insertion points on a two-dimensional plane. This reduces the number of independent cross ratios to four, while the Mellin representation is oblivious to it. Therefore our strategy is to rewrite the Mellin amplitude ansatz as a sum of $D$-functions and then implement the chiral algebra twist in position space. However we should note that the rewriting is not unique. Different expressions in terms of $D$-functions with the {\it same} Mellin amplitude may differ in position space by a rational function or a logarithmic term.\footnote{These ambiguities correspond to different choices of the integration contours.} On the other hand, the part with transcendental degree 2 does not suffer from such ambiguities. Therefore, we only use the chiral algebra constraints from the coefficient functions of the box diagrams.\footnote{One might wonder if the chiral algebra conditions are now much weaker. In the position space method, we observed that the conditions from the coefficients of the logarithms do not lead to new constraints in addition to the ones from the box diagram coefficients. }  Finally, the correlator satisfies the condition (\ref{5ptSO6twisted}) imposed by the $SO(6)$ twist. We also implement this condition in position space and focus on the pieces with transcendental degree 2.  

Solving the above constraints fixes the ansatz up to an overall normalization. The leftover degree of freedom is expected because the twisted five-point functions in (\ref{5ptchiralalgebratwist}) and (\ref{5ptSO6twisted}) are rational, and do not contribute to the box diagram coefficients. The conditions from the chiral algebra twist and the $SO(6)$ twist are therefore homogenous and do not allow us to determine the overall coefficient. We can fix the remaining coefficient by, for example, looking at the factorization of the five-point Mellin amplitude on a scalar exchange.

The final result for the Mellin amplitude takes the following form
\begin{equation}\label{MellinResult}
\mathcal{M}=\mathcal{M}_{sim}+\mathcal{M}_{sing}+\mathcal{M}_{reg} \;,
\end{equation}
where $\mathcal{M}_{sim}$ are the simultaneous poles
\begin{equation}
\begin{split}
\mathcal{M}_{sim}={}&\frac{2\sqrt{2}}{(\gamma_{12}-1)(\gamma_{34}-1)}\bigg(A_{(125)(34)}\gamma_{45} \gamma_{35} + A_{(345)(12)}\gamma_{15} \gamma_{25}-2 A_{(12543)} \gamma_{15} \gamma_{35}\\
{}&-2 A_{(12345)} \gamma_{25} \gamma_{35}-2 A_{(12534)} \gamma_{15} \gamma_{45}-2 A_{(12435)} \gamma_{25} \gamma_{45}\bigg)+perm\;,
\end{split}
\end{equation}
$\mathcal{M}_{sing}$ are the single poles
\begin{equation}
\begin{split}
{}&\mathcal{M}_{sing}=\frac{1}{8\sqrt{2}(\gamma_{12}-1)}\bigg((A_{(345)(12)}-2A_{(12534)}-2A_{(12435)})\gamma_{45}\\
{}&\quad+(A_{(345)(12)}-2A_{(12543)}-2A_{(12345)})\gamma_{35}+(A_{(345)(12)}-2A_{(12354)}-2A_{(12453)})\gamma_{34}\bigg)+perm
\end{split}
\end{equation}
and $\mathcal{M}_{reg}$ is the regular piece
\begin{equation}
\begin{split}
\mathcal{M}_{reg}=\frac{3(8A_{(14325)}-5A_{(345)(12)})}{10\sqrt{2}}+perm\;.
\end{split}
\end{equation}
The R-symmetry structures $A_{(ijklm)}$, $A_{(ijk)(lm)}$ were defined in (\ref{RstructuresA}). Note that $\mathcal{M}_{reg}$ does not contain terms linear in the Mandelstam variables. Moreover, one can show $\mathcal{M}_{reg}$ can be absorbed into $\mathcal{M}_{sing}$. This corresponds to our observation in position space that there are no intrinsic contact interactions.

\section{OPE analysis}\label{CFTdata}

In this section we analyze the short-distance behavior of the supergravity five-point function and use the Euclidean OPE to extract new CFT data of strongly coupled $\mathcal{N}=4$ SYM. To simplify the analysis, we restrict our attention to only the singular and the leading non-singular behavior of the correlator. A complete analysis of the supergravity five-point function is left to the future. 

In Section \ref{defs} we discuss the kinematics of the Euclidean OPE. We discuss the decomposition of five-point functions in conformal blocks, and also explain how to take a single OPE to obtain four-point functions. This part can be read independently, and applies to generic CFTs with and without supersymmetry. In Section \ref{ops} we introduce all the operators up to dimension four that contribute to the OPE of two $\mathbf{20'}$ operators. The reader interested solely in the results of the OPE analysis might skip directly to Section \ref{data}, where we present the new data obtained.

\subsection{Euclidean OPE limit}\label{defs}
The information of the CFT is encoded in five-point functions according to the principle of operator product expansion. By leveraging this expansion in different ways, we can extract various information from the five-point functions.

To extract the CFT data, it is most straightforward to use OPE in two different channels. The five-point function essentially becomes a sum of products of three-point functions, analogous to the case of four-point functions. More precisely, we send the points  $x_1$, $x_3$ and $x_4$ to  $0$, $1$ and $\infty$ respectively, by using the global conformal symmetry. The Euclidean double coincidence limit (in the $12$, $35$ channel) is then obtained by taking both $x_{12}$ and $x_{35}$ to approach zero, in which case \eqref{G5cr} becomes
\begin{equation}
\lim_{x_4 \rightarrow \infty} x_4^4 \,G_5 = \frac{1}{x_{12}^4 x_{35}^4} \mathcal G_5(V_i;t_i) \,.
\end{equation}
In Euclidean kinematics we have two small parameters, $s_1$ and $s_2$, and three angle variables $\xi_1$, $\xi_2$ and $\xi_3$ defined by
\begin{align}\label{DoubleOPE}
s_1&=|x_{12}| \,, & \xi_1&= \frac{x_{12}\cdot x_{13}}{|x_{12}|} =\cos\theta_1 \,,& \xi_3 &= \frac{x_{12}\cdot x_{35}-2 x_{12}\cdot x_{13} \,x_{13}\cdot x_{35}}{|x_{12}||x_{35}|}\,,\nonumber\\
s_2&=|x_{35}| \,, & \xi_2&= \frac{x_{13}\cdot x_{35}}{|x_{35}|} =\cos \theta_2\,.
\end{align}
In these variables, the cross ratios defined in \eqref{CrossRatios} become\footnote{Note that if all the five points are restricted to the plane then only four of the five cross ratios are independent, as $\xi_3 =-\cos(\theta_1+ \theta_2)$,   and $V_5$ simplifies to
\begin{equation}
V_5 =\frac{ s_1^2 s_2^2 }{(1-s_1 e^{i\theta_1}+s_2 e^{i\theta_2})(1-s_1 e^{-i\theta_1}+s_2 e^{-i\theta_2})} \,.
\end{equation}}
\begin{align}
&V_1 = s_1^2\,, \qquad  V_2 = 1+s_1^2-2 s_1 \xi_1\,,\qquad V_3 = s_2^2  \,, \qquad  V_4=1+ s_2^2 +2 s_2 \xi_2 \,,\nonumber\\
&V_5= s_1^2 s_2^2 (1+s_1^2 +s_2^2 -2s_1 \xi_1 + 2 s_2 \xi_2 - 2 s_1  s_2 (\xi_3+2 \xi_1\xi_2))^{-1} \,.
\end{align}
Operator product expansion dictates that the five-point function can be expanded in terms of conformal blocks 
\begin{align}
\mathcal{G}_{5}(V_i,t_i) = \sum_{(\Delta_k,J),(\Delta_{k'},J')} \sum_p C_{\mathcal{O}_{\textrm{\bf{20}}'}\mathcal{O}_{\textrm{\bf{20}}'}\mathcal{O}_{k}}C_{\mathcal{O}_{\textrm{\bf{20}}'}\mathcal{O}_{\textrm{\bf{20}}'}\mathcal{O}_{k'}}C_{\mathcal{O}_{\textrm{\bf{20}}'}\mathcal{O}_{k}\mathcal{O}_{k'}}^pG_{k,k'}^{p}(s_i,\xi_i)\;.
\end{align}
The five-point conformal block $G_{k,k'}^{p}(s_i,\xi_i)$ encodes all the contribution of the exchanged primaries $\mathcal{O}_k$, $\mathcal{O}_{k'}$, as well as their conformal descendants. The label $p$ is associated with the different structures of a three-point function with two spinning operators. We will refrain from giving here the explicit expressions for the conformal blocks. They will be given in Appendix \ref{Conformal blocks}, where we discuss how to compute them as series expansions in both $s_1$ and $s_2$.

Similarly, we can apply a single OPE and obtain information about the {\it full} four-point functions.  To achieve this let us consider the OPE of two external scalar operators
\begin{align}
\mathcal{O}_1(x_1)\mathcal{O}_2(x_2)=\sum_{k}\frac{C_{12k}}{(x_{12}^2)^{\frac{\Delta_1+\Delta_2-\Delta_k+J}{2}}}\big[F^{(12k)}(x_{12},\partial_{x_1},D_{z})\mathcal{O}_{k,J}(x_1,z)\big]\,,
\end{align}
where the function $F^{(12k)}(x,\partial_y,D_z)$  and the derivative $D_z$ are defined in Appendix \ref{Conformal blocks}. The exact coefficients in this expansion can be fixed by imposing the consistency of the OPE with the conformal structure of the three-point function.
Applying $F^{(ijk)}$ on the spinning four-point function gives its contribution to the single OPE of the five-point function
\begin{equation}\label{singleF}
\frac{F^{12k}(x_{12},\partial_{x_1},D_{z})}{(x_{12}^2)^{\frac{\Delta_1+\Delta_2-\Delta_k+J}{2}}}\langle \mathcal{O}_{k,J}(x_1,z)\mathcal O_3(x_3)\mathcal{O}_{4}(x_4) \mathcal O_5 (x_5)\rangle \,.
\end{equation}
To proceed, we show how we can distinguish operators with different spins. Four-point functions with an external leg of spin $J$ have $J+1$ conformal structures, where  the coefficient of each structure is a function of the cross ratios
\begin{equation}
\lim_{x_4 \rightarrow \infty} x_4^{2\Delta_4} \langle\mathcal{O}_{k,J}(x_1,z)\mathcal O_3(x_3)\mathcal{O}_{4}(x_4) \mathcal O_5 (x_5)\rangle = \sum_{p=0}^J \alpha_l^{(p)}(w, \bar w) \left(\frac{z\cdot x_{13}}{x_{13}^2} \right)^{p} \left(\frac{z\cdot x_{15}}{x_{15}^2} \right)^{J-p} \,.
\end{equation}
For simplicity we set $x_{13}^2=1$ and  rewrote the cross ratios  defined in \eqref{CrossRatios} with complex variables
\begin{equation}
V_3 = (1-w)(1-\bar w)  \,,\qquad\qquad V_4 = w \bar w\,.
\end{equation}
The derivatives from the OPE expansion \eqref{singleF} act not only on the conformal structures, but also on its coefficients $\alpha_l^{(p)}(w,\bar w)$.
This effect is important for the subleading terms in the expansion of $s_1$.
The leading spin one and spin two contributions to the five-point function are therefore given by
\begin{align}
\mathcal G_{\Delta_k,1}&= s1^{\Delta_k} \left( \frac{ \xi}{w \bar w} \alpha_1^{(0)} + \xi_1 \alpha_1^{(1)}\right) \,,\\
\mathcal G_{\Delta_k,2} &=s1^{\Delta_k} \left( \frac{4\xi^2-w \bar w}{4 w^2 \bar w^2} \alpha_2^{(0)} +\frac{8 \xi \xi_1-w-\bar w}{8 w \bar w}\alpha_2^{(1)} + \frac{4 \xi_1^2-1}{4} \alpha_2^{(2)}\right)\,,
\end{align}
where we introduce the new angle variable
\begin{equation}
\xi=\frac{ x_{12} \cdot x_{15} } {|x_{12}|}\,.
\end{equation}
The dependence on $\xi$ allows us to disentangle the different spinning four-point tensor structures in the single OPE of the scalar five-point function, just like $\xi_3$ parametrizes the contribution of different three-point tensor structures in the double OPE limit. 
Finally, when glueing the three- and four-point spinning correlators of 4d $\mathcal{N}=4$ SYM into five-point function contributions, we also need to perform the contractions of the R-symmetry structures. The details of this procedure can be found in Appendix \ref{Rsymmpoly}.

\subsection{Low-lying operators}\label{ops}

From the representation theory of the 4d $\mathcal{N}=4$ superconformal algebra, we know that the tensor product of two stress tensor multiplets takes the following schematic form \cite{Eden:2001ec,Nirschl:2004pa}
\begin{align}
\mathcal{B}^{\frac{1}{2},\frac{1}{2}}_{[0,2,0],(0,0)}\times \mathcal{B}^{\frac{1}{2},\frac{1}{2}}_{[0,2,0],(0,0)}{}\to  &\;\mathbf{1}+\mathcal{B}^{\frac{1}{2},\frac{1}{2}}_{[0,2,0],(0,0)}+\mathcal{B}^{\frac{1}{2},\frac{1}{2}}_{[0,4,0],(0,0)}+\mathcal{B}^{\frac{1}{4},\frac{1}{4}}_{[2,0,2],(0,0)}+\sum_{J=0}^\infty\mathcal{C}^{1,1}_{[0,0,0],(j,j)} \nonumber\\
{}&+\sum_{J=0}^\infty\mathcal{C}^{\frac{1}{2},\frac{1}{2}}_{[0,2,0],(j,j)}+\sum_{J=0}^\infty\mathcal{C}^{\frac{1}{4},\frac{1}{4}}_{[1,0,1],(j,j)}+\sum_{J=0}^\infty\mathcal{A}^{\Delta}_{[0,0,0],(j,j)}\;.
\end{align}
Here we use the notation $\mathcal{X}^{\frac{s}{4},\frac{\bar{s}}{4}}_{[d_1,d_2,d_3](j,\bar{j})}$ to denote the supermultiplets, where $[d_1,d_2,d_3]$ is the R-symmetry Dykin label of the super primary and $(j,\bar{j})$ are the Lorentz spins. We will also use $J=2j$ when $j=\bar{j}$, as the spin of the superconformal primary. The multiplets $\mathcal{B}$ and $\mathcal{C}$ are short (semi-short) multiplets satisfying $(b,\bar{b})$ and $(c,\bar{c})$ type shortening conditions, while $\mathcal{A}$ are generic long multiplets which do not satisfy any shortening condition. We refer the reader to \cite{Dolan:2002zh} for details of the classification of superconformal multiplets. The multiplets $\mathcal{C}^{1,1}_{[0,0,0],(j,j)}$ contain higher-spin currents and therefore should not appear in an interacting CFT. Moreover, the restriction to the singular and leading regular part of the Euclidean OPE leaves us with only a handful of contributing operators. Below we list the operators which appear in the OPE at the supergravity limit, and enumerate their properties. 

The operators responsible for the singular contributions are: 
\begin{itemize}
	\item The operator $\mathcal{O}_{\mathbf{20'}}$ from the 1/2-BPS multiplet $\mathcal{B}^{\frac{1}{2},\frac{1}{2}}_{[0,2,0],(0,0)}$. It has $\Delta=2$, $J=0$ and $\mathcal{R}=[0,2,0]$
	\begin{equation}\label{O20}
	\mathcal O_{\mathbf{20'}}^{IJ}(x)=\mathrm{Tr}\big(\Phi^{\{I} \Phi^{J\}} \big)(x) \,.
	\end{equation}
	\item The R-symmetry current operator $\mathcal{J}_\mu$ from the 1/2-BPS multiplet $\mathcal{B}^{\frac{1}{2},\frac{1}{2}}_{[0,2,0],(0,0)}$. It has $\Delta=3$, $J=1$ and $\mathcal{R}=[1,0,1]$. 
\end{itemize}
Notice the identity operator contribution is singular as well, but it does not appear in the connected component of the five-point function. For the leading regular contribution, we have
\begin{itemize}
	\item The stress tensor operator $\mathcal{T}_{\mu\nu}$ from the 1/2-BPS multiplet $\mathcal{B}^{\frac{1}{2},\frac{1}{2}}_{[0,2,0],(0,0)}$. It has $\Delta=4$, $J=2$ and $\mathcal{R}=[0,0,0]$.
	\item The bottom component of the 1/2-BPS multiplet $\mathcal{B}^{\frac{1}{2},\frac{1}{2}}_{[0,4,0],(0,0)}$. It has $\Delta=4$, $J=0$ and $\mathcal{R}=[0,4,0]$, and its OPE coefficients with operators of short multiplets are also protected. In the free theory, the 1/2-BPS operator can either be realized as a single-trace operator, or as a double-trace operator of the $\mathbf{20'}$ operators projected to the $[0,4,0]$ representation. Requiring that the operators should have orthonormal two-point functions enforces the single-trace operator to appear in a linear combination with the double-trace operator, and the latter is suppressed by an $\mathcal{O}(1/N)$ coefficient. In the bulk supergravity description, this state is dual to a scalar field which sits at the next level of the KK tower and, by construction, it has a vanishing coupling with two $\mathbf{20'}$ scalar fields. Therefore the dimension-4 1/2-BPS operator which appears in the OPE of the five-point function corresponds to the double-trace operator
	\begin{equation}
	(\mathcal{O}_{\mathbf{105}}^{DT})^{IJKL}= :\mathcal O_{\mathbf{20'}}^{\{IJ} \mathcal O_{\mathbf{20'}}^{KL\}}:\,.
	\end{equation}
	\item The bottom component of the 1/4-BPS multiplet $\mathcal{B}^{\frac{1}{4},\frac{1}{4}}_{[2,0,2],(0,0)}$. This operator has $\Delta=4$, $J=0$ and $\mathcal{R}=[2,0,2]$. It is realized as a double-trace operator plus a single-trace operator with a coefficient of order $\mathcal{O}(1/N)$ \cite{Ryzhov:2001bp,DHoker:2003csh}. For simplicity we write down the operator with a specific choice of the polarization 
	\begin{equation}
	\mathcal \mathcal \mathcal{Q}= \mathrm{Tr} (Z^2)\mathrm{Tr}(X^2) - \mathrm{Tr} (ZX)\mathrm{Tr} (ZX) + \frac{1}{N} \mathrm{Tr}([ZX][ZX]) \,,
	\end{equation}
	where $Z$ and $X$ are two complex scalar fields defined as $Z=\Phi^1+i\Phi^2$, $X=\Phi^3+i\Phi^4$. Note that Wick contraction of the two scalar fields vanishes, so that the operator $\mathcal Q$ is completely traceless.
	The OPE coefficients of 1/4-BPS and 1/2-BPS operators are also protected \cite{DHoker:2001jzy}.
	\item The bottom component of the multiplet $\mathcal{C}^{\frac{1}{2},\frac{1}{2}}_{[0,2,0],(0,0)}$. It has $\Delta=4$, $J=0$, $\mathcal{R}=[0,2,0]$, and is also realized as a double-trace operator (see, {\it e.g.}, \cite{Bianchi:2001cm})
	\begin{equation}
	\mathcal C^{IJ} =: \mathcal O_{\mathbf{20'}}^{IK} \mathcal O_{\mathbf{20'}}^{JK}: -\frac{1}{6} \delta^{IJ} :\mathcal O_{\mathbf{20'}}^{KL}  \mathcal O_{\mathbf{20'}}^{KL}:  \,.
	\end{equation}
	\item The bottom component of the long multiplet $\mathcal{A}^{\Delta}_{[0,0,0],(0,0)}$. This operator has $\Delta\approx4$, $J=0$ and $\mathcal{R}=[0,0,0]$. It is realized as a double-trace operator.
\end{itemize}
Note that the bottom component of $\mathcal{C}^{\frac{1}{4},\frac{1}{4}}_{[1,0,1],(0,0)}$ is also a scalar operator with $\Delta=4$, but it has $\mathcal{R}=[1,0,1]$. The total parity of R-symmetry and spacetime spins of this operator is therefore odd, which forbids it to appear in the OPE.

\subsection{Extracting CFT data }\label{data}

We can already make some qualitative predictions about the result after taking the Euclidean OPE. For example, the only unprotected operator appearing in the OPE is the super primary of the long multiplet, which is an R-symmetry singlet. Therefore none of the other representations should contribute to the logarithmic singularities which are associated with anomalous dimensions. 
Even without decomposing into conformal blocks, an R-symmetry projection of the correlator expanded to order $\mathcal O(s_1^4 s_2^4)$ confirms this to be a feature of our supergravity five-point function. 

Let us now consider the double OPE limit in more detail. In the following we always write the OPE coefficients for {\it normalized operators}, and we strip off the R-symmetry structures which are defined explicitly in Appendix \ref{Rsymmpoly}. This notation follows naturally from the decomposition of the five-point function into the R-symmetry polynomials defined in \eqref{pols}.\footnote{By contrast if we want to directly compute the normalized OPE coefficients in the free theory, we need to evaluate both two- and three-point functions where the former set the normalizations.}
To begin, let us first project the five-point function into the channel in the $\mathbf{20'}$ representation and with $\Delta=2$. This simply corresponds to the intermediate operator being the operator $\mathcal{O}_{\mathbf{20'}}$. We obtain several OPE coefficients with two chiral primaries, which were known previously from the analysis of the $\mathbf{20'}$ supergravity four-point function \cite{Arutyunov:2000ku}
\begin{align}\label{OPEdata4pt}
 C_{\mathcal O_{\mathbf{20'}} \mathcal O_{\mathbf{20'}} \mathcal A}&=\frac{1}{\sqrt{10}} \left(1+ \frac{19}{15 N^2} \right) \,,& C_{\mathcal O_\mathbf{20'}\mathcal O_\mathbf{20'}\mathcal Q} &=\frac{2\sqrt 2}{\sqrt 3} \left(1- \frac{3}{2 N^2} \right) \,,\nonumber\\
C_{\mathcal O_\mathbf{20'} \mathcal O_\mathbf{20'}\mathcal O_{\mathbf{105}}^{DT}} &=\sqrt 2\left(1+ \frac{1}{ N^2} \right)  \,, &
C_{\mathcal O_\mathbf{20'} \mathcal O_\mathbf{20'} \mathcal C} &=\frac{\sqrt 6}{\sqrt 5} \left(1+ \frac{1}{6 N^2} \right) \,.
\end{align}
Note that our results have different normalizations as we use the tensor structures defined in Appendix \ref{Rsymmpoly}. Except for $C_{\mathcal O_{\mathbf{20'}} \mathcal O_{\mathbf{20'}} \mathcal A}$, the three-point functions above are protected, and so they coincide with their free field theory values.\footnote{Three-point functions of half-BPS operators are known to be independent of the coupling, thanks to the non-renormalization theorems \cite{Freedman:1998tz,Lee:1998bxa,Intriligator:1998ig,Intriligator:1999ff,Eden:1999gh,Petkou:1999fv,Howe:1999hz,Heslop:2001gp,Baggio:2012rr}, while three-point functions mixing half- and quarter-BPS operators were shown to be protected in \cite{DHoker:2001jzy}. Meanwhile, the non-renormalization of $C_{\mathcal O_\mathbf{20'} \mathcal O_\mathbf{20'} \mathcal C}$ was observed in \cite{Arutyunov:2000ku}, and proved in \cite{Heslop:2003xu} using superspace techniques.}

Focusing now on intermediate operators of dimension $4$ in both channels, we are able to extract three-point functions which could not be obtained from the four-point function of single-trace operators. In particular, we find the following OPE coefficients
\begin{align}
C_{\mathcal O_\mathbf{20'} \mathcal C \mathcal O_{\mathbf{105}}^{DT}} &=\frac{4 \sqrt 2}{\sqrt{15} N} \left(1+\frac{5}{6N^2}  \right)   \,,&  C_{\mathcal O_\mathbf{20'} \mathcal O_{\mathbf{105}}^{DT} \mathcal O_{\mathbf{105}}^{DT}} &= \frac{4 \sqrt 2}{N}\,\nonumber\\
C_{\mathcal O_\mathbf{20'} \mathcal C \mathcal Q} &=- \frac{2 \sqrt{10}}{3 N} \left(1-\frac{5}{3N^2}  \right) \,, & C_{\mathcal O_\mathbf{20'} \mathcal Q \mathcal Q} &=\frac{8 \sqrt 2}{N} \,.
\end{align}
which match exactly with their free theory values. The three-point functions $C_{\mathcal O_\mathbf{20'} \mathcal O_{\mathbf{105}}^{DT} \mathcal O_{\mathbf{105}}^{DT}} $, $C_{\mathcal O_\mathbf{20'} \mathcal Q \mathcal Q} $ and $C_{\mathcal O_\mathbf{20'} \mathcal C \mathcal O_{\mathbf{105}}^{DT}}$ are known to be protected. We reproduced these three-point functions from supergravity calculations, which gives nontrivial checks of our results. Moreover, the precise match of $C_{\mathcal O_\mathbf{20'} \mathcal C \mathcal Q} $ with the free theory value also strongly indicates that the three-point function is protected, supporting the claim from \cite{Heslop:2003xu} using superspace arguments.

We also extract  the OPE coefficient of one $\mathcal{O}_{\mathbf{20'}}$  with two $\mathcal{C}$ operators at strong coupling, which reads
\begin{equation}\label{OCC}
C_{\mathcal O_{\mathbf{20'}} \mathcal C \mathcal C}  = \frac{9\sqrt 2}{5 N} \left(1+\frac{10}{81 N^2}  \right)\,.\\
\end{equation}
We find that this OPE coefficient does not match the free field theory computation, indicating that this type of three-point function is unprotected.\footnote{At first sight our result seems to contradict the protected nature of the chiral algebra. However, unlike the $\mathcal{B}$-type multiplets where the Schur operators are the super primary, Schur operators in the $\mathcal{C}$-type  are superconformal descendants. We strongly suspect that the three-point functions for one $\mathcal{B}$-type multiplet and two $\mathcal{C}$-type multiplets have more than one superstructure in superspace. The protected chiral algebra three-point function and $\langle \mathcal O_{\mathbf{20'}} \mathcal C \mathcal C\rangle$ are in different superstructures which are unrelated by the action of supercharges. The non-renormalization theorem applies only to the former case. We thank Carlo Meneghelli for discussions on this point.}
Further support for our claim can be obtained from perturbation theory. In \cite{Drukker:2008pi} the authors obtained the five-point function at one loop, and a decomposition in conformal blocks reveals that the OPE coefficient receives a one-loop correction\footnote{The weak coupling analysis is more subtle, as there could be more operators appearing in the OPE. Fortunately, at one loop there is no new scalar operator with dimension 4 and in the $[0,2,0]$ representation, as we can see in the conformal block decomposition of the one-loop four-point function of $O_{\mathbf{20'}}$.}
\begin{equation}
C^{\mathrm{pert}}_{\mathcal O_{\mathbf{20'}} \mathcal C \mathcal C} =\frac{9\sqrt 2  }{5 N} \left(1+ \frac{20(1-15 \lambda)}{27 N^2} \right) \,.
\end{equation}
Finally, we consider the singlet and $\mathbf{20'}$ R-symmetry channels, from which we derive a new OPE coefficient involving the unprotected operator $\mathcal A$ and a semi-short operator $\mathcal{C}$
\begin{equation}\label{coefA}
C_{\mathcal O_{\mathbf{20'}}  \mathcal A \mathcal C} = \frac{2 \sqrt 2}{\sqrt 3 N} \left( 1- \frac{521}{90 N^2} \right)\,.
\end{equation}

While the machinery developed in Appendix \ref{Conformal blocks} 
makes it convenient to directly extract the CFT data, there is still much to gain by performing just one single OPE. It allows us to obtain the complete four-point functions from the five-point function.
To start with, we reproduce the known four-point functions of single-trace operators. We found that the projection of the singular part of the correlator on the $[0,2,0]$ channel is exacly reproduced by the scalar four-point function $\langle \mathcal{O}_\mathbf{20'}\mathcal{O}_\mathbf{20'}\mathcal{O}_\mathbf{20'}\mathcal{O}_\mathbf{20'}\rangle$. The projection of the singular part into the $[0,1,0]$ channel is matched by the four-point function \eqref{corrJD2} of three chiral primaries and one R-symmetry current. Moreover, the $[0,0,0]$ spin two component of the regular part is matched by the four-point function \eqref{corrTD} of three $\mathcal{O}_{\mathbf{20'}}$ and one $\mathcal{T}_{\mu\nu}$.
Once we have removed the contribution of the stress-tensor, we can use the single OPE to extract the correlator of the unprotected double-trace operator with three chiral primaries, which is a new result for strongly coupled planar $\mathcal{N}=4$ SYM
\begin{align}\label{new4pt}
\langle \mathcal A \,\mathcal O_\mathbf{20'} \mathcal O_\mathbf{20'} \mathcal O_\mathbf{20'} \rangle =& \frac{2 \,t_{23} t_{24} t_{34}}{\sqrt 5 \,x_{12}^2 x_{13}^4 x_{14}^2 x_{24}^2 } \Bigg(\frac{ u+v + u \,v }{ u \,v \,N} +\frac{1}{N^3} \bigg( 44 \,\bar D_{2224} +36 \,(1+u+v)\bar D_{2222}  \nonumber\\
&\qquad+\frac{u+v +u \,v}{u \,v}\Big(\frac{139}{15}-8 \,\bar D_{2112}-8 \,v\, \bar D_{2121}\Big)-\frac{28}{3}\bar D_{1111} \bigg)\Bigg)
\end{align}
Much new information is encoded in this correlator, with the OPE coefficient of \eqref{coefA} being just an example of the type of data that can be extracted. Also note that there are many possible rewritings of the correlator \eqref{new4pt} in terms of $D$-functions, and we have only presented the simplest expression. It is also possible to write an expression which requires only $D$-functions of total conformal dimension 10.

\section{Discussion and outlook}\label{discussion}
In this paper, we developed new systematic methods to compute five-point functions from $AdS_5\times S^5$ IIB supergravity. We also obtained five-point conformal blocks in series expansions, which allowed us to perform conformal block decompositions for five-point correlators. As a concrete example, we computed the five-point function of the $\mathbf{20'}$ operator. We performed a number of consistency checks on the $\mathbf{20'}$ five-point function  and extracted new CFT data at strong coupling. 

There are many directions which one can pursue in the future. 
\begin{itemize}
\item First of all, an immediate interesting extension is to apply our methods to more general five-point functions. As the complexity of the correlators grows with the extremality, the best starting point is  correlators with the same extremality as the $\mathbf{20'}$ five-point function. These correlators should have very similar structures, which is particularly manifest in Mellin space. Work in this direction is in progress and we hope to report the results in the near future.
\item Second, we would like to better understand the general structure of the five-point correlation functions dictated by superconformal symmetry. For four-point functions, superconformal constraints boil down to the partial non-renormalization theorem of \cite{Eden:2000bk}. This theorem  reduces the correlators to a free part and a ``quantum correction'' part, which has a much simpler form than the full correlator. For five-point functions the pressing issue is to  find and solve the {\it full} set of constraints from superconformal symmetry, and the solution will constitute the five-point analogue of the ``partial non-renormalization theorem''. Such a solution will give us a more compact way to write the five-point function. 
\item Relatedly, it has recently been observed that the correction part in four-point functions exhibits a hidden ten dimensional conformal symmetry \cite{Caron-Huot:2018kta}. Using this hidden symmetry, one can lift the lowest-weight four-point function into a generating function. Establishing the five-point ``partial non-renormalization theorem'' will be extremely useful for identifying the action of the hidden symmetry at the level of five-point functions. It should then also be possible to write down a generating function which gives five-point functions of arbitrary conformal dimensions. 

\item From our analysis, it is clear that there are close analogies between holographic correlators and flat space scattering amplitudes. For example, factorization in Mellin space played a crucial role in our position space approach of computing the five-point function. We also showed that the $\mathbf{20'}$ five-point function has no intrinsic five-point contact interaction. This seems to suggest certain ``constructibility'' of the holographic correlators. It would be extremely interesting to develop such constructive approaches further and extend them to higher points, perhaps in the form of Mellin recursion relations similar to the famous BCFW relation \cite{Britto:2005fq}. 

\item One aspect which we have not considered in detail is the flat space limit. We would like to examine this limit more carefully in the future. The flat space limit will also be important when we consider higher-derivative (stringy) corrections to the five-point functions, as has been emphasized in the four-point function case by, {\it e.g.}, \cite{Goncalves:2014ffa,Chester:2018aca,Binder:2018yvd,Alday:2018pdi,Alday:2018kkw,Binder:2019jwn}.

\item Finally, the technology developed in this paper can be readily applied to eleven dimensional supergravity on $AdS_7\times S^4$. The chiral algebra in six dimensions \cite{Beem:2014kka} places strong constrains on the five-point functions. However, it is not clear if a twist similar to the $SO(6)$ twist of \cite{Drukker:2009sf} exists for the $(2,0)$ theories. It may be necessary to resort to the flat space limit, which gives extra constraints on contact interactions. It would be interesting to compute five-point correlators for this background, and extract new information about the $(2,0)$ theory in six dimensions. 
\end{itemize}

\acknowledgments
We thank Fernando Alday, Paul Heslop, Tristan McLoughlin, Carlo Meneghelli, Wolfger Peelaers, Joao Penedones, Yanliang Shi and especially Sergey Frolov and Leonardo Rastelli for many useful discussions and/or comments on the draft. We thank the International Institute of Physics for hospitality during the workshop ``Nonperturbative Methods for Conformal Theories''. X.Z. also thanks the organizers of the Pollica Summer Workshop for their hospitality during the final stage of this project. The Pollica Summer Workshop was supported in part by the Simons Foundation (Simons Collaboration on the Non-perturbative Bootstrap) and in part by the INFN. The work of V.G. is supported by FAPESP grant 2015/14796-
7.  The work of X.Z. is supported in part by the Simons Foundation Grant No. 488653.  R.P. is supported by SFI grant 15/CDA/3472.

\appendix
\section{Integrating out an internal line}\label{integratingout}
In this appendix we extend the method of \cite{DHoker:1999mqo} to higher-point Witten diagrams with more than one internal line, and evaluate the various diagrams that appear in Section \ref{5ptcorrelator}. The key point is that integrating out an internal line replaces the integrated cubic vertex by a sum of contact vertices. When the quantum numbers are fine-tuned to satisfy certain conditions (such as in $\mathcal{N}=4$ SYM and the 6d $(2,0)$ theory), the sum in the contact vertices truncates to finitely many terms. Repeated use of the vertex identities then allows us to write an exchange Witten diagram in terms of a finite sum of contact diagrams. 

\subsection{Vertex identities}
In \cite{DHoker:1999mqo} the consequence of integrating out a bulk-to-bulk propagator for a four-point exchange Witten diagram was worked out. The upshot is that the exchange Witten diagram can be expressed in terms of a sum of four-point contact diagrams. For our purpose, we want to extract from their result the vertex identities that relate an integrated cubic vertex to a sum of contact vertices.

\subsubsection{Scalar exchange}
Let us start with the scalar internal line. The integrated cubic interaction is
\begin{equation}
I_{\rm scalar}\equiv \int \frac{d^{d+1}z}{z_0^{d+1}} G_{B\partial}^{\Delta_1}(z,x_1)G_{B\partial}^{\Delta_2}(z,x_2)G_{BB}^{\Delta}(z,y)\;.
\end{equation}
Using the result in \cite{DHoker:1999mqo}, we can express this integral as
\begin{equation}
I_{\rm scalar}=\sum_{k=k_{\rm min}}^{k_{\rm max}} a_k (x_{12}^2)^{k-\Delta_2} G_{B\partial}^{k+\Delta_1-\Delta_2}(y,x_1)\;G_{B\partial}^{k}(y,x_2)\;
\end{equation}
where
\begin{equation}
\begin{split}
{}&k_{\rm min}=(\Delta-\Delta_{1}+\Delta_2)/2\;,\;\;\;\;\; k_{\rm max}=\Delta_2-1\;,\\
{}&a_{k-1}= \frac{(k-\frac{\Delta}{2}+\frac{\Delta_{1}-\Delta_2}{2})(k-\frac{d}{2}+\frac{\Delta}{2}+\frac{\Delta_{1}-\Delta_2}{2})}{(k-1)(k-1-\Delta_{1}+\Delta_2)}a_k\;,\\
{}& a_{\Delta_2-1}=\frac{1}{4(\Delta_1-1)(\Delta_2-1)}\;,
\end{split}
\end{equation}
with $\Delta_1+\Delta_2-\Delta$ being a positive even integer.

\subsubsection{Graviphoton exchange}
We now consider the integral involving the exchange of a vector field of general dimension $\Delta$. When $\Delta=d-1$, the vector field is a massless gauge field and couples to a conserved current. Denoting $\Delta_1=\Delta_2=\Delta_{\rm ext}$, we will consider the coupling of the vector field to a conserved current
\begin{equation}
I_{\rm vector}^\mu\equiv \int \frac{d^{d+1}z}{z_0^{d+1}}  \left(G_{B\partial}^{\Delta_{\rm ext}}(z,x_1) {\overset{\leftrightarrow}{\triangledown}}_\nu G_{B\partial}^{\Delta_{\rm ext}}(z,x_2)\right) G_{BB}^{\Delta,1,\mu\nu}(z,y)\;.
\end{equation}
where $G_{BB}^{\Delta,1,\mu\nu}(z,y)$ is the vector bulk-to-bulk propagator. This integral can be evaluated as a sum of contact vertices 
\begin{equation}
I_{\rm vector}^\mu=-\sum_{k=k_{\rm min}}^{k_{\rm max}} \frac{a_k}{2k}(x_{12}^2)^{-\Delta_{\rm ext}+k} g^{\mu\nu}(y) \left(G_{B\partial}^{k}(y,x_1) {\overset{\leftrightarrow}{\triangledown}}_\nu G_{B\partial}^{k}(y,x_2)\right)
\end{equation}
where
\begin{equation}
   \begin{split}
      k_{\rm min} = {}& \frac{d-2}{4}+\frac{1}{4}\sqrt{(d-2)^2+4(\Delta-1)(\Delta-d+1)}\;, \\
      k_{\rm max}=   {}& \Delta_{\rm ext}-1\;, \\
      a_{k-1}= {}& \frac{2k(2k+2-d)-(\Delta-1)(\Delta-d+1)}{4(k-1)k}a_k\;,\\
      a_{\Delta_{\rm ext}-1}={}&\frac{1}{2(\Delta_{\rm ext}-1)}\;.
   \end{split}
\end{equation}
The truncation requires that $k_{\rm max}-k_{\rm min}$ is a non-negative integer. Notice that in evaluating the cubic integral, vanishing divergence of the source is not required. Therefore this result holds even when the source coupled to $I_{\rm vector}^\mu$ is not conserved.

\subsubsection{Graviton exchange}\label{vertexidgraviton}
Finally we consider the cubic integral involving a graviton field. Let $\Delta_1=\Delta_2=\Delta_{\rm ext}$, the cubic integral is 
\begin{equation}
\begin{split}
I_{\rm graviton}^{\mu\nu}={}&\int \frac{d^{d+1}z}{z_0^{d+1}} G^{\Delta=d-2,\ell=2,\;\mu\nu;\rho\sigma}_{BB}(z,y)\times\bigg(\triangledown_{\rho} G^{\Delta_{\rm ext}}_{B\partial}(z,x_1) \triangledown_{\sigma} G^{\Delta_{\rm ext}}_{B\partial}(z,x_2)\\
{}&-\frac{1}{2}g^{\rho\sigma}(z)(\triangledown^\kappa G^{\Delta_{\rm ext}}_{B\partial}(z,x_1)  \triangledown_\kappa G^{\Delta_{\rm ext}}_{B\partial}(z,x_2)+m^2 G^{\Delta_{\rm ext}}_{B\partial}(z,x_1) G^{\Delta_{\rm ext}}_{B\partial}(z,x_2))\bigg)\;.
\end{split}
\end{equation}
Using the result of \cite{DHoker:1999mqo}, we find that this integral reduces to the following sum of contact vertices
\begin{equation}
\begin{split}
I_{\rm graviton}^{\mu\nu}={}&\sum_{k=k_{\rm min}}^{k_{\rm max}}a_k (x_{12}^2)^{-\Delta_{\rm ext}+k}\bigg( \frac{g^{\mu\nu}(y)}{d-1}G^{k}_{B\partial}(y,x_1) G^{k}_{B\partial}(y,x_2)\\
{}&+\frac{1}{k(k+1)}\left(D^\mu D^\nu G^{k}_{B\partial}(y,x_1)+k g^{\mu\nu}(y)G^{k}_{B\partial}(y,x_1)\right)G^{k}_{B\partial}(y,x_2)\bigg)
\end{split}
\end{equation}
where
 \begin{equation}
   \begin{split}
      k_{\rm min} = {}& \frac{d}{2}-1\;, \\
      k_{\rm max}=   {}& \Delta_{\rm ext}-1\;, \\
      a_{k-1}= {}& \frac{k+1-\frac{d}{2}}{k-1}a_k\;,\\
      a_{\Delta_{\rm ext}-1}={}&-\frac{\Delta_{\rm ext}}{2(\Delta_{\rm ext}-1)}\;.
   \end{split}
\end{equation}

For the above expression to be valid, $I^{\mu\nu}_{graviton}$ must be coupled to a conserved current. This is because in the derivation of \cite{DHoker:1999mqo} total derivative terms in $I^{\mu\nu}_{graviton}$ are assumed to drop out, which is consistent only when coupled to divergence-less sources. On the other hand, when $I^{\mu\nu}_{graviton}$ is coupled to a non-conserved source, there is a nonzero contribution from the total derivative terms in $I^{\mu\nu}_{graviton}$. The contribution of these terms cannot be determined using the techniques of \cite{DHoker:1999mqo}.

\subsection{Five-point exchange Witten diagrams}
Using the vertex identities, we can evaluate the exchange diagrams that we encountered in Section \ref{5ptcorrelator}. We record here their explicit expressions. 
\subsubsection{$W^{s_{[12]},s_{[34]}}$}
The double-exchange Witten diagram $W^{s_{[12]},s_{[34]}}$ is defined by (\ref{Wss}). It evaluates to 
\begin{equation}
W^{s_{[12]},s_{[34]}}=\frac{D_{11112}}{16x_{12}^2x_{34}^2}\;.
\end{equation}
\subsubsection{$W^{s_{[12]}}_{\rm 0-der}$}
The scalar single-exchange diagram with a zero-derivative quartic vertex is defined by 
\begin{equation}
W^{s_{[12]}}_{\rm 0-der}=\int \frac{dz^5}{z_{0}^5} \frac{dy^5}{y_{0}^5}G^{\Delta=2}_{B\partial}(z;x_1) G^{\Delta=2}_{B\partial}(z;x_2)G^{\Delta=2}_{BB}(z;y)G^{\Delta=2}_{B\partial}(y;x_3) G^{\Delta=2}_{B\partial}(y;x_4)G^{\Delta=2}_{B\partial}(y;x_5)\;.
\end{equation}
It has the value
\begin{equation}
W^{s_{[12]}}_{\rm 0-der}=\frac{D_{11222}}{4x_{12}^2}\;.
\end{equation}
\subsubsection{$W^{s_{[12]},(5)}_{\rm 2-der}$}
The scalar single-exchange diagram with a two-derivative quartic vertex in which the two derivatives are on the external legs 3 and 4 is defined by
\begin{equation}
\begin{split}
W^{s_{[12]},(5)}_{\rm 2-der}=\int \frac{dz^5}{z_{0}^5} \frac{dy^5}{y_{0}^5}{}&G^{\Delta=2}_{B\partial}(z;x_1) G^{\Delta=2}_{B\partial}(z;x_2)G^{\Delta=2}_{BB}(z;y)G^{\Delta=2}_{B\partial}(y;x_5)\\
{}&\times \triangledown_{y,\mu} G^{\Delta=2}_{B\partial}(y;x_3) \triangledown_y^\mu G^{\Delta=2}_{B\partial}(y;x_4)\;.
\end{split}
\end{equation}
Using the identity 
\begin{equation}
\triangledown^\mu G_{B\partial}^{\Delta_1}\triangledown_\mu G_{B\partial}^{\Delta_2}=\Delta_1\Delta_2(G_{B\partial}^{\Delta_1} G_{B\partial}^{\Delta_2}-2x_{12}^2 G_{B\partial}^{\Delta_1+1}G_{B\partial}^{\Delta_2+1})\;,
\end{equation}
we find
\begin{equation}
W^{s_{[12]},(5)}_{\rm 2-der}=\frac{1}{x_{12}^2}\left(D_{11222}-2x_{34}^2D_{11332}\right)\;.
\end{equation}
\subsubsection{$W^{V_{[12]},V_{[34]}}$}
The double-exchange diagram $W^{V_{[12]},V_{[34]}}$ is defined by (\ref{Wvv}). Using the vertex identities, we have
\begin{equation}\label{Wvvvalue}
\begin{split} 
W^{V_{[12]},V_{[34]}}=\frac{1}{2x_{12}^2x_{34}^2}{}&\bigg(-x_{24}^2D_{12122}+x_{23}^2D_{12212}+x_{14}^2D_{21122}-x_{13}^2D_{21212}\\
{}&+2(x_{13}^2x_{24}^2-x_{14}^2x_{23}^2)D_{22222}\bigg)\;.
\end{split}
\end{equation}
\subsubsection{$W^{V_{[12]},s_{[34]}}$}
The double-exchange diagram $W^{V_{[12]},s_{[34]}}$ is defined in (\ref{Wvs}) and it evaluates to
\begin{equation}\label{Wvsvalue}
\begin{split}
W^{V_{[12]},s_{[34]}}=\frac{1}{8 x_{12}^2x_{34}^2}{}&\bigg(-2 x_{25}^2D_{12113}+x_{24}^2D_{12122}+x_{23}^2D_{12212}+2x_{15}^2 D_{21113}\\
{}&-x_{14}^2D_{21122}-x_{13}^2D_{21212}\bigg)\;.
\end{split}
\end{equation}

\subsubsection{$W^{\varphi_{[12]},s_{[34]}}$}
The definition of the double-exchange diagram $W^{\varphi_{[12]},s_{[34]}}$ is given by (\ref{Wgs}).  As we commented before, the vertex identity in Section \ref{vertexidgraviton} does not hold because the source (\ref{IT5}) is not conserved. A naive application of the identities leads to a wrong answer since the dropped total derivative terms have nonzero contributions. On the other hand, the total coupling to the graviton field is conserved when we sum up all the diagrams, and the extra contributions due to the ignored total derivative terms will vanish in the sum. Therefore it does not matter that we use the vertex identities of Section \ref{vertexidgraviton} to evaluate the diagrams so long as all the diagrams are added up correctly at the end of the day. With this caveat, we find that
\begin{equation}\label{Wgsvalue}
\begin{split}
W^{\varphi_{[12]},s_{[34]}}\overset{\bullet}{=}\frac{1}{3x_{12}^2x_{34}^2}{}&\bigg(2D_{11112}-3(x_{14}^2D_{21122}+x_{13}^2D_{21212})\\
{}&+6x_{15}^2(D_{21113}-x_{14}^2D_{31123}-x_{13}^2D_{31213})\bigg)
\end{split}
\end{equation}
where we used $\overset{\bullet}{=}$ to remind us that this expression only makes sense in the sum of all diagrams.

\subsubsection{$W^{\varphi_{[12]}}$}
The single-exchange diagram $W^{\varphi_{[12]}}$ is defined by (\ref{WgSE}). The coupling to the graviton is also non-conserved, but we will evaluate it with the same caveat for $W^{\varphi_{[12]},s_{[34]}}$. Using the vertex identities we have
\begin{equation}\label{WgSEvalue}
W^{\varphi_{[12]}}\overset{\bullet}{=}-\frac{2}{3x_{12}^2}D_{11222}\;.
\end{equation}

\section{R-symmetry polynomials}\label{Rsymmpoly}
An R-symmetry basis can be obtained by solving the two-particle Casimir equations in two compatible channels.  In terms of the null vectors, the $SO(6)_R$ generators take the form
\begin{equation}
L^{(i)}_{IJ}=t_{i,I}\frac{\partial}{\partial t_i^J}-t_{i,J}\frac{\partial}{\partial t_i^I}\;.
\end{equation}
The two-particle Casimir operator, say for 1 and 2, is constructed from $L^{(1)}_{IJ}$ and $L^{(2)}_{IJ}$
\begin{equation}
\mathcal{C}^{(1,2)}=\frac{1}{2}\left(L^{(1)}_{IJ}+L^{(2)}_{IJ}\right)\left(L^{(1),IJ}+L^{(2),IJ}\right)\;.
\end{equation}
When acting on the five-point correlator, which is a polynomial of $t_{ij}=t_i\cdot t_j$, the two-particle Casimir $\mathcal{C}^{(1,2)}$ can be written as
\begin{equation}
\begin{split}
\mathcal{C}^{(1,2)}
={}&-\sum_{i,j=3,4,5}(\mathbb{D}_{t_{1i}}\mathbb{D}_{t_{1j}})-(d_R-2)\sum_{i=3,4,5}\mathbb{D}_{t_{1i}}\\
{}&-\sum_{i,j=3,4,5}(\mathbb{D}_{t_{2i}}\mathbb{D}_{t_{2j}})-(d_R-2)\sum_{i=3,4,5}\mathbb{D}_{t_{2i}}\\
{}&+2\sum_{i,j=3,4,5}\big(t_{12} t_{ij}-t_{1i}t_{2j}\big)\frac{\partial}{\partial t_{1j}}\frac{\partial}{\partial t_{2i}}
\end{split}
\end{equation}
where $d_R=6$ for $SO(6)$ and $\mathbb{D}_x\equiv x\frac{\partial}{\partial x}$. Other two-particle Casimir operators $\mathcal{C}^{(i,j)}$ are similarly defined and can be obtained from $\mathcal{C}^{(1,2)}$ by permuting the labels. 

\begin{figure}[htbp]
\begin{center}
\includegraphics[width=0.5\textwidth]{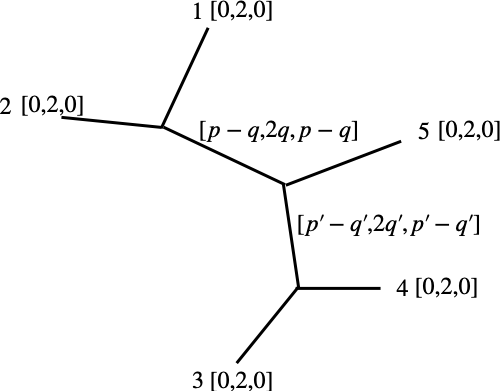}
\caption{An R-symmetry channel of the $\mathbf{20'}$ five-point function where the representation $[p-q,2q,p-q]$ is exchanged in 12 and the representation $[p'-q',2q',p'-q']$ is exchanged in 34.}
\label{Rsymchans}
\end{center}
\end{figure}

We now consider the solution to the following Casimir equations 
\begin{eqnarray}
\mathcal{C}^{(a,b)}\circ R^{(p,q),(p',q')}_{ab|cd}&=&-2(p(p+d_R-3)+q(q+1)) R^{(p,q),(p',q')}_{ab|cd}\;,\\
\mathcal{C}^{(c,d)}\circ R^{(p,q),(p',q')}_{ab|cd}&=&-2(p'(p'+d_R-3)+q'(q'+1)) R^{(p,q),(p',q')}_{ab|cd}\;
\end{eqnarray}
where $a$, $b$, $c$, $d$ are different points. When the quantum numbers $\{p,p',q,q'\}$ are such that the solution is nontrivial, there exists a process where the $[0,2,0]$ representations at points $a$ and $b$ merge into the representation $[p-q,2q,p-q]$, while the tensor product of points $c$ and $d$ produces the representation $[p'-q',2 q',p'-q']$. This process is illustrated  by Figure \ref{Rsymchans}. The solution  $R^{(p,q),(p',q')}_{ab|cd}$ is the R-symmetry structure associated with the process.  R-symmetry selection rules at the vertices with $a$, $b$ and $c$, $d$ require $0\leq q\leq p\leq 2$ and $0\leq q'\leq p'\leq 2$, and the solutions are further restricted by the selection rule at the vertex with the remaining external point $e$. There are in total 22 solutions to the equations, which are in correspondence with the 22 R-symmetry structures and form a complete basis. 
Notice that when the two intermediate representations are $[1,2,1]$ there are two solutions to the Casimir equations. This is because $[1,2,1]$ appears twice in the tensor product of $[1,2,1]$ with $[0,2,0]$.

The OPE coefficients are meaningfully defined only when we set the conventions for the three-point $SU(4)$ tensor structures. In the context of a double OPE analysis it is useful to fix the normalization of the polynomials $R^{(p,q),(p',q')}_{ab|cd}$ such that they correspond to the product of three such structures. 
In order to do so, let us first introduce the tensors associated with each of the representations $[p-q,2q,p-q]$. For $p=2$ we encode the representations by traceless orthonormal tensors of rank 4, denoted by
\begin{equation}
(C^2_q)^J_{ijkl}\,,
\end{equation}
where $i,j,k,l$ are $SO(6)$ vector indices, and $J$ parametrizes the degrees of freedom of the representation. 
$C^2_2$ is completely symmetric, $C^2_1$ is symmetric in both $i \leftrightarrow j$ and $k \leftrightarrow l$  but antisymmetric in the exchange $(ij) \leftrightarrow (kl)$, while $C^2_0$ is antisymmetric in both $i \leftrightarrow k$ and $j \leftrightarrow l$, symmetric in the exchange $(ik) \leftrightarrow (jl)$ and also obeys $\epsilon^{ijklmn} (C^2_0)^J_{klmn}=0$. Similarly, for $p=1$ we encode  representations by traceless orthonormal tensors of rank 2, which we write as
\begin{equation}
(C^1_q)^{J}_{ij}\,,
\end{equation}
where $C^1_1$ is symmetric and $C^1_0$ is antisymmetric.

We can now set our definitions for the tensor structures arising in all relevant three-point functions. When two of the representations are the $\mathbf{20'}$, all possible tensor contractions can be related to the following three-point structures
\begin{align}
T_{(2,q)}^{I_1 I_2 J} &=(C^1_1)^{I_1}_{ij}  (C^1_1)^{I_2}_{kl} (C^2_q)^{J}_{ijkl} \,,\\
T_{(1,q)}^{I_1 I_2 J} &=(C^1_1)^{I_1}_{ik}  (C^1_1)^{I_2}_{jk}  (C^1_q)^{J}_{ij} \,,\\
T_{(0,0)}^{I_1 I_2 } &=  (C^1_1)^{I_1}_{ij}  (C^1_1)^{I_2}_{ij} = \delta^{I_1 I_2}\,.
\end{align}
If only one of the representations is the $\mathbf{20'}$, then the independent three-point structures can be chosen as
\begin{align}
T_{(2,q),(2,q'),1}^{J I K} &=  (C^2_q)^{J}_{iklm}  (C^1_1)^{I}_{ij}  (C^2_{q'})^{K}_{jklm}\,,\\
T_{(2,1),(2,1),2}^{J I K} &=  (C^2_1)^{J}_{iklm}  (C^1_1)^{I}_{ij}  (C^2_1)^{K}_{jlkm}\,,\\
T_{(1,0),(2,1)}^{J I K} &=  (C^1_0)^{J}_{ij}  (C^1_1)^{I}_{kl}  (C^2_{1})^{K}_{ikjl}\,,\\
T_{(1,0),(1,0)}^{J I K} &=  (C^1_0)^{J}_{ik}  (C^1_1)^{I}_{ij}  (C^1_{0})^{K}_{jk}\,.
\end{align}
Note that when both $(p,q)$ and $(p',q')$ are equal to $(2,1)$ there are two possible tensor structures, which reflects the fact that there are two independent singlets in the tensor product $[1,2,1]\times[1,2,1]\times[0,2,0]$.

It is natural to write the  basis of four-point structures in the $(ab)$ channel as the product of three-point structures introduced above
\begin{equation}
T^{I_a I_b | I_c I_d}_{(p,q)}= T^{I_a I_b J}_{(p,q)} \, T^{I_c I_d J}_{(p,q)}  \,,
\end{equation}
which can be written as polynomials in $t_{ij}$ by performing the following contraction
\begin{equation} \label{R4def}
	R^{(pq)}_{ab|cd} =T^{I_a I_b | I_c I_d}_{(p,q)} (C^1_1)^{I_a}_{i_1 j_1} t_a^{i_1} t_a^{j_1} \ldots (C^1_1)^{I_d}_{i_4 j_4} t_d^{i_4} t_d^{j_4}\,.
\end{equation}
For example, the basis suitable for the OPE in the 12 channel of the four-point function would be
\begin{align}\label{R4pt}
R^{(0,0)}_{12|34} &= t_{12}^2 t_{34}^2  \,,\nonumber\\
R^{(1,0)}_{12|34} &= \frac{1}{2} t_{12} t_{34} (t_{13} t_{24}-t_{14}t_{23} )\,,\nonumber\\
R^{(1,1)}_{12|34} &= \frac{1}{2} t_{12} t_{34} (t_{13} t_{24}+t_{14}t_{23} )- \frac{1}{6} t_{12}^2 t_{34}^2\,, \nonumber\\
R^{(2,0)}_{12|34} &=\frac{1}{4}( t_{13}^2 t_{24}^2+ t_{14}^2 t_{23}^2 ) - \frac{1}{2} t_{13} t_{14} t_{23} t_{24}-\frac{1}{8}(t_{12} t_{14}t_{23} t_{34}+ t_{12}t_{13} t_{24} t_{34} ) +\frac{1}{40} t_{12}^2 t_{34}^2 \,,\nonumber\\
R^{(2,1)}_{12|34} &= \frac{1}{2}( t_{13}^2 t_{24}^2- t_{14}^2 t_{23}^2 ) +\frac{1}{4}(t_{12} t_{14}t_{23} t_{34}- t_{12}t_{13} t_{24} t_{34} ) \,,\nonumber\\
R^{(2,2)}_{12|34} &=\frac{1}{6}( t_{13}^2 t_{24}^2+ t_{14}^2 t_{23}^2 ) + \frac{2}{3} t_{13} t_{14} t_{23} t_{24}-\frac{2}{15}(t_{12} t_{14}t_{23} t_{34}+ t_{12}t_{13} t_{24} t_{34} ) +\frac{1}{60} t_{12}^2 t_{34}^2\,.
\end{align}
Analogously, the natural basis of five-point tensor structures in the double OPE analysis in $(ab)$ and $(cd)$ channels is then given by the following product of three three-point structures
\begin{equation}
T^{I_a I_b | I_c I_d | I_e}_{(p,q),(p',q'),i}= T^{I_a I_b J}_{(p,q)} \, T_{(p,q),(p',q'),i}^{J I_e K} \,T^{I_c I_d K}_{(p',q')}  \,.
\end{equation}
With these definitions, the normalization of the polynomials $R^{(p,q),(p',q')}_{ab|cd}$ is fixed by requiring that
\begin{equation}
R^{(pq),(p',q')}_{ab|cd,i} =T^{I_a I_b | I_c I_d | I_e}_{(p,q),(p',q'),i} (C^1_1)^{I_a}_{i_1 j_1} t_a^{i_1} t_a^{j_1} \ldots (C^1_1)^{I_e}_{i_5 j_5} t_e^{i_5} t_e^{j_5}\,.
\end{equation}
Performing the contractions on the right-hand side we obtain the following basis suitable for a double OPE in the 12 and 34 channels
\begin{align}\label{pols}
R^{(0,0),(1,1)}_{12|34} &= \mathcal H \nonumber\,,
\\
R^{(1,0),(1,0)}_{12|34} &= -\frac{\mathcal A_1 -\mathcal A_2-\mathcal A_3 +\mathcal A_4}{4} \,,\nonumber
\\
R^{(1,0),(1,1)}_{12|34} &= \frac{\mathcal A_1 -\mathcal A_2+\mathcal A_3 -\mathcal A_4}{4}  \,,\nonumber
\\
R^{(1,1),(1,1)}_{12|34} &= \frac{\mathcal A_1 +\mathcal A_2+\mathcal A_3 +\mathcal A_4}{4}  -\frac{\mathcal H}{6}-\frac{\mathcal I}{6}  \,,\nonumber
\\
R^{(1,0),(2,1)}_{12|34} &= \frac{\mathcal A_1 -\mathcal A_2-\mathcal A_3 +\mathcal A_4}{16}  -\frac{\mathcal D_1 - \mathcal D_2}{2} \,,\nonumber
\\
R^{(1,1),(2,0)}_{12|34} &= -\frac{\mathcal A_1 +\mathcal A_2+\mathcal A_3 +\mathcal A_4}{16}-\frac{\mathcal D_1+ \mathcal D_2}{4}   + \frac{\mathcal E_1+\mathcal E_2}{4}  +\frac{\mathcal H}{8}  +\frac{ \mathcal I}{40} \,,\nonumber
\\
R^{(1,1),(2,1)}_{12|34}  &= \frac{\mathcal A_1 +\mathcal A_2-\mathcal A_3 -\mathcal A_4}{8}   -\frac{\mathcal E_1 - \mathcal E_2}{2} \,,\nonumber
\\
R^{(1,1),(2,2)}_{12|34}  &= -\frac{\mathcal A_1 +\mathcal A_2+\mathcal A_3 +\mathcal A_4}{15} +\frac{\mathcal D_1+\mathcal D_2}{3}+ \frac{\mathcal E_1+\mathcal E_2}{6}    -\frac{ \mathcal H}{15} + \frac{ \mathcal I}{60} \,,\nonumber
\\
R^{(2,0),(2,0)}_{12|34}  &= -3\frac{\mathcal A_1 +\mathcal A_2+\mathcal A_3 +\mathcal A_4}{128} -\frac{\mathcal B_1+ \mathcal B_2+\mathcal B_3+\mathcal B_4}{16} + \frac{\mathcal C_1+ \mathcal C_2+\mathcal C_3+\mathcal C_4}{16} \nonumber\\
&\quad -\frac{\mathcal D_1 + \mathcal D_2}{32}+ \frac{\mathcal E_1 + \mathcal E_2}{32} -\frac{\mathcal F_1 + \mathcal F_2}{32} + \frac{\mathcal G_1 + \mathcal G_2}{32} +\frac{\mathcal H}{64}  + \frac{\mathcal I}{64} \,, \nonumber
\\
R^{(2,0),(2,1)}_{12|34}  &= -\frac{\mathcal A_1 +\mathcal A_2-\mathcal A_3 -\mathcal A_4}{32}  -\frac{\mathcal B_1 +\mathcal B_2-\mathcal B_3 -\mathcal B_4}{8} - \frac{\mathcal C_1 +\mathcal C_2-\mathcal C_3 -\mathcal C_4}{8} +\frac{\mathcal E_1 - \mathcal E_2}{16} \,, \nonumber
\\
R^{(2,1),(2,1)}_{12|34,1}  &= 3\frac{\mathcal A_1 -\mathcal A_2-\mathcal A_3 +\mathcal A_4}{64}  +\frac{\mathcal C_1 -\mathcal C_2-\mathcal C_3 +\mathcal C_4}{4} +\frac{\mathcal D_1 - \mathcal D_2}{8} + \frac{\mathcal F_1 - \mathcal F_2}{8} \,,  \nonumber
\\
R^{(2,1),(2,1)}_{12|34,2}  &= 5\frac{\mathcal A_1 -\mathcal A_2-\mathcal A_3 +\mathcal A_4}{128} -\frac{\mathcal B_1 -\mathcal B_2-\mathcal B_3 +\mathcal B_4}{4}  -\frac{\mathcal D_1 - \mathcal D_2}{16} -\frac{\mathcal F_1 -\mathcal F_2}{16} \,,\nonumber 
\\
R^{(2,1),(2,2)}_{12|34}  &= -\frac{\mathcal A_1 +\mathcal A_2-\mathcal A_3 -\mathcal A_4}{30} +\frac{\mathcal B_1 + \mathcal B_2 - \mathcal B_3 -\mathcal B_4}{6} + \frac{\mathcal C_1 +\mathcal C_2-\mathcal C_3 -\mathcal C_4}{12} +\frac{\mathcal G_1-\mathcal G_2}{30}  \,,\nonumber
\\
R^{(2,2),(2,2)}_{12|34}  &=-\frac{\mathcal A_1 +\mathcal A_2+\mathcal A_3 +\mathcal A_4}{50}+\frac{\mathcal B_1 + \mathcal B_2 + \mathcal B_3 +\mathcal B_4}{6}+\frac{\mathcal C_1 +\mathcal C_2+\mathcal C_3 +\mathcal C_4}{12} \nonumber\\
&\quad-\frac{\mathcal D_1 + \mathcal D_2}{15} - \frac{\mathcal E_1 + \mathcal E_2}{30}-\frac{\mathcal F_1 + \mathcal F_2}{15}  - \frac{\mathcal G_1 + \mathcal G_2}{30} +\frac{\mathcal H}{75}  + \frac{\mathcal I}{75}  \,,
\end{align}
where we introduced the following short-hand notation for the monomials
\begin{align}
\mathcal{A}_1 &= t_{12} t_{23} t_{34} t_{45} t_{51} \,, &\mathcal{A}_2 &= t_{21} t_{13} t_{34} t_{45} t_{52} \,,  &\mathcal{A}_3 &= t_{12} t_{24} t_{43} t_{35} t_{51}  \,, &\mathcal{A}_4 &= t_{21} t_{14} t_{43} t_{35} t_{52}\,,  \nonumber\\
\mathcal{B}_1 &= t_{13} t_{32} t_{24} t_{45} t_{51}\,,  &\mathcal{B}_2 &= t_{23} t_{31} t_{14} t_{45} t_{52}\,,  &\mathcal{B}_3 &= t_{14} t_{42} t_{23} t_{35} t_{51} \,, &\mathcal{B}_4 &= t_{24} t_{41} t_{13} t_{35} t_{52}  \,,\nonumber\\
\mathcal{C}_1 &= t_{13} t_{35} t_{51} t_{24}^2  \,,&\mathcal{C}_2 &= t_{23} t_{35} t_{52} t_{14}^2   \,,&\mathcal{C}_3 &= t_{14} t_{45} t_{51} t_{23}^2 \,, &\mathcal{C}_4 &= t_{24} t_{45} t_{52} t_{13}^2   \,,\nonumber\\
\mathcal{D}_1 &=t_{12} t_{23} t_{35} t_{54} t_{41} \,,&\mathcal{D}_2 &= t_{12} t_{24} t_{45} t_{53} t_{31}\,,  &\mathcal{E}_1 &= t_{12} t_{23} t_{31} t_{45}^2 \,, &\mathcal{E}_2 &= t_{12} t_{24} t_{41} t_{35}^2  \,, \nonumber\\
\mathcal{F}_1 &=t_{34} t_{41} t_{15} t_{52} t_{23} \,,&\mathcal{F}_2 &= t_{34} t_{42} t_{25} t_{51} t_{13} \,, &\mathcal{G}_1 &= t_{34} t_{41} t_{13} t_{25}^2  \,,&\mathcal{G}_2 &= t_{34} t_{42} t_{23} t_{15}^2 \,,  \nonumber\\
\mathcal{H} &=t_{34} t_{45} t_{53} t_{12}^2 \,,&\mathcal{I} &= t_{12} t_{25} t_{51} t_{34}^2  \,.
\end{align}

When performing the single OPE, it is useful to know how the structures of the four-point function contract with the three-point structure, so that we can recognize their contribution to the five-point function. We will now see that we can do this easily with the  knowledge of four-point R-symmetry polynomials from \eqref{R4pt}.
If the intermediate operator is in the $\mathbf{15}$ or the $\mathbf{20'}$, then the tensor contraction is of the form
\begin{equation}
 T^{I_a I_b | I_c I_d I_e} =(C^1_1)^{I_a}_{mn} (C^1_1)^{I_b}_{np}  (C^1_q)^{J}_{mp} \times (C^1_q)^J_{ij}  (C^1_1)^{I_c}_{jk} (C^1_1)^{I_d}_{kl} (C^1_1)^{I_e}_{li}\,.
\end{equation}
It is useful to rewrite this tensor in terms of the variables $t_i$, which we do as follows
\begin{align}
R^{(1,q)}_{ab|cde}&=C^{I_a}_{i_1 j_1} t^{i_1}_a t^{j_1}_a \ldots C^{I_e}_{i_5 j_5} t^{i_5}_e t^{j_5}_e \;T^{I_a I_b | I_c I_d I_e} \nonumber\\
&=t_{ab} t_{cd} t_{de}  \,t_a^m t_b^p (C^1_1)^{I_J}_{mp} (C^1_q)^J_{ij} t_{c}^j t_e^i \,.
\end{align}
It is possible to recognize part of the definition for the four-point polynomials in this expression, from which we obtain
\begin{equation}
R^{(1,q)}_{ab|cde}=\frac{t_{cd} t_{de}}{t_{ce}}  R^{(1,q)}_{ab|ce}\,.
\end{equation}

\section{Spinning correlators}\label{spinningcorrelator}

In order to use factorization we must obtain the Mellin representation for correlators of the chiral operator $\mathcal{O}_{\mathbf{20'}}$ with a single insertion of the $R$-current $\mathcal{J}_\mu$ or the stress-tensor $\mathcal{T}_{\mu\nu}$, whose $AdS$  duals are the graviphoton $V_\mu$ and graviton $\varphi_{\mu\nu}$, respectively. Three-point functions of these fields are protected, and their Mellin transform is a constant, but factorization requires also the knowledge of the Mellin representation for the following four-point functions
\begin{align}\label{corrJ}
 \langle \mathcal{J}_\mu \mathcal O_{\mathbf{20'}} \mathcal O_{\mathbf{20'}}  \mathcal O_{\mathbf{20'}} \mathcal  \rangle \,,\\
 \langle \mathcal{T}_{\mu\nu}  \mathcal O_{\mathbf{20'}}  \mathcal O_{\mathbf{20'}}  \mathcal O_{\mathbf{20'}} \mathcal \rangle \,. \label{corrT}
\end{align}
It is useful to think of these spinning operators as different components of the superfield
\begin{equation}
\mathcal T(x_\mu, \theta^a_{\alpha}, \bar \theta_{\dot a}^{\dot \alpha}) \,,
\end{equation}
which depends only on four chiral and four antichiral Grassmann variables, due to a shortening condition. Therefore, the four-point function
\begin{equation}\label{supercorr}
\mathcal G_4=\langle \mathcal T(1) \ldots \mathcal T(4)\rangle
\end{equation}
depends on 16 chiral and 16 antichiral variables, which exactly matches the number of supercharges in $\mathcal N=4$ SYM. In \cite{Belitsky:2014zha} the superconformal symmetry was used to relate all elements of \eqref{supercorr} to the lowest component, {\it i.e.}, the four-point function of chiral primaries. The chiral primary four-point function can be split into a free part and an ``anomalous'' part which depends on the coupling
\begin{equation}
\langle\mathcal O_{\mathbf{20'}}  (x_1,t_1) \ldots \mathcal O_{\mathbf{20'}}  (x_4,t_4) \rangle=  \mathcal G_4 \big|_{\theta_i=\bar\theta_i=0} = G^{free}(x_i,t_i) +R(x_i, t_i) \frac{\Phi(u,v)}{x_{13}^2x_{24}^2} \,,
\end{equation}
where $u$ and $v$ are the four-point conformal cross ratios
\begin{equation}
u= \frac{x_{12}^2 x_{34}^2}{x_{13}^2 x_{24}^2} \,,\qquad v= \frac{x_{14}^2 x_{23}^2}{x_{13}^2 x_{24}^2} \,,
\end{equation}
and the prefactor of the anomalous correlator is defined as
\begin{align}
R(x_i, t_i) &= d_{12}^2 d_{34}^2 x_{12}^2 x_{34}^2 + d_{14}^2 d_{23}^2 x_{14}^2 x_{23}^2  + d_{13}^2 d_{24}^2 x_{13}^2 x_{24}^2 \nonumber\\
&\quad+d_{12} d_{13} d_{24} d_{34} (x_{12}^2 x_{34}^2 + x_{13}^2x_{24}^2 - x_{14}^2 x_{23}^2)\nonumber\\
&\quad+d_{12} d_{14} d_{23} d_{34} (x_{12}^2 x_{34}^2 + x_{14}^2 x_{23}^2 - x_{13}^2x_{24}^2)\nonumber\\
&\quad+d_{13} d_{14} d_{24} d_{23} ( x_{13}^2x_{24}^2 + x_{14}^2 x_{23}^2-x_{12}^2 x_{34}^2 ) \,,
\end{align}
with the propagator $d_{ij}= t_{ij}/x_{ij}^2$. By superconformal symmetry, the four-point functions $\langle \mathcal{J}_\mu\mathcal{O}_{\mathbf{20'}} \mathcal{O}_{\mathbf{20'}} \mathcal{O}_{\mathbf{20'}} \rangle$ and $\langle \mathcal{T}_{\mu\nu}\mathcal{O}_{\mathbf{20'}} \mathcal{O}_{\mathbf{20'}} \mathcal{O}_{\mathbf{20'}} \rangle$ have similar structures, namely, they can also be expressed as the sum of a free piece and an anomalous piece.

Using the results of \cite{Belitsky:2014zha}, the anomalous component of the correlators \eqref{corrJ} and \eqref{corrT} can be written in terms of the function $ \Phi(u,v)$ in the following way
\begin{align}\label{JT}
\langle J_{\alpha\dot{\alpha},a \dot a} \mathcal O_{\mathbf{20'}} \ldots  \rangle &= \frac{ (\partial_{x_1})_{\dot \alpha}^\beta}{4} \left((y_{23}^2 y_{34}^2 Y_{124}-u y_{23}^2 y_{24}^2 Y_{134}-v y_{24}^2 y_{34}^2 Y_{123})_{a \dot a} \frac{[X_{124},X_{134}]_{(\alpha \beta)}}{x_{23}^2 x_{24}^2 x_{34}^2}  \Phi(u,v)\right) \,,\nonumber\\
\langle T_{\alpha\dot{\alpha},\beta\dot \beta} \mathcal O_{\mathbf{20'}}  \ldots  \rangle &= \frac{ (\partial_{x_1})_{\dot \alpha}^\gamma (\partial_{x_1})_{\dot \beta}^\delta}{4} \left([X_{124},X_{134}]_{(\alpha \beta} [X_{124},X_{134}]_{\gamma \delta)} \frac{x_{12}^2 x_{14}^2}{x_{24}^2}\frac{y_{23}^2 y_{24}^2 y_{34}^2}{x_{23}^2 x_{24}^2 x_{34}^2} \Phi(u,v)\right)\,,
\end{align}
where the tensor structures are defined as
\begin{equation}
(X_{ijk})_{\alpha \dot \beta}= \frac{ (x_{ij})_{\alpha \dot \alpha} (x_{jk})^{\dot \alpha \beta} (x_{ki})_{\beta \dot \beta} }{x_{ij}^2 x_{ki}^2}\,,\qquad (Y_{ijk})_{a \dot b} = (y_{ij})_{a \dot a} (y_{jk})^{\dot a b} (y_{ki})_{b \dot b}\,.
\end{equation}
By contracting $(X_{ijk})_{\alpha \dot \beta}$ with $\frac{1}{2}(\sigma^\mu)^{\dot \beta \alpha}$ we recognize it as the building block of correlators with a spinning operator
\begin{equation}
X^\mu_{ijk} = \frac{x_{ik}^\mu}{x_{ik}^2} - \frac{x_{ij}^\mu}{x_{ij}^2}\,.
\end{equation}
Note however that a four-point function depends only on two such structures, due to the identity
 \begin{equation}
 X^\mu_{123}-X^\mu_{124}+X^\mu_{134}=0\,.
\end{equation} 

Similarly, we can rewrite equations \eqref{JT} with vector indices by contracting with Pauli matrices and performing the traces. In position space the expressions are quite lenghty, so here we present the results only schematically
\begin{align}\label{Jpos}
 \langle J_\mu   \mathcal O_{\mathbf{20'}}  \mathcal O_{\mathbf{20'}}  \mathcal O_{\mathbf{20'}} \mathcal\rangle&=\frac{1}{x_{13}^4x_{24}^4} \left( \alpha^{(2)}(u,v) X_{124}^\mu + \alpha^{(3)}(u,v) X_{134}^\mu \right) \,, \\
  \langle T_{\mu\nu}   \mathcal O_{\mathbf{20'}}  \mathcal O_{\mathbf{20'}}  \mathcal O_{\mathbf{20'}} \mathcal\rangle&=\frac{1}{x_{13}^4x_{24}^4} \left( \beta^{(2,2)}(u,v) X_{124}^\mu X_{124}^\nu+\beta^{(3,3)}(u,v) X_{134}^\mu X_{134}^\nu \right. \nonumber\\
  &\qquad\qquad\qquad\left.+ \beta^{(2,3)}(u,v) X_{123}^{(\mu} X_{134}^{\nu)}\right)  + \frac{x_{23}^2 x_{34}^2}{x_{13}^8 x_{24}^6} \gamma(u,v) \delta^{\mu\nu}\,,\label{Tpos}
\end{align}
where the functions $\alpha^{(i)}(u,v)$, $\beta^{(i,j)}(u,v)$ and $\gamma(u,v)$ are linear combinations of $\Phi(u,v)$  and its derivatives, with  coefficients given by Laurent polynomials of the cross ratios.
It is not manifest in \eqref{JT} that the result can be expressed in terms of $X_{124}^\mu$ and $X^\mu_{134}$ alone, so the fact that \eqref{Jpos} and \eqref{Tpos} have this form provides a good consistency check of the result. Furthermore, $\gamma(u,v)$ is such that the expression for the stress-tensor is traceless, and we observe also that, as expected, both correlators satisfy the equations for conserved currents
\begin{equation}
\frac{\partial}{\partial x_1^\mu}  \langle J_\mu   \mathcal O_{\mathbf{20'}}  \mathcal O_{\mathbf{20'}}  \mathcal O_{\mathbf{20'}} \mathcal\rangle=0 \,,\qquad\qquad \frac{\partial}{\partial x_1^\mu}  \langle T_{\mu\nu}   \mathcal O_{\mathbf{20'}}  \mathcal O_{\mathbf{20'}}  \mathcal O_{\mathbf{20'}} \mathcal\rangle= 0\,.
\end{equation}

The expressions for the spinning four-point functions simplify greatly when we use the spinning Mellin formalism \cite{Goncalves:2014rfa}. Here we can ignore the free piece because they are rational functions which do not contribute to the Mellin amplitude \cite{Rastelli:2017udc}. The Mellin amplitudes therefore come exclusively from the anomalous piece computed above. By inverting \eqref{eq:MellindefinitionSpin}, we can extract the Mellin amplitudes by performing the following Mellin transforms
\begin{align}\label{spintransforms}
  M^{ab}(s,t) &= \frac{1}{ \prod_{i=2}^4 \Gamma(\gamma_i + \delta_i^a+ \delta_i^b)  \prod_{i<j}^4 \Gamma(\gamma_{ij}) }\int_0^\infty \mathrm d u \int_0^\infty \mathrm d v \, u^{1-\frac{s}{2}} v^{1-\frac{t}{2}} \beta^{(a,b)}(u,v) \,,\nonumber\\
   M^{a}(s,t) &= \frac{1}{ \prod_{i=2}^4 \Gamma(\gamma_i + \delta_i^a)  \prod_{i<j}^4 \Gamma(\gamma_{ij}) }\int_0^\infty \mathrm d u \int_0^\infty \mathrm d v \, u^{1-\frac{s}{2}} v^{1-\frac{t}{2}} \alpha^{(a)}(u,v) \,,
\end{align}
where  $\gamma_i$ are fixed in \eqref{gammas} and we define the remaining Mellin variables as
\begin{equation}\label{stu}
\gamma_{23}=2 -\frac{t}{2} \,,  \qquad \gamma_{24}= \frac{s+t}{2} - 2 \,,\qquad \gamma_{34}= 2-\frac{s}{2} \,.
\end{equation}
Since the functions $\alpha^{(i)}(u,v)$ and $\beta^{(i,j)}(u,v)$ are linear combinations of $\Phi(u,v)$  and its derivatives,
it is useful to consider the relation
\begin{align}\label{shifts}
&\int_0^\infty \mathrm d u \mathrm d v\,  u^{1-\frac{s}{2}} v^{1-\frac{t}{2}} u^m v^n \frac{\partial^p}{\partial u^p} \frac{\partial^q}{\partial v^q} \Phi(u,v)= \nonumber\\
&=\Big(\frac{s}{2}-m-1 \Big)_p \Big(\frac{t}{2}-n-1\Big)_q\;\int_0^\infty \mathrm d u  \mathrm d v\, u^{-\frac{s-2m+2p-2}{2}} v^{-\frac{t-2n+2q-2}{2}}  \Phi(u,v) \,,
\end{align}
and
\begin{equation}\label{PhiinMellin}
\int_0^\infty \mathrm d u  \mathrm d v\, u^{-\frac{s}{2}} v^{-\frac{t}{2}}  \Phi(u,v) =\frac{32\,\Gamma(2-\frac{s}{2})^2 \Gamma(2-\frac{t}{2})^2 \Gamma(\frac{s+t}{2})^2}{(s-2)(t-2)(2-s-t)}\,.
\end{equation}
Putting equations \eqref{spintransforms}, \eqref{shifts} and \eqref{PhiinMellin} together, we obtain the following Mellin representation for the R-current correlator \eqref{corrJ}
\begin{align}\label{MellinJ}
M^{2}&=- \frac{2(t-4)^2 y_{24}^2 y_{34}^2 Y_{123}}{ (s-2) (s+t-6)} + \frac{2(t+s-4)^2 y_{23}^2 y_{34}^2 Y_{124}}{ (s-2) (t-2)} -\frac{2(s-4)(2t+s-8) y_{23}^2 y_{24}^2 Y_{134}}{ (t-2) (s+t-6)}\,, \nonumber\\
M^{3}&=\frac{2(t-4)^2 y_{24}^2 y_{34}^2 Y_{123}}{ (s-2)(s+t-6)} + \frac{2(s-t)(s+t-4) y_{23}^2 y_{34}^2 Y_{124}}{ (s-2)(t-2)} - \frac{2(s-4)^2 y_{23}^2 y_{24}^2 Y_{134}}{ (t-2)(s+t-6)} \,,
\end{align}
while for \eqref{corrT} we get
\begin{align}\label{MellinT}
M^{22}&= \frac{4(s-4) (t-4)(s+t-4) }{3 (s-2)(t-2)(s+t-6)} y_{23}^2 y_{24}^2 y_{34}^2 \,,\nonumber\\
M^{23}&= \frac{ 4(t-4)((s-4)(s+t-4) +6(t-2) ) }{3 (s-2)(t-2)(s+t-6)} y_{23}^2 y_{24}^2 y_{34}^2\,,\nonumber\\
M^{33}&= \frac{4(s-4) (t-4)(s+t-4) }{3 (s-2)(t-2)(s+t-6)} y_{23}^2 y_{24}^2 y_{34}^2\,.
\end{align}
We also obtained the other Mellin components, $M^{4}$ and $M^{a4}$, and verified that  they satisfy transversality 
\begin{align}
\sum_a \gamma_a M^{a} &=0 \,,\nonumber\\
\sum_a (\gamma_a + \delta_a^b) M^{ab} &=0\label{eq:transversalityraulappendix}\,.
\end{align}

As a side comment, we observe that the transversality condition (\ref{eq:transversalityraulappendix}) imposes non-trivial constraints on the Mellin amplitudes. For example, the Mellin amplitudes with one external stress tensor can be cast in the following form 
\begin{align}
M^{ab}(s,t) = \frac{c_{ab,1}}{s-2}+\frac{c_{ab,2}}{t-2}+\frac{c_{ab,3}}{s+t-6}+c_{ab,4}.
\end{align}
Imposing transversality fixes all the coefficients $c_{ab,i}$ in terms of just one, say $c_{22,1}$.

It is now instructive to go back to position space once again. This is a subtle procedure, but when performed correctly we are able to recover both the anomalous and free pieces of the correlator. From the definition of the spinning Mellin amplitude \eqref{eq:MellindefinitionSpin}, the integration variables must sit in the domain
\begin{equation}\label{FundDomain}
\{ (s_0, t_0)|\;\mathfrak{R}(s)<4,\mathfrak{R}(t)<4,\mathfrak{R}(s)+\mathfrak{R}(t)>4\}\,.
\end{equation}
In order to translate our Mellin amplitudes \eqref{MellinJ} and \eqref{MellinT} into $D$-functions, we use the definition
\begin{equation}
D_{\Delta_1 \ldots \Delta_n} =\frac{\pi^2 \Gamma(\Sigma-2)}{\prod_i \Gamma(\Delta_i)} \int [\mathrm d \tilde \gamma] \prod_{i<j} \Gamma(\tilde \gamma_{ij}) \,x_{ij}^{-2\tilde \gamma_{ij}} \,,
\end{equation}
with
\begin{equation}
\sum_{j\neq i} \tilde \gamma_{ij}=\Delta_i \,,
\end{equation}
whose contour integral is defined for Mellin variables in the domain 
\begin{equation}
\{ (s_0, t_0)|\; \tilde\gamma_{ij}(\mathfrak R(s),\mathfrak R(t))>0\}\,.
\end{equation}
In order to have a faithful position space representation, all integration domains for the $D$-functions must overalap with the fundamental domain \eqref{FundDomain}. Having this in mind we obtain the following R-current correlator
\begin{align}\label{corrJD2}
\langle J^\mu_{a \dot a} \mathcal O_{\mathbf{20'}}  \mathcal O_{\mathbf{20'}}  \mathcal O_{\mathbf{20'}} \rangle=&\frac{2 y_{24}^2 y_{34}^2 (Y_{123})_{a \dot a}}{ u x_{13}^4 x_{24}^4}
\left(-\left(\bar D_{2134} +u\bar D_{2224}  \right) X_{124}^\mu +\left(\bar D_{2224} +u\bar D_{2314}  \right) X_{134}^\mu \right) \nonumber\\
&+\frac{2y_{23}^2 y_{34}^2 (Y_{124})_{a\dot a}}{ u x_{13}^4 x_{24}^4}
\left(-\left(\bar D_{2143} +u\bar D_{2242}  \right) X_{124}^\mu +\left(\bar D_{2233} -u\bar D_{2332}  \right) X_{134}^\mu \right)  \nonumber\\
&+\frac{2u y_{23}^2 y_{24}^2 (Y_{134})_{a\dot a}}{ x_{13}^4 x_{24}^4}
\left(\left(\bar D_{2323} -\bar  D_{2332}  \right) X_{124}^\mu -\left(\bar D_{2413} + \bar D_{2422}  \right) X_{134}^\mu \right) \,.
\end{align}
The contributions with the ladder integral and logarithms are exactly the same as in \eqref{Jpos}, and so we are able to extract the free correlator
\begin{equation}
\langle J_{a\dot a}^\mu \mathcal O_{\mathbf{20'}}   \mathcal O_{\mathbf{20'}}   \mathcal O_{\mathbf{20'}} \rangle^{free}= -\frac{y_{23}^2 y_{34}^2 (Y_{124})_{a\dot a} X_{124}^\mu+ u y_{23}^2 y_{24}^2 (Y_{134})_{a\dot a} X_{134}^\mu+v y_{24}^2 y_{34}^2 (Y_{123})_{a\dot a} X_{123}^\mu}{ u  v x_{13}^4 x_{24}^4} \,.
\end{equation}
Meanwhile, for the stress-tensor we obtain
\begin{align}\label{corrTD}
\langle T^{\mu\nu} \mathcal O_{\mathbf{20'}}  \mathcal O_{\mathbf{20'}}  \mathcal O_{\mathbf{20'}} \rangle= \frac{4 y_{23}^2 y_{24}^2 y_{34}^2 }{3 v x_{13}^4 x_{24}^4}&\left(u^2\left(\bar D_{4312}+ v \bar D_{4321}+ v \bar D_{4411} + v \bar D_{4422}\right)X_{124}^\mu X_{124}^\nu \right. \nonumber\\
& \;+ 2 u\left(\bar D_{4222}-2v \bar D_{4231}- 2v \bar D_{4321} + v \bar D_{4332}\right)X_{124}^{(\mu} X_{134}^{\nu)} \nonumber\\
&\left.\;+ \left(\bar D_{4132}+ v \bar D_{4231}+ v \bar D_{4141} + v \bar D_{4242}\right)X_{134}^\mu X_{134}^\nu\right)\,,
\end{align}
which differs from \eqref{Tpos} by 
\begin{equation}
\langle T^{\mu \nu}\mathcal O_{\mathbf{20'}}   \mathcal O_{\mathbf{20'}}   \mathcal O_{\mathbf{20'}} \rangle^{free}=\frac{4y_{23}^2 y_{24}^2 y_{34}^2 \left( (1+v)X_{124}^\mu X_{124}^\nu+ (u+v) X_{134}^\mu X_{134}^\nu-2v X_{124}^{(\mu} X_{134}^{\nu)}\right)}{3 u v x_{13}^4 x_{24}^4}\,.
\end{equation}

\section{Properties of $D$-functions}\label{propertyDfunction}
In this appendix, we summarize some basic properties of the $D$-functions which we encountered in this paper. A general $D$-function with $n$-external points is defined as an integral in $AdS_{d+1}$
\begin{equation}
D_{\Delta_1,\ldots,\Delta_n}=\int\frac{dz_0d^dz}{z_0^{d+1}}\prod_{i=1}^n\left(\frac{z_0}{z_0^2+(\vec{z}-\vec{x}_i)^2}\right)^{\Delta_i}\;.
\end{equation}
After some standard manipulations, the $D$-function can be written as a Feynman integral
\begin{equation}
\frac{\pi^{d/2}\Gamma(\Sigma-\frac{d}{2})\Gamma(\Sigma)}{2\prod_i\Gamma(\Delta_i)}\int_0^1 \prod_{j=1}^n da_j \delta(1-\sum_ja_j)\frac{\prod_j a_j^{\Delta_j-1}}{(\sum_{i<j}a_ia_jx_{ij}^2)^\Sigma}
\end{equation}
where $\Sigma=\frac{1}{2}\sum_{i=1}^n\Delta_i$. One should notice that the $d$-dependence only appears in the overall factor $\Gamma(\Sigma-\frac{d}{2})$, and therefore $D$-functions defined in all $AdS_{d+1}$ are essentially the same. From the Feynman representation, we can derive a useful relation
\begin{equation}\label{diffrecurofD}
D_{\Delta_1,\ldots,\Delta_i+1,\ldots,\Delta_j+1,\ldots,\Delta_n}=\frac{d/2-\Sigma}{\Delta_i\Delta_j}\frac{\partial}{\partial x_{ij}^2}D_{\Delta_1,\ldots,\Delta_n}\;,
\end{equation}
which relates $D$-functions of higher weights.

Let us now focus on the special case with $n=5$ and $(\Delta_1,\Delta_2,\Delta_3,\Delta_4,\Delta_5)=(1,1,1,1,2)$. This seed function $D_{11112}$ and its permutations generate all other $D$-functions in this paper via the differential recursion relations (\ref{diffrecurofD}). To explicitly evaluate $D_{11112}$, we need to compute the following integral
\begin{equation}\label{D11112Feynman}
\int_0^1 \prod_{j=1}^5 da_j \delta(1-\sum_ja_j)\frac{a_5}{(\sum_{i<j}a_ia_jx_{ij}^2)^3}\;.
\end{equation}
In fact (\ref{D11112Feynman}) is a special case of a more general class of integrals
\begin{equation}
I_n[P(\{a_i\})]=\Gamma(n-2)\int_0^1 \prod_{j=1}^n da_j \delta(1-\sum_ja_j)\frac{P(\{a_i\})}{(\sum_{i<j}a_ia_jx_{ij}^2)^{n-2}}
\end{equation}
where $P(\{a_i\})$ is a polynomial of $a_i$. These integrals can be evaluated in terms of the scalar one-loop box integral $I_4[1]$ \cite{Bern:1992em,Bern:1993kr}.  The result is
\begin{equation}
D_{11112}=\frac{4\pi^2}{x_{14}^2x_{35}^2x_{25}^2}\sum_{i=1}^5\frac{\eta_{i5}\hat{I}^{(i)}_4}{N_5}\;.
\end{equation}
Here $N_5$ and $\eta_{i5}$ are defined via a matrix $\rho$
\begin{equation}
\rho=N_n\eta^{-1}\;, \quad\quad N_n=2^{n-1}\det \rho\;,
\end{equation}
where
\begin{equation}
\rho=\left(\begin{array}{ccccc}0 & V_4 & 1 & 1 & V_3 \\V_4 & 0 & V_5 & 1 & 1 \\1 & V_5 & 0 & V_1 & 1 \\1 & 1 & V_1 & 0 & V_2 \\V_3 & 1 & 1 & V_2 & 0\end{array}\right)
\end{equation}
with
\begin{equation}
V_1=\frac{x_{25}^2 x_{34}^2}{x_{24}^2 x_{35}^2}\;,\quad V_2=\frac{x_{31}^2 x_{45}^2}{x_{35}^2 x_{14}^2}\;,\quad V_3=\frac{x_{24}^2 x_{15}^2}{x_{14}^2 x_{25}^2}\;,\quad V_4=\frac{x_{12}^2 x_{35}^2}{x_{25}^2 x_{13}^2}\;,\quad V_5=\frac{x_{14}^2 x_{23}^2}{x_{13}^2 x_{24}^2}\;.
\end{equation}
The function $\hat{I}^{(i)}_4$ is the scalar one-loop box diagram where the point $i$ is omitted from the set of five. For example,
\begin{equation}
\hat{I}^{(5)}_4=\Phi(V_1V_4,V_5)
\end{equation}
with
\begin{equation}
\Phi(V_1V_4,V_5)=\frac{1}{z-\bar{z}}\left(2Li_2(z)-2Li_2(\bar{z})+\log(z\bar{z})\log\frac{1-z}{1-\bar{z}}\right)\;,
\end{equation}
and
\begin{equation}
V_1V_4=\frac{x_{12}^2x_{34}^2}{x_{13}^2x_{24}^2}=z \bar{z}\;,\quad V_5= \frac{x_{14}^2x_{23}^2}{x_{13}^2x_{24}^2}=(1-z)(1-\bar{z})\;.
\end{equation}
The one-loop box diagrams satisfy the following differential recursion relations \cite{Eden:2000bk}
\begin{eqnarray}
\nonumber&&\partial_z\Phi=\frac{\Phi}{\bar{z}-z}+\frac{\log (1-z)(1-\bar{z})}{z(\bar{z}-z)}+\frac{\log(z\bar{z})}{(z-1)(z-\bar{z})}\;,\\ \label{Phidiffrecur}
&&\partial_{\bar{z}}\Phi=\frac{\Phi}{z-\bar{z}}+\frac{\log (1-z)(1-\bar{z})}{\bar{z}(z-\bar{z})}+\frac{\log(z\bar{z})}{(\bar{z}-1)(\bar{z}-z)}\;.
\end{eqnarray}

\newcommand{\floor}[1]{\lfloor #1 \rfloor}

\section{Five-point conformal blocks}\label{Conformal blocks}
\begin{figure}[h]
\centering
\includegraphics[width=0.6\textwidth]{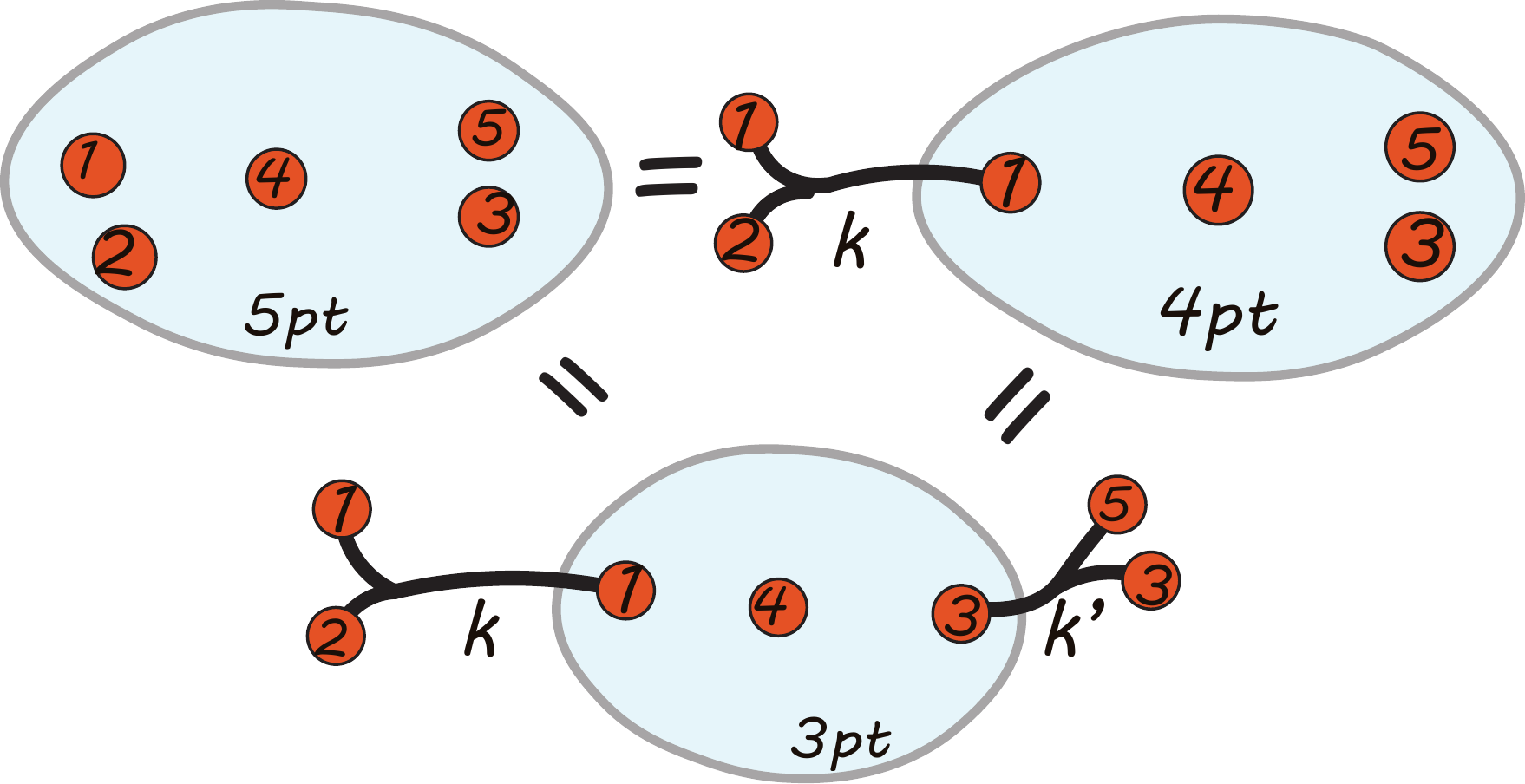} 
\caption{A five-point function can be written in terms of sum of four-point function where one of the operators appears in the OPE of $(12)$ or in terms of a sum of three-point function where now one of the operator appears in OPE of $(12)$ and the other in the OPE of $(35)$. }
\label{fig:MultiOPE}
\end{figure}

\begin{figure}[h]
\centering
\includegraphics[width=0.3\textwidth]{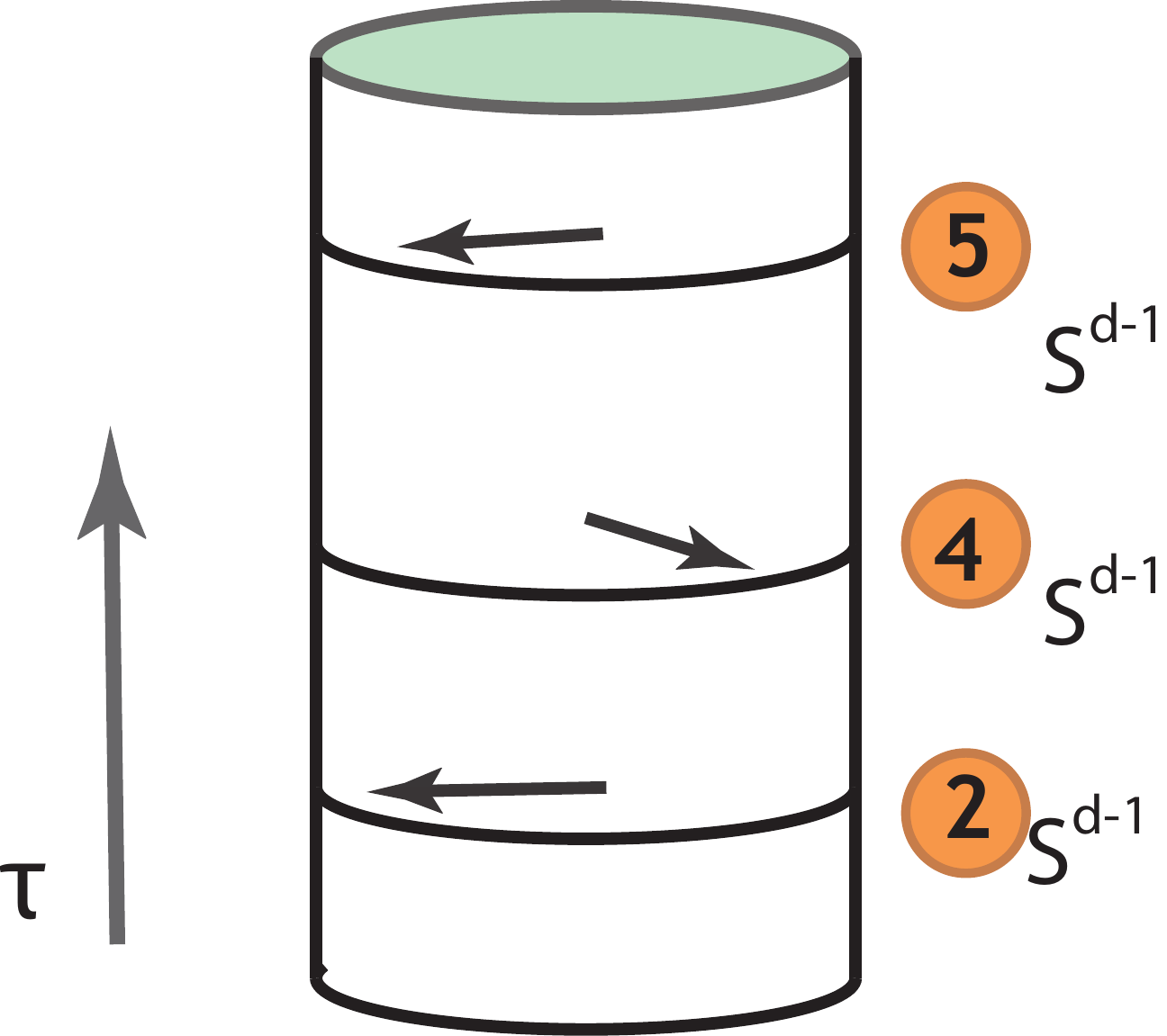} 
\caption{The five-point correlation function in the plane $R^d$ can be mapped through a Weyl transformation to the cylinder $R\times S^{d-1}$.  }
\label{fig:Cylinder}
\end{figure}

A CFT correlator contains information about lower-point functions, which can be accessed through the OPE. For example, a five-point function can be either written in terms  of a sum of four-point functions or a double sum of three-point functions, as illustrated in Figure \ref{fig:MultiOPE}
\begin{align}
\langle \mathcal{O}_1(x_1)\dots \mathcal{O}_5(x_5)\rangle &= \sum_{k}\frac{C_{\mathcal{O}_1\mathcal{O}_2\mathcal{O}_k}}{(x_{12}^2)^{\frac{\Delta_1+\Delta_2-\Delta_k+J}{2}}}\langle \mathcal{O}_{k,J}(x_1) \mathcal{O}_3(x_3)\mathcal{O}_4(x_4) \mathcal{O}_5(x_5)\rangle \\
&=\sum_{k}\frac{C_{\mathcal{O}_1\mathcal{O}_2\mathcal{O}_k}C_{\mathcal{O}_3\mathcal{O}_5\mathcal{O}_{k'}}}{(x_{12}^2)^{\frac{\Delta_1+\Delta_2-\Delta_k+J}{2}}(x_{35}^2)^{\frac{\Delta_3+\Delta_5-\Delta_{k'}+J'}{2}}}\langle \mathcal{O}_{k,J}(x_1) \mathcal{O}_{k',J'}(x_3)\mathcal{O}_4(x_4) \rangle \,,\nonumber
\end{align}
where the sum is over both primary and descendant operators\footnote{We chose to perform the OPE in the $(12)$ and $(35)$ channels.}. The conformal algebra fixes the contribution of the descendants in terms of the primary operators, in what is usually called the conformal block. In the following we will be interested in obtaining this kinematical contribution to the double OPE channel $(12)(35)$ of the five-point function. This can also be viewed by inserting a complete basis of states labeled by their dimension and spin
\begin{align}
\langle 0  |\mathcal{O}_1\mathcal{O}_2\mathcal{O}_3\mathcal{O}_4\mathcal{O}_5| 0\rangle = \sum_{E,E'}s_1^{E}s_2^{E'} \langle \mathcal{O}_1| \mathcal{O}_2| E\rangle \langle E| \mathcal{O}_4 |E' \rangle \langle E'|\mathcal{O}_5 |\mathcal{O}_3 \rangle \label{eq:CylinderPicture}
\end{align}
where $s_1=e^{-(\tau_2-\tau_4)}$ and $s_2=e^{-(\tau_4-\tau_5)}$ are two cross ratios and  we have used the cylinder picture of Figure \ref{fig:Cylinder}, which is obtained with a Weyl transformation that maps $R^d$ to $R\times S^{d-1}$. The goal of this appendix is to lay out the strategy to obtain the conformal blocks for five-point functions in the double OPE\footnote{Conformal blocks for $n$-point function have been recently obtained in \cite{Rosenhaus:2018zqn} for $d=1,2$. See also \cite{Fortin:2019dnq} for recent results on higher-point conformal blocks.  } in the channels $(12)$ and $(35)$ as an expansion in terms of powers of $s_1$ and $s_2$. The method used here is an adaptation of the one already implemented for four-point functions in \cite{Dolan:2003hv,Hogervorst:2013sma}.

\begin{figure}[h]
\centering
\includegraphics[width=0.4\textwidth]{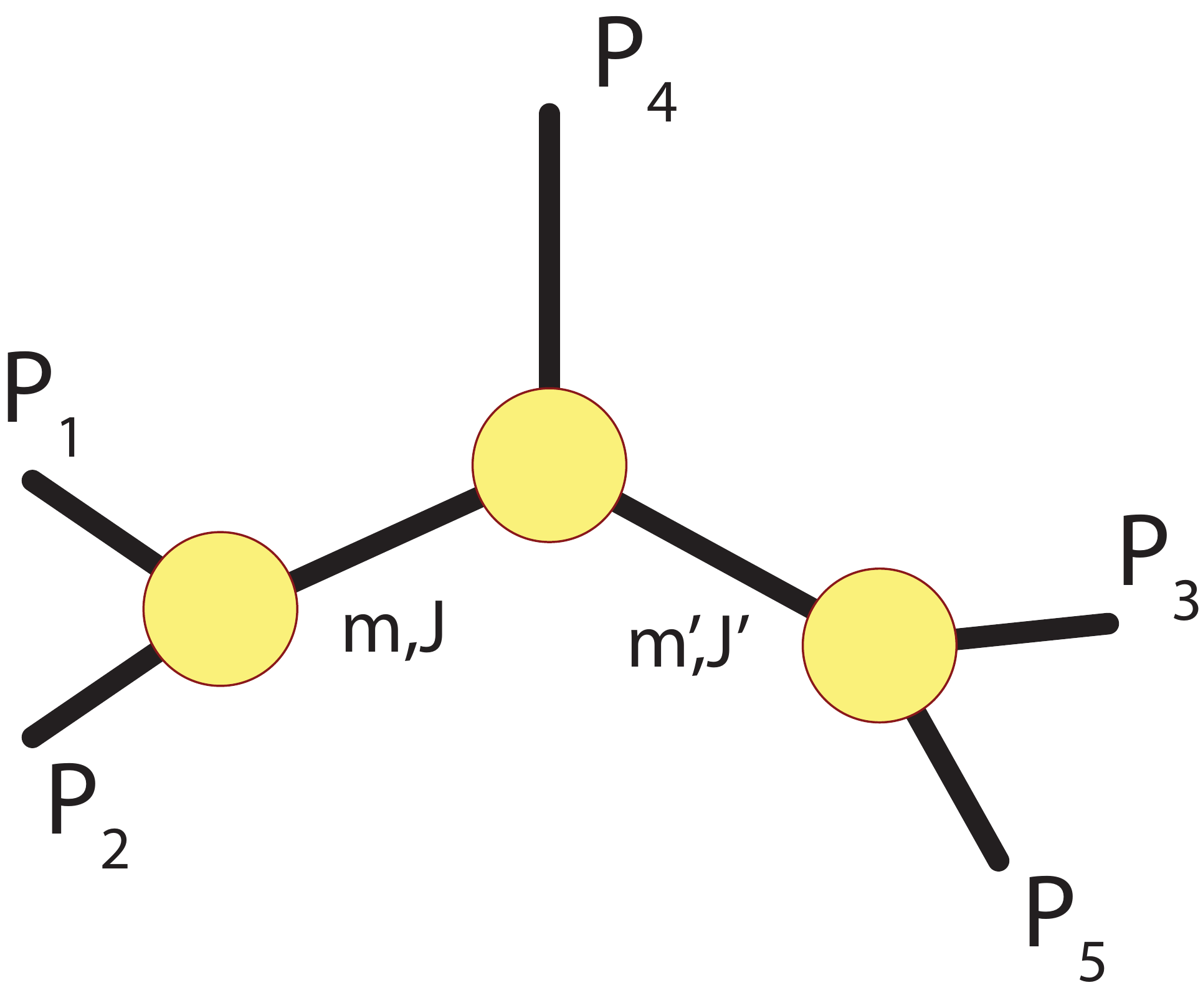} 
\caption{A five-point scattering amplitude can also be decomposed by inserting a complete basis of states labeled by the mass of the intermediate particles and their spin.  }
\label{fig:PartialWaveScattering}
\end{figure}

The double OPE in the cylinder picture (\ref{eq:CylinderPicture}) will not be explicitly used in the derivation of the conformal  blocks but its schematic form and simplicity makes it very appealing to explain the key points in the derivation. First notice that there are three three-point functions in this decomposition with two of them of the form scalar-scalar-spin while the third  has two spinning operators and one scalar
\begin{align}
\langle \mathcal{O}_1| \mathcal{O}_2| E,\mu_1\dots \mu_J\rangle,\, \, \ \ \  \langle E',\nu_1\dots \nu_{J'}|\mathcal{O}_5 |\mathcal{O}_3 \rangle,\, \ \ \ \ \langle E,\mu_1\dots \mu_J| \mathcal{O}_4 |E',\nu_1\dots \nu_{J'} \rangle^{(m)}. 
\end{align} 
The spinning operators must transform in a symmetric traceless representation since they appear in the OPE of two scalar operators. A three-point function $scalar-scalar-spin$ has only one structure, but on the other hand a three-point function with $scalar-spin-spin$ has two fundamental structures \cite{Costa:2011mg}
and the upper index $m$ is used to label this property. The spinning operators carry with them Lorentz indices that should be contracted among each of the three-point functions. The most efficient way to do this contraction of indices is to introduce null polarization vectors $z_1$ and $z_3$ that are contracted with the indices of the $J-J'-\textrm{scalar}$ three-point function and use the differential operator
\begin{align}
D_z=\left(\frac{d}{2}-1+z\cdot \frac{\partial}{\partial z}\right)\frac{\partial}{\partial z^{\mu}}-\frac{1}{2}z^{\mu}\frac{\partial^2}{\partial z\cdot \partial z}
\end{align}
in order to recover the tensor structures.
The final formula for the contraction is  given by (see \cite{Costa:2011mg} for more details)
\begin{align}
\displaystyle {\frac{\big(x_{12}\!\cdot\! D_{z_1}\big)^J\big(x_{35}\!\cdot\! D_{z_3}\big)^{J'}(z_1\!\cdot\! z_3 x_{13}^2-2z_1\!\cdot\! x_{13}z_3\!\cdot\! x_{13})^m(z_1\!\cdot\! x_{31})^{J-m}(z_3\!\cdot\! x_{13})^{J'-m}}{J!J'!(h-1)_J(h-1)_{J'}\,(x_{12}^2)^{\frac{J}{2}}(x_{35}^2)^{\frac{J'}{2}}}=\mathcal{H}_{J,J',m}(\xi_i)}\label{eq:angularPartContraction}
\end{align}
where $h=d/2$, and $z_3\!\cdot\! x_{13}, z_1\!\cdot\! x_{13}$ and $z_1\!\cdot\! z_3 x_{13}^2-2z_1\!\cdot\! x_{13}z_3\!\cdot\! x_{13}$ are related with the fundamental structures $V_{i,jk}$ and $H_{ij}$ of spinning three-point functions
\begin{align}
H_{ij}=z_i\cdot z_j x_{ij}^2+ z_i\cdot x_{ij}z_j\cdot x_{ij},\, \ \ \ V_{i,jk}=\frac{z_i\cdot x_{ij} x_{ik}^2-z_i\cdot x_{ik} x_{ij}^2}{x_{jk}^2}\,.
\end{align}
In the formula above we used conformal symmetry to put the point $x_4$ to infinity and $x_{13}^2=1$ to simplify the computation. The right-hand side of (\ref{eq:angularPartContraction}) depends only on the angles
\begin{align}
&\xi_1=\frac{x_{12}\cdot x_{13}}{(x_{12}^2x_{13}^2)^{\frac{1}{2}}},\,  \ \ \xi_2=\frac{x_{13}\cdot x_{35}}{(x_{35}^2x_{13}^2)^{\frac{1}{2}}}, \ \ \ \ \xi_3=\frac{x_{12}\cdot x_{35}x_{13}^2-2x_{12}\cdot x_{13}x_{13}\cdot x_{35}}{(x_{12}^2)^{\frac{1}{2}}(x_{35}^2)^{\frac{1}{2}}}\label{eq:AngleEquationcrossratios}.
\end{align}
This is equivalent to the contraction of unit vectors in the cylinder picture. The steps to obtain the explicit form of $\mathcal{H}_{J,J',m}(\xi_i)$ are lengthy but follow by straightfoward application of the derivatives in (\ref{eq:angularPartContraction}). We omit the details and present only the final result\footnote{We also include this formula in an auxiliary file. }
\begin{align}
\!\!\mathcal{H}_{J,J',m}(\xi_1,\xi_2,\xi_3)=&\sum_{k=0}^{\floor{\frac{J}{2}}}\sum_{k'=0}^{\floor{\frac{J'}{2}}}\sum_{n_1=0}^{m}\sum_{n_2=0}^{J'-n_1}\frac{(-1)^{J+k'+k-n_1}}{2^{2k+2k'-m+n_1}}\frac{\xi_1^{J-2k-n_1}\xi_3^{J'-2k'-n_2}\xi_2^{n_2}{{J'-2k'}\choose{n_2}}J!J'! (k!)}{{{J}\choose{m}}{{J'}\choose{J'-n_1}}(J-2k)!k!k'!(J'-2k')!n_1!m!}\nonumber\\
&\times \frac{(h-1)_{J-k}(h-1)_{J'-k'}(2k-J)_{n_1}(-m)_{n_1}\,C_{J'-n_2-n_1}^{-k'}(\xi_1)}{(k-m+n_1)!(h-1)_J(h-1)_{J'}}\,.\label{eq:HangularexplicitComputation}
\end{align}
As a side remark notice that this angular function $\mathcal{H}_{J,J',m}(\xi_1,\xi_2,\xi_3)$ should also appear in the double partial decomposition of a five-point scattering amplitude, see Figure \ref{fig:PartialWaveScattering}. It would be interesting to make this relation more precise and try to apply it in the context of the S-matrix bootstrap. 

The index contraction in  (\ref{eq:CylinderPicture}) is already taken into account by $\mathcal{H}_{J,J',m}(\xi_i)$ even if the operator in the three-point function is a descendant. In fact we can say more, the dimension of each descendant differs from the corresponding primary operator by an integer $n$ and $n'$, with  their spins $j$ and $j'$ in the range \cite{Hogervorst:2013sma}
\begin{align}
j=J+n,J+n-2,\dots, \textrm{max}(J-n,J+n \,\,\textrm{mod } 2).
\end{align}
This analysis shows that the conformal block can be written in terms of a double expansion in $s_1$ and $s_2$
\begin{align}
\!\!\!\!\!\mathcal{G}_{\Delta_k,\Delta_k,J,J',m}(s_1,s_2,\xi_i) &=\sum_{n=n'=0}^{\infty}\sum_{j,j'}\sum_{m'=0}^{\textrm{min}(j,j')}a_{n,n',j,j',m'}s_1^{\Delta_k+n}s_2^{\Delta_{k'}+n'}\mathcal{H}_{jj'm'}(\xi_1,\xi_2,\xi_3).\label{eq:ConformalblockAnsatzHfunction}\\
s_1^2&=\frac{x_{12}^2x_{34}^2}{x_{13}^2x_{24}^2}\,,  \ \ \ \ \ \ \ s_2^2=\frac{x_{35}^2x_{14}^2}{x_{13}^2x_{45}^2}
\end{align}
The coefficients $a_{n,n',j,j',m'}$ are fixed by conformal symmetry alone and they could be obtained by working out precisely how the contribution of a primary operator is related with the corresponding descendant in (\ref{eq:CylinderPicture}). However we find it easier to use the property that the conformal block $\mathcal{G}$ satisfies two eigenvalue equations coming from applying the quadratic Casimir of the conformal group
\begin{align}
&\left[\frac{1}{2}\left(L^{1}_{AB}+L_{AB}^{2}\right)\left(L^{1,AB}+L^{2,AB}\right)-C_{\Delta_k,J}\right]\frac{1}{(x_{12}^2x_{35}^2)^{\Delta_{\mathcal{O}}}}\left(\frac{x_{13}^2}{x_{14}^2x_{34}^2}\right)^{\Delta_{\mathcal{O}}}\mathcal{G} \,,\\
&\left[\frac{1}{2}\left(L^{3}_{AB}+L_{AB}^{5}\right)\left(L^{3,AB}+L^{5,AB}\right)-C_{\Delta_{k'},J'}\right]\frac{1}{(x_{12}^2x_{35}^2)^{\Delta_{\mathcal{O}}}}\left(\frac{x_{13}^2}{x_{14}^2x_{34}^2}\right)^{\Delta_{\mathcal{O}}}\mathcal{G}\,,
\end{align}
where $L^{i}_{AB}$ are generators of the conformal group acting on the operator at position $i$ and $C_{\Delta,J}=\Delta(\Delta-d)+J(J+d-2)$ is the Casimir eigenvalue.

This differential equation can be written in terms of the cross ratios by acting on the conformal block $\mathcal{G}_{\Delta_k,\Delta_k,J,J',m}(s_1,s_2,\xi_1,\xi_2,\xi_3)$ with
\begin{align}
\big[\mathcal{D}_{12}^{(0)}+\mathcal{D}_{12}^{(1)}-C_{\Delta_k,J}\big]\mathcal{G}=0 \,,\label{eq:CasimirEquation1}\\
\big[\mathcal{D}_{35}^{(0)}+\mathcal{D}_{35}^{(1)}-C_{\Delta_{k'},J'}\big]\mathcal{G}=0\,,
\end{align}
where
\begin{align}
&\displaystyle {\mathcal{D}_{12}^{(0)} =s_1^2\partial_{s_1}^2+(\xi_1^2-1)\partial_{\xi_1}^2+2(\xi_1 \xi_3+\xi_2)\partial_{\xi_1}\partial_{\xi_3}+(\xi_3^2-1)\partial_{\xi_3}^2-(2h-1)\big[s_1\partial_{s_1}-\xi_1\partial_{\xi_1}-\xi_3\partial_{\xi_3}\big],}\\
&\displaystyle {\mathcal{D}_{12}^{(1)}=-d_{12}^{(1)}s_1^2\partial_{s_1}+d_{12}^{(2)}(s_1 s_2\partial_{s_2}-s_1\Delta_1)+d_{12}^{(3)}s_1+s_1\xi_1(s_1^2\partial_{s_1}^2+s_1\partial_{s_1}(s_2\partial_{s_2}-(\Delta_1-1)))}\\
&d_{12}^{(1)}=(\xi_1\xi_2+\xi_3)\partial_{\xi_2}-3(\xi_1\xi_3+\xi_2)\partial_{\xi_3}+2(1-\xi_1^2)\partial_{\xi_1},\,d_{12}^{(2)}=(\xi_3\xi_1+\xi_2)\partial_{\xi_3}-(1-\xi_1^2)\partial_{\xi_1}\nonumber\\
&d_{12}^{(3)}=2(\xi_1\xi_3-2(h-1)\xi_2)\partial_{\xi_3}+(\xi_1^2+2(h-1))\partial_{\xi_1}+(1-\xi_3^2-\xi_1\xi_3\xi_2-\xi_2^2)\partial_{\xi_3}\partial_{\xi_2}\nonumber\\
&+2\xi_1(\xi_3^2-1)\partial_{\xi_3}^2 - \xi_1(\xi_3+\xi_1\xi_2)\partial_{\xi_1}\partial_{\xi_2}+3(\xi_1\xi_3+\xi_2)\xi_1\partial_{\xi_1}\partial_{\xi_3}+\xi_1(\xi_1^2-1)\partial_{\xi_1}^2.
\end{align}
The differential equation for $\mathcal{D}_{35}^{(0)}$ and $\mathcal{D}_{35}^{(1)}$ can be obtained  with the replacement $\xi_1\leftrightarrow \xi_2$. Notice that the differential operators $\mathcal{D}_{12}^{(0)},\mathcal{D}_{35}^{(0)}$ keep the degree of the cross ratios $s_1$ and $s_2$, while the other differential operators raise the degree of cross ratios by one.

The angular function $\mathcal{H}_{J,J',m}(\xi_i)$ plays an analogous role in the conformal block as the Gegenbauer polynomial for the case of four-point functions \cite{Hogervorst:2013sma}. In particular it has to satisfy two eigenvalue equations coming from the leading order limit of (\ref{eq:CasimirEquation1})
\begin{align}
&\displaystyle {\!\!\!\!\big[(1-\xi_1^2)\partial_{\xi_1}^2+(1-\xi_3^2)\partial_{\xi_3}^2-(2h-1)(\xi_1\partial_{\xi_1}+\xi_3\partial_{\xi_3})-2(\xi_1\xi_3+\xi_2)\partial_{\xi_1}\partial_{\xi_3}+C_J\big]\mathcal{H}=0}\label{eq:AngulardifferentialEquation}\\
&\displaystyle {\!\!\!\!\big[(1-\xi_2^2)\partial_{\xi_2}^2+(1-\xi_3^2)\partial_{\xi_3}^2-(2h-1)(\xi_2\partial_{\xi_2}+\xi_3\partial_{\xi_3})-2(\xi_2\xi_3+\xi_1)\partial_{\xi_2}\partial_{\xi_3}+C_{J'}\big]\mathcal{H}=0}\nonumber
\end{align} 
with $C_J=J(J+2h-2)$. The solution, $\mathcal{H}_{J,J',m}(\xi_i)$, is a polynomial of degree $J,J'$ and $m$ in $\xi_1,\xi_2$ and $\xi_3$ respectively. It is natural to consider an expansion of $\mathcal{H}$ in powers of $\xi_3$ 
\begin{align}
\mathcal{H}_{J,J',m}(\xi_i) =\sum_{m'=0}^{m}\xi_3^{m-m'} f_{m'}(\xi_1,\xi_2)\label{eq:AngularexpansionH}. 
\end{align}
The action of the differential operators (\ref{eq:AngulardifferentialEquation}) will transform (\ref{eq:AngularexpansionH}) into a differential recurrence relation. For example the first line in (\ref{eq:AngularexpansionH}) becomes
\begin{align}
&\xi_3^m\left[(1-2 h-2 m)\xi_1 \partial_{\xi_1}f_0-\left(\xi_1^2-1\right) \partial_{\xi_1}^2f_0+(J-m) (2 h+J+m-2) f_0\right]\nonumber\\
&+\xi_3^{m-1}\left[\xi_1 (3-2 h-2 m) \partial_{\xi_1}f_1-2 m \xi_2 \partial_{\xi_1}f_0-\left(\xi_1^2-1\right) \partial_{\xi_1}^2f_1\right.\nonumber\\
&\qquad\qquad\left.+(J-m+1) (2 h+J+m-3) f_1\right]+\dots
\end{align}
where the $\dots$ represent subleading terms in $\xi_3$ and there is also a similar equation coming from the other channel $(35)$. The leading order differential equation can be recognized as the equation for the Gegenbauer polynomial in one variable $\xi_1$ with spin $J-m$
\begin{align}
C_{J-m}^{h-1+m}\big(\xi_1\big).
\end{align}
The solution from the equation of the channel $(35)$ is of the same form with the replacement $\xi_1\rightarrow \xi_2, J\rightarrow J'$. Obviously the differential equation does not fix the normalization of the solution. Comparison with (\ref{eq:HangularexplicitComputation}) imposes the normalization to be
\begin{align}
f_0(\xi_1,\xi_2)= \frac{(-1)^{J+m}(J-m)!(J'-m)!}{2^{J+J'-2m}(h-1+m)_{J-m}(h-1+m)_{J'-m}}C_{J-m}^{h-1+m}\big(\xi_1\big)C_{J'-m}^{h-1+m}\big(\xi_2\big).
\end{align} 
The leading order solution $f_0$ will enter as a non-homogeneous term in the differential equation for the the subleading order. The homogeneous solution to the differential equation in the subleading order is solved by the Gegenbauer polynomial
\begin{align}
C_{J-m+1}^{h+m-2}(\xi_1),
\end{align}
while the non-homogeneous part is also solved by a Gegenbauer polymial but with other indices. This indicates that the generic solution is of the form
\begin{align}
\mathcal{H}_{J,J',m}(\xi_i)=\sum_{m'=0}^{m}\xi_3^{m-m'}\sum_{a,b=0}^{m'}r_{a,b,m'}C_{J-m+m'}^{h+m-a-1}(\xi_1)C_{J'-m+m'}^{h+m-b-1}(\xi_2).
\end{align}
We did not try to find the coefficients $r_{a,b,m'}$ in full generality since we have an alternative representation for  $\mathcal{H}$ given by (\ref{eq:HangularexplicitComputation}). However it would be interesting to pursue this further and also try to apply the same ideas to the angular functions relevant to higher-point functions.

Now we notice that all the differential operators $d_{12}^{(i)}$ depend only on the angles (\ref{eq:AngleEquationcrossratios}) and moreover their action on the function $\mathcal{H}$ is simple
\begin{align}
\xi_1\mathcal{H}_{J, J',m}(\xi_i) &= -\mathcal{H}_{J+1,J',m}(\xi_i)+\frac{(m-J)(J+m+2h-3)}{4(h+J-2)(h+J-1)}\mathcal{H}_{J-1,J',m}(\xi_i)\nonumber\\
&\qquad -\frac{m(m+h-2)}{2(h+J-2)(h+J-1)}\mathcal{H}_{J-1,J',m-1}(\xi_i)\,,\\
d_{12}^{(i)}\mathcal{H}_{J, J',m}(\xi_i)&=\sum_{n_1,n_2,n_3=-1}^{1}c_{n_1 n_2 n_3}^{(i)}\mathcal{H}_{J+n_1,J'+n_2,m+n_3}(\xi_i)\,,
\end{align}
where the non-zero $c_{n_1 n_2 n_3}^{(i)}$ are given by
\begin{align}
&{\scriptstyle  c_{100}^{(1)}=(2J-J'+2m),\, \ \ c_{-100}^{(1)}=-\frac{(J - m) (12 + 8 h^2 + 2 J^2 + J' - J (10 + J' - 4 m) - 4 m + 3 J' m - 4 m^2 + 4 h (-5 + 2 J + m))}{4(h+J-1)(h+J-2)}}\,,\nonumber\\
&{\scriptstyle c_{-10-1}^{(1)}=-\frac{m(6 h + 4 J + J' - 2 m-6)( h + m-2)}{2(h+J-1)(h+J-2)},\, \ \ \ c_{-101}^{(1)}=-\frac{(J' - m) (m-J) (1 - J + m)}{4(h+J-1)(h+J-2)}}\,,\nonumber\\
&{\scriptstyle c_{101}^{(1)}=(J'-m),\, \ \ c_{100}^{(2)}=-J,\, \ \ \ c_{-100}^{(2)}=\frac{(2h+J-2)(J-m)(2h+J+m-3)}{4(h+J-1)(h+J-2)}},\nonumber\\
&{\scriptstyle c_{-10-1}^{(2)}=\frac{m(h+m-2)(2h+J-2)}{2(h+J-1)(h+J-2)},\, \ \ \ c_{100}^{(3)}=J(J'-J-2m),\, \ \ \ c_{101}^{(3)}=J(m-J')},\nonumber\\
&{\scriptstyle c_{-101}^{(3)}=\frac{(2 h + J-2) (J' - m) (J - m) (1 - J + m)}{4(h+J-1)(h+J-2)}\,, \ \ \ c_{-10-1}^{(3)}=-\frac{(2 h + J-2) (4 h + 3 J + J' - 2 m-4) m (h + m-2)}{2(h+J-1)(h+J-2)}},\\
&{\scriptstyle c_{-100}^{(3)}=\frac{(2 h + J-2) (m-J) (6 - 10 h + 4 h^2 - 5 J + 4 h J + J^2 + J' - 
   J J' - 2 m + 2 h m + 3 J m + 3 J' m - 4 m^2)}{4(h+J-1)(h+J-2)} \,.}
\end{align}
Obviously there are similar relations for the differential operators $d_{35}^{(i)}$. These are obtained by replacing $J\rightarrow J'$.  

These properties of the $\mathcal{H}$ function make the action of the Casimir differential equation on the ansatz (\ref{eq:ConformalblockAnsatzHfunction}) particularly simple
\begin{align}
0=&\sum\big[\mathcal{D}_{12}^{(0)}+\mathcal{D}_{12}^{(1)}-C_{\Delta_k,J}\big]a_{n,n',j,j',m'}s_1^{\Delta_k+n}s_2^{\Delta_{k'}+n'}\mathcal{H}_{jj'm'}(\xi_1,\xi_2,\xi_3)\nonumber\\
=&\sum a_{n,n',j,j',m'}s_2^{\Delta_{k'}+n'}s_1^{\Delta_{k}+n}\big[2 s_1  d_{12}^{(3)}\mathcal{H}-2 (\Delta +n) s_1  d_{12}^{(1)}\mathcal{H}-2 s_1 (\Delta_1-\Delta'-n')  d_{12}^{(2)}\mathcal{H}+ \nonumber\\
&\qquad -(\mathcal{C}_{\Delta +n,J'}-\mathcal{C}_{\Delta ,J})\mathcal{H}+2 \xi_1 (\Delta +n) s_1^{n+1}  (\Delta -\Delta_1+\Delta'+n+n')\mathcal{H}\big] 
\end{align}
Now one can use that $\mathcal{H}$ functions with different indices $jj'm'$ are orthogonal to each other to write a recurrence relation between the unknown coefficients $a_{n,n',j,j',m'}$ in (\ref{eq:ConformalblockAnsatzHfunction}). 

We verified that this method gives the same result as the one where we use the formal expression for the OPE  twice on the five-point function\footnote{The coefficients $a_{n,m,q}$ associated with a given primary in this expression for the OPE can be obtained by solving the equation 
\begin{align}
\langle\mathcal{O}_1(x_1)\mathcal{O}_2(x_2)\mathcal{O}_k(x_3,z_3) \rangle = \frac{F^{(12k)}(x_{12},\partial_{x_1},D_{z_1})}{(x_{12}^2)^{\frac{\Delta_1+\Delta_2-\Delta_k+J}{2}}}\langle \mathcal{O}_k(x_1,z_1) \mathcal{O}_k(x_3,z_3)\rangle .
\end{align}
}
\begin{align}
&\mathcal{O}_1(x_1)\mathcal{O}_2(x_2) \approx \sum_{k}\frac{C_{12k}}{(x_{12}^2)^{\frac{\Delta_1+\Delta_2-\Delta_k+J}{2}}}\big[F^{(12k)}(x_{12},\partial_{x_1},D_{z_1})\mathcal{O}_{k,J}(x_1,z_1)\big]\,,\\
&F^{(12k)}(x_{12},\partial_{x_1},D_{z_1})= \sum_{n,m=0}^{\infty}\sum_{q=0}^la_{n,m,q}(x\cdot D)^{l-q}(x^2\partial_{y}\cdot D)^q (x\cdot \partial_{y})^n(x^2\partial_{y}^2)^m\,, \nonumber
\end{align}
with $a_{0,0,0}=l!(h-1)_l$.

\bibliography{20prime5ptfunction} 
\bibliographystyle{utphys}

\end{document}